\documentclass[a4paper]{cas-sc}

\usepackage[authoryear,longnamesfirst]{natbib}
\usepackage{amsthm}
\usepackage{algorithm}
\usepackage{algpseudocode}
\usepackage{caption}
\usepackage{subcaption}
\usepackage{rotating}
\usepackage[super]{nth}
\usepackage{float}

\ExplSyntaxOn
\cs_set:Npn \__first_footerline:
{
  \group_begin:
  \small
  \sffamily
  \ifnum\theblind>0\relax
  \else
    \__short_authors: :~
  \fi
  { \rmfamily \itshape Preprint }
  \group_end:
}
\ExplSyntaxOff

\newcommand{\densityfigw}{0.24\textwidth}
\newcommand{\densityfigh}{0.18\textheight}
\newcommand{\rowlab}[1]{\raisebox{0.062\textwidth}{\rotatebox{90}{\small #1}}}

\newtheorem{theorem}{Theorem}
\newtheorem{corollary}[theorem]{Corollary}
\newtheorem{lemma}[theorem]{Lemma}
\newtheorem{remark}[theorem]{Remark}

\definecolor{orange}{rgb}{1,0.5,0}
\definecolor{purple}{rgb}{0.55,0.2,0.90}

\newcommand{\RR}{\mathbb{R}}

\begin{document}
\let\WriteBookmarks\relax
\def\floatpagepagefraction{1}
\def\textpagefraction{.001}

\shorttitle{Bandwidth-free density estimation for grouped data}
\shortauthors{Danisman, Jankowski, and de Souza}

\title[mode=title]{Bandwidth-free nonparametric density estimation for grouped data}

\author[1]{Furkan Danisman}
\cormark[1]
\fnmark[1]
\ead{furkan.danisman@mail.utoronto.ca}
\credit{Conceptualization, Methodology, Software, Writing}

\affiliation[1]{organization={University of Toronto},
            city={Toronto},
            state={ON},
            country={Canada}}

\author[2]{Hanna Jankowski}
\credit{Conceptualization, Methodology, Supervision, Writing}

\affiliation[2]{organization={York University},
            city={Toronto},
            state={ON},
            country={Canada}}

\author[3]{Camila P. E. de Souza}
\credit{Conceptualization, Methodology, Supervision, Writing}

\affiliation[3]{organization={Western University},
            city={London},
            state={ON},
            country={Canada}}

\cortext[1]{Corresponding author}
\fntext[1]{Also affiliated with the Vector Institute for Artificial Intelligence, Toronto, ON, Canada.}

\begin{abstract}
In some situations, data is collected under systematical and technical constraints due to uncertainty in experimental reports, intermittent measurements, confidentiality, and non-detects. For this reason, it might not be possible to retrieve or receive the data in a conventional format but rather in a grouped form where only the number of occurrences is known within intervals. The challenge is to estimate the density of the underlying ungrouped data based on the observed grouped data with no information regarding the underlying distribution. To overcome this problem, this study introduces a mean-adjusted log-concave (MALC) density estimation method for univariate grouped data, aiming to provide a bandwidth-free non-parametric approach that does not rely on specific distributional assumptions. The performance of the MALC method is evaluated through simulations across various distributions with different sample sizes and grid widths. The results demonstrate the robustness and effectiveness of the MALC approach in grouped data analysis, offering a broader range of applications over traditional methods.
\end{abstract}



\maketitle

\section{Introduction}
\label{sec:intro}

In some cases, the collected data may not fully meet the researcher's goals due to systematic or technical limitations in the data collection process, such as uncertainty in experimental reports, intermittent measurements, confidentiality, and non-detects  \citep{ferson2004summary, lindsey1998methods, osegueda2002non, wieringa2021data, zhang2010interval}. Consequently, under such circumstances, data may be retrieved or received only in a grouped manner, i.e., when for a particular variable, only the intervals and the frequency of observations falling into each interval are known.

Density estimation for grouped data has been studied from both parametric and nonparametric perspectives. Under a parametric approach, the problem reduces to estimating the parameters from a given distribution, such as Gaussian, exponential, Weibull, or log-normal \citep{tallis1967approximate,crafford2007statistical,xiao2016estimation,Zahra2022}. Second to the choice of parametric family, there are various types of estimation methods. These include, for example,  maximizing the log-likelihood via numerical techniques \citep{tallis1967approximate}, maximum likelihood via the Expectation-Maximization (EM) algorithm \citep{mclachlan1988fitting,cadez2002maximum,teimouri2021algorithm} and Monte Carlo EM algorithms \citep{Zahra2022}, Bayesian inference \citep{heitjan1989inference,heitjan1991ignorability,rubio2011inference}, bootstrap-based estimation \citep{velez2015bootstrap}, robust modification to maximum likelihood \citep{victoria1997robust}, and moment estimation \citep{xiao2016estimation}.

Within the non-parametric framework, \citet{rizzi2015efficient} proposed an adaptable method for ungrouping data by maximizing a penalized likelihood. When individual observations are fully available, a widely used nonparametric tool for density estimation is the classical Parzen–Rosenblatt kernel estimator \citep{rosenblatt1956remarks,parzen1962estimation}. Several authors have extended this estimator to handle grouped data \citep{titterington1983kernel,scott1985kernel,reyes2016nonparametric}. In further kernel-based developments,  \citet{BlowerKelsall2002} proposed a nonlinear kernel estimator by replacing the unobserved within-bin observations with their expected kernel contributions. \citet{binnednp2019} considered density estimation from the grouped empirical distribution. \citet{KernSmoothPackage} approximate the ordinary kernel estimator by replacing observations with bin summaries before smoothing. A key challenge of kernel-based approaches is their dependence on the bandwidth, which affects the smoothness and accuracy of the estimate. For grouped data, this issue has been studied by \citet{reyes2017bandwidth}. More recently, \citet{lee2024tuning} proposed a maximum-entropy density estimator for grouped data to avoid bandwidth selection altogether. This highlights the need for additional nonparametric methods for grouped data, particularly those that do not rely on tuning parameters.

Log-concave density estimation has emerged in recent decades as a robust alternative to kernel density estimation with similar convergence properties but without the need to select a bandwidth.   For a nice review, we refer the reader to \citet{revwalt} or \citet{revsam}.   Log-concave density estimation has been studied in both the continuous setting \citep{dumbgen2009logcondens, Cule08, chenSam2013} and the discrete setting \citep{JankowskiDiscr2013}.  Here, we consider applying the log-concave paradigm to discretized data, which lies somewhere in between these two setups. To this end, we propose a nonparametric, adjusted log-concave density estimation approach for univariate continuous grouped data.  The advantage of using the log-concave assumption is that we do not need to choose a specific parametric family, and the class is sufficiently broad that it encompasses many popular densities.   Furthermore, the log-concave density maximum-likelihood estimator is fully automatic, thereby avoiding the need for bandwidth selection.   


This paper is organized as follows. Section \ref{method} gives the proposed mean-adjusted log-concave density estimation (MALC) algorithm for univariate grouped data.  In this section, we also establish some basic properties of the method, including consistency.    Section~\ref{sec:methods} provides the necessary background to Section~\ref{method}.  Here, we provide necessary definitions and discuss the problem at hand, noting in particular the lack of identifiability in the nonparametric paradigm. We also show how we can use the standard EM algorithm under a Gaussian assumption to recover the true population mean with a high level of accuracy, even when the true population is not Gaussian. This section contains our main theoretical results.   All proofs appear in the Supplementary Material B.

In Section \ref{sec:results}, we present the results of a numerical simulation study as well as an application of our method to a human mortality dataset.  In our simulation study, we compare the proposed MALC to the kernel-based methods proposed by \citet{BlowerKelsall2002}, \citet{binnednp2019}, and \citet{KernSmoothPackage}. Notably, we consider scenarios in which the true density is log-concave as well as those in which it is not.  As observed previously in \citet{Cule08}, the log-concave assumption is robust, and can even outperform other nonparametric methods under misspecification for small to medium sample sizes.   A more complete comparison and summary of our findings is given in Section~\ref{sec:discussion}.    

\section{Assumptions and preliminary results}
\label{sec:methods}

Assume that $X_1, \ldots, X_n$ are $n$ independent and identically distributed (IID) random variables such that each variable has density $f$. Let $k$ denote a fixed integer and $a_1, \ldots, a_{k+1}$ represent the group boundaries such that $a_1 < a_2 < \ldots < a_{k+1}$.   We assume that we do not observe the variables $X_1, \ldots, X_n$ directly. Instead, we observe their grouped counts, $n_j= \sum_{i=1}^n \mathbb{I}_{[a_j, a_{j+1})}(X_i)$, where $\mathbb{I}_A(x)$ denotes the standard indicator function, equal to one when $x\in A$ and zero otherwise.   Thus, the values $n_j$ denote the number of observations falling within the interval $[a_j, a_{j+1})$ for $1 \leq j \leq k$. Note that we have $\sum_{j=1}^{k} n_j = n$.  Although in practice the exact boundaries of the intervals may not be closed/open as depicted, this should not make a difference in the analysis, as we assume that the data is generated by a continuous density. 

We may therefore describe our observations as $n_1, \ldots, n_{k}$ from a multinomial distribution with parameter $n$ and probabilities $p_1, \ldots, p_k,$ with 
\begin{eqnarray}\label{line:def_problem}
p_j &=& \int_{a_j}^{a_{j+1}} f(x)dx, \ \ \ \mbox{for} \; j=1,\ldots, k.
\end{eqnarray}
We will use the notation $p_{n,j}=n_j/n$ to denote the proportions observed in each interval $[a_j, a_{j+1})$.  Notably, we assume that the grid width is fixed and does not vary with the sample size. 

Our goal is to recover the density $f,$ assuming that $f$ belongs to the class of log-concave densities in $\RR$. Recall that a probability density function $f$ on $\mathbb{R} $ is log-concave  if:
\begin{eqnarray*}
f(x) &=& \exp(\varphi(x)) \ \ \ x\in \RR,
\end{eqnarray*}
where $\varphi$ is a concave function  $\varphi: \RR \to [-\infty, \infty)$.  If the variables $X_1, \ldots, X_n$ are observed, the density $f$ is typically estimated using maximum likelihood.  As the observations are IID in our setup, the maximum likelihood estimator (MLE) of the log-concave density $f$, $\widehat f_n$, is given by $\exp(\widehat \varphi_n)$ where $\widehat \varphi_n$ is the concave function which maximizes the criterion 
\begin{eqnarray*}
    l(\varphi) &=& \frac{1}{n} \sum_{i=1}^{n} \varphi(X_i) - \int e^{\varphi(x)} dx,
\end{eqnarray*}
over all concave functions on $\RR$.  One can show that this modified likelihood criterion is maximized at a proper log-concave density function. 

This estimator was first studied in \citet{dumbgen2009logcondens}, where existence and uniqueness of the MLE, as well as a computational algorithm, were established. The authors also show that the MLE $\widehat f_n$ is a consistent estimator of the true log-concave density $f$. Further asymptotic properties, including local and global convergence rates of order $n^{-2/5}$ under certain assumptions, are studied in \citet{brwlogcondens} and \citet{doss1}. For a broader overview of log-concave density estimation, see \citet{revsam, revwalt}.

As we assume that the variables $X_1, \ldots, X_n$ are unobservable, we need to first discuss the log-concave model under this assumption.

\subsection{Estimation of a log-concave mass function and model identifiability for grouped data}

\noindent Our goal is to find a non-parametric estimator of the density $f$ in \eqref{line:def_problem}, given the empirical estimates of the probabilities $p_j, j=1, \ldots, k.$  If we assume that the density $f$ belongs to a parametric family, the problem should be identifiable.   However, without such restrictions, the problem will in general be non-identifiable.   In this section, we show that this continues to be the case if $f$ belongs to the class of log-concave densities.   As this question is closely linked to definition of a log-concave mass function, we begin by reviewing this concept first.

\citet{JankowskiDiscr2013} studied the log-concave mass function estimator on the univariate grid.   A probability mass function (PMF) $\{p_j, j \in \mathbb N\}$ is discrete log-concave if $\varphi_j = \log p_j$ has a negative discrete Laplace transform.   Recall that the discrete Laplace transform, $(\Delta \varphi)_j$, is defined as:
\begin{eqnarray*}
(\Delta \varphi)_j &=& \varphi_{j+1}+\varphi_{j-1}-2\varphi_j\\
&=& \left\{\varphi_{j+1}-\varphi_j\right\}-\left\{\varphi_j-\varphi_{j-1}\right\}.
\end{eqnarray*}
\citet{JankowskiDiscr2013} derive the maximum likelihood estimator of a log-concave mass function (assuming random sampling) and establish consistency and rates of convergence for this estimator. They also show that the MLE is unique and derive an algorithm, based on the active set algorithm, to compute the MLE. See \citet{activeset} for details on the active set algorithm. The algorithm has been implemented in the package \texttt{logConDiscr()}. 

From the work in \citet{JankowskiDiscr2013}, the following then follows:  suppose that the empirical probabilities $p_{n,j}$, $j=1, \ldots, k$ have been observed.   Then the closest (in the Kullback-Leibler sense) log-concave PMF exists and is unique.   We denote this closest PMF as $\widehat{p}_n$.   Practically, $\widehat p_n$ can be found using the R package \texttt{logConDiscr()}.  Note that it follows from \citet{JankowskiDiscr2013} that $\sum_{j=1}^k p_{n,j} a_j=\sum_{j=1}^k \widehat p_{n,j} a_j.$  Another key result from \citet{JankowskiDiscr2013}  is the following lemma. 

\begin{lemma}
Suppose that the grid values (or interval boundaries) $a_1, \ldots, a_{k+1}$ are such that
\begin{eqnarray*}
a_{j+1}-a_j &=& \delta >0
\end{eqnarray*}
for some $\delta \in \RR_+$ and for all $j=1, \ldots, k$ (where we can allow for $k=\infty$ in this statement).   We refer to this as a ``uniform grid". If $f$ is a log-concave density with
$p_j=\int_{a_j}^{a_{j+1}}f(x)dx$  and $\sum_j p_j = 1$ then $\{p_j\}$ is a discrete log-concave probability mass function.
\end{lemma}
It follows that, if we observed only the grouped version of IID log-concave data, then, as long as the grouping is on a uniform grid, the resulting probabilities $p_i$ will be discretely log-concave.   Of course, the reverse is not true.   That is, if the grouped $p_i$ is discrete log-concave, we cannot guarantee that the density prior to grouping is also log-concave.   More importantly, if we assume that the ungrouped density is log-concave, given the grouped probability mass function, we cannot uniquely recover the ungrouped density function. 

\begin{lemma}\label{lem:unID}
Suppose that $a_1 < \ldots < a_{k+1}$ is a uniform grid.   Let 
\begin{eqnarray}\label{line:defrelationship}
p_j&=&\int_{a_j}^{a_{j+1}} f(x)dx
\end{eqnarray}
where $p=\{p_j\}$ is a discrete log-concave probability mass function.   Then, for a fixed $p,$ there can be more than one log-concave density $f$ satisfying $\eqref{line:defrelationship}$.
\end{lemma}
The above lemma is proved by example.  Define two densities $f_1$ and $f_2$ 
\begin{eqnarray*}
f_1(x) &=& -\frac{8}{59}\, x^2 + \frac{20}{59}\,x + \frac{41}{177} \ \ \ \ \text{if } \ \ 0\leq x \leq 3,\\\\
f_2(x) &=& \left\{
\begin{array}{ll}
     \frac{14}{59} \,x + \frac{15}{59} & \ \ \text{if } \ \ 0 \leq x < 1 \\\\
    -\frac{8}{59}\,x + \frac{37}{59}  & \ \ \text{if } \ \ 1 \leq x < 2 \\\\
    -\frac{1}{59}\,x + \frac{53}{59} & \ \ \text{if } \ \ 2 \leq x \leq 3
\end{array} \right.
 \end{eqnarray*}
Clearly, the densities have the same support and both are log-concave.   Furthermore, a quick calculation reveals that $p_1=\int_0^1 f_i(x)dx = 21/59, p_2=\int_1^2 f_i(x)dx = 25/59$ and $p_3=\int_2^3 f_i(x)dx = 13/59$ for both $i=1$ and $i=2.$

Lemma~\ref{lem:unID} implies that our stated problem of recovering the log-concave density $f$ from the multinomial observations $n_1, \ldots, n_k$ is unidentifiable.  We will return to this issue in Section~\ref{sec:consistent}.  {Again, it is important to note that the lack of identifiability is present for all nonparametric methods for grouped data.} 

A key component of our approach is that, as a first step, we use the raw data to estimate the true population mean.  This step is discussed in the next section.

\bigskip

\subsection{Mean Recovery}

In order to estimate the true mean of the unobserved data, we assume that the data follow a Gaussian distribution and applying the EM algorithm, as in \citet{mclachlan1988fitting,teimouri2021algorithm,Zahra2022}. Although Gaussian densities are log-concave, we expect the Gaussian assumption to be incorrect in general. Nevertheless, the motivation for using this approach is that, under the Kullback--Leibler (KL) divergence, the Gaussian density $g$ closest to a fixed density $f_0$ is the Gaussian density with mean $\mu_0=\int x f_0(x)\,dx$. Recall that the KL divergence is defined as
\begin{eqnarray*}
\rho_{KL}(g\vert f_0) &=& \int f_0(x) \log f_0(x) \, dx - \int f_0(x) \log g(x) \, dx.
\end{eqnarray*}
For a log-concave density, the first term $\int f_0 \log f_0 \, dx$ exists. Therefore, minimizing the KL divergence over some class of densities $g\in \mathcal G$ is equivalent to minimizing the second term. Taking $g=g_\mu$ to be a Gaussian density with mean $\mu$ and fixed variance $\sigma^2>0,$ we can calculate
\begin{eqnarray}\label{line:KL}
-2 \int f_0(x) \log g_{\mu}(x)dx &=& -\frac{1}{\sigma^2}\int (x-\mu)^2f_0(x)dx - \log \sigma^2 +c,
\end{eqnarray}
for some constant $c.$ Minimizing this expression with respect to $\mu$ yields the result. We assume that the variance is known and therefore suppress its dependence in the notation. Because the data are grouped, the above relationship no longer holds exactly, even asymptotically. However, the error due to grouping is quantifiable as we demonstrate in what follows. 

The log-likelihood for the grouped data under IID sampling and assuming a Gaussian density is given by 
\begin{eqnarray}\label{line:loglik}
\ell_n(\mu) &=& \sum_{j=1}^k p_{n,j} \ \log \int_{a_j}^{a_{j+1}}  g_{\mu}(x)dx. 
\end{eqnarray}
Note also the mild abuse of notation as the log-likelihood depends on the data (via the empirical weights $\{p_{n,j}\}$) as well as the pre-defined grid $\{a_j\}$.  The partial derivatives of $\ell_n(\mu)$ can be calculated easily as 
\begin{eqnarray*}
\partial_\mu \ell_n(\mu) & = & \sum_{j=1}^k p_{n,j} \frac{\int_{a_j}^{a_{j+1}}  \frac{x-\mu}{\sigma} g_{\mu}(x)dx}{\sigma  \int_{a_j}^{a_{j+1}}  g_{\mu}(x)dx} \ \ = \ \ \sum_{j=1}^k p_{n,j} \frac{\int_{\alpha_j}^{\alpha_{j+1}}  u \phi(x)dx}{ \int_{\alpha_j}^{\alpha_{j+1}}  \phi(x)dx},
\end{eqnarray*}
where $\alpha_j=(a_j-\mu)/\sigma$ for $j=1, \ldots, k+1.$  We also use $\phi$ to denote the density of the standard normal random variable. 
Let $f_0$ denote the true log-concave density of the unobserved data $X_1, \ldots, X_n.$  Our first result is the following.

\begin{remark}\label{rem:grid}
In order to consider any large sample behaviour, we assume that, in general, the grid need not be finite.  That is, let $a_j, j\in \mathbb Z$ denote a countable set of strictly increasing values in $\RR$ such that $\cup_j [a_j, a_{j+1}) \subseteq \RR$.  As such, the number of intervals $k$ need not be finite.  However, for a finite sample size $n$ or for finite support, we can use $k<\infty,$ by considering only the non-empty intervals.   In general, we make no such assumption. 
\end{remark}

\begin{theorem}\label{thm:mle_exists}
Assume that $\delta<2\sigma$.  The maximizer $\widehat \mu_n$ of the log-likelihood \eqref{line:loglik} exists and is unique. Furthermore, as $n\rightarrow\infty,$ $\widehat \mu_n\rightarrow \widehat \mu_0,$ where $\widehat \mu_0$ is the unique solution to the equation
\begin{eqnarray*}
s_0(\mu) \ \ \equiv \ \   \sum_{j=1}^k p_{0,j} \frac{\int_{a_j}^{a_{j+1}}  \frac{x-\mu}{\sigma} g_{\mu}(x)dx}{\int_{a_j}^{a_{j+1}}  g_{\mu}(x)dx} &=& \sum_{j=1}^k p_{0,j} \frac{\int_{\alpha_j}^{\alpha_{j+1}}  u \phi(x)dx}{\sigma  \int_{\alpha_j}^{\alpha_{j+1}}  \phi(x)dx} \ \ = \ \ 0,
\end{eqnarray*}
where $p_{0,j}=\int_{a_j}^{a_{j+1}}f_0(x)dx.$
\end{theorem}
From \eqref{line:loglik}, we obtain the (modified) score function
\begin{eqnarray*}
s_n(\mu)&=& \partial_\mu \ell(\mu).
\end{eqnarray*}

It is possible to verify that the equation in the above theorem is simply the score function used to minimize the Kullback-Liebler divergence between the probability mass function $p_{0,j}$ and $\int_{a_j}^{a_{j+1}}g_{\mu}(x)dx$.

\begin{theorem}\label{thm:mean_reco}
Let $a_j, j=1, \ldots, k+1$ denote a uniform partition with $\delta=a_{j+1}-a_j,$ for all $j.$  Then, as $\delta\rightarrow 0,$ we have that $\widehat \mu_0 \rightarrow \mu_0.$ Furthermore, for a fixed uniform partition, we have that there exists an $N=N(\omega),$ such that for all $n\geq N(\omega),  \ \ \vert\mu_0 - \widehat \mu_n\vert \ \ \leq \ \ \delta.$   Furthermore, 
\begin{eqnarray*}
\vert\mu_0 - \widehat \mu_0\vert \ \ \leq \ \ \delta. 
\end{eqnarray*}
\end{theorem}

This last result is key to our mean recovery method.  We will use the EM algorithm to find the MLE associated with \eqref{line:loglik}. The two theorems above state that this MLE exists and is unique, and converges to the true population mean (without grouping)  as the sample size grows and the grouping becomes more refined.  

Lastly, we note that the EM algorithm works as expected and finds the unique MLE.

\begin{lemma}\label{lem:mle_em}
The EM algorithm maximizes the log-likelihood in \eqref{line:loglik}.
\end{lemma}

\section{Main Method}\label{method}

The proposed MALC density estimator combines the mean recovery approach with log-concave distribution estimation.  As above, let $\widehat \mu_n$ denote the mean estimated using the EM algorithm under the Gaussian assumption.  We also smooth the empirical probabilities $p_{n,j}$ to obtain a discrete log-concave distribution instead. Once we have these two components, we generate continuous random variables with mean $\widehat \mu_n$ that have a ``discretized" distribution matching the log-concave smoothed $p_{n,j}$ probabilities.  Details are given below.

\begin{figure}[p]
    \centering
\includegraphics[width=0.3\textwidth, height=0.22\textheight]{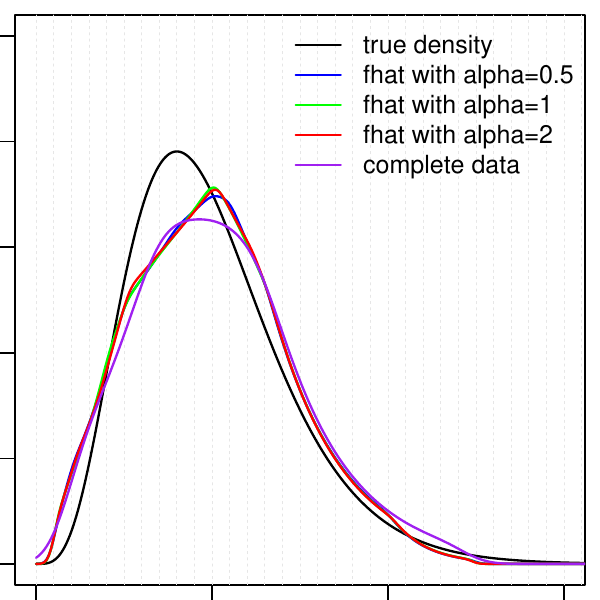}
\includegraphics[width=0.3\textwidth, height=0.22\textheight]{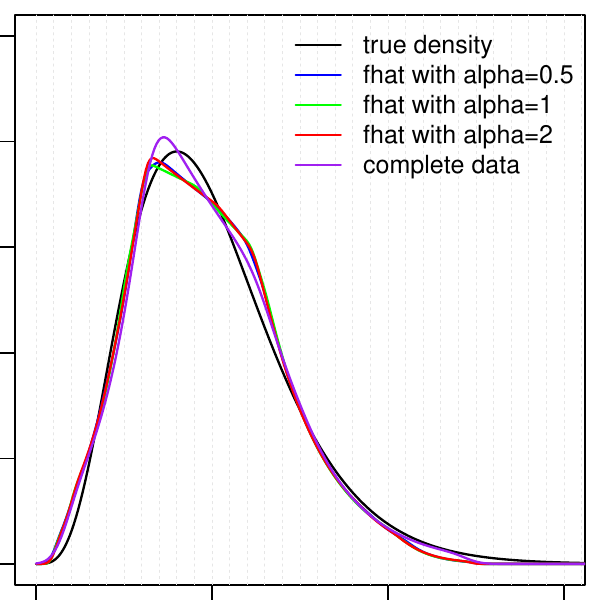}
\includegraphics[width=0.3\textwidth, height=0.22\textheight]{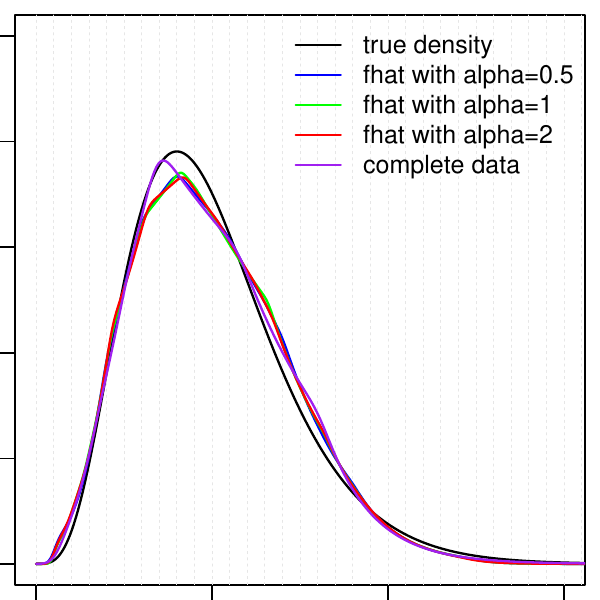}\\
\includegraphics[width=0.3\textwidth, height=0.22\textheight]{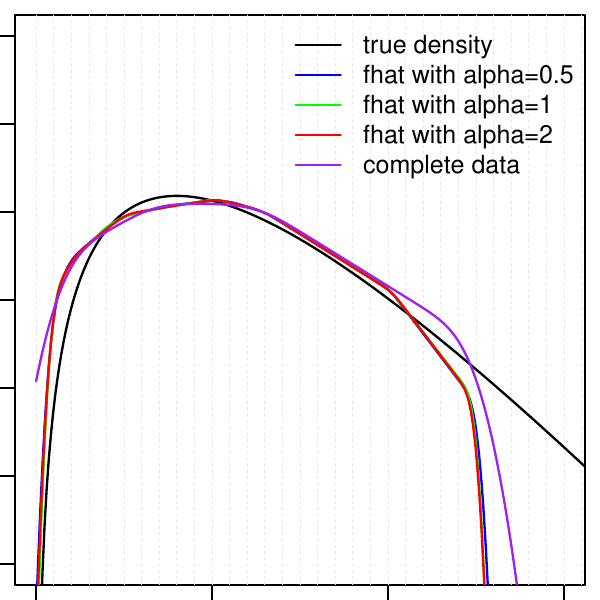}
\includegraphics[width=0.3\textwidth, height=0.22\textheight]{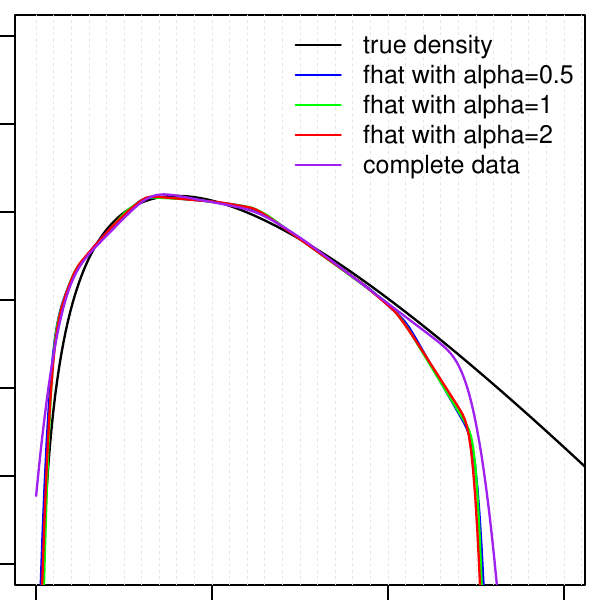}
\includegraphics[width=0.3\textwidth, height=0.22\textheight]{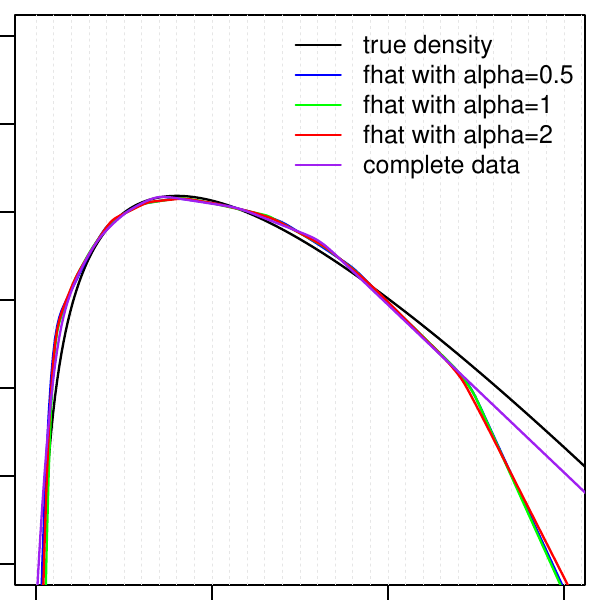}\\
\includegraphics[width=0.3\textwidth, height=0.22\textheight]{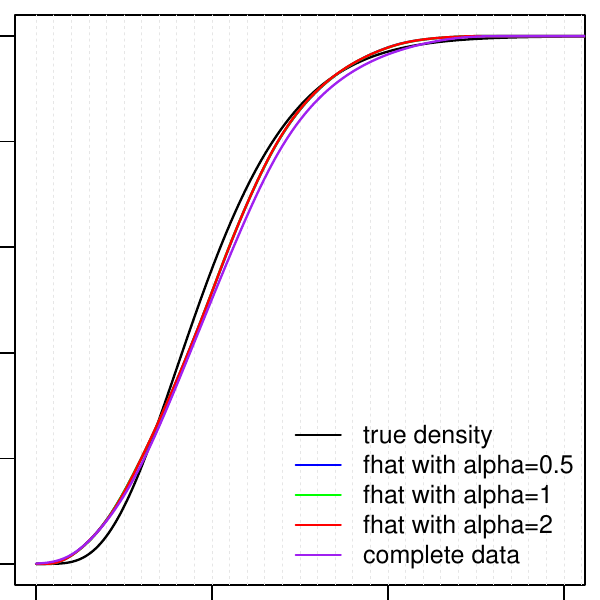}
\includegraphics[width=0.3\textwidth, height=0.22\textheight]{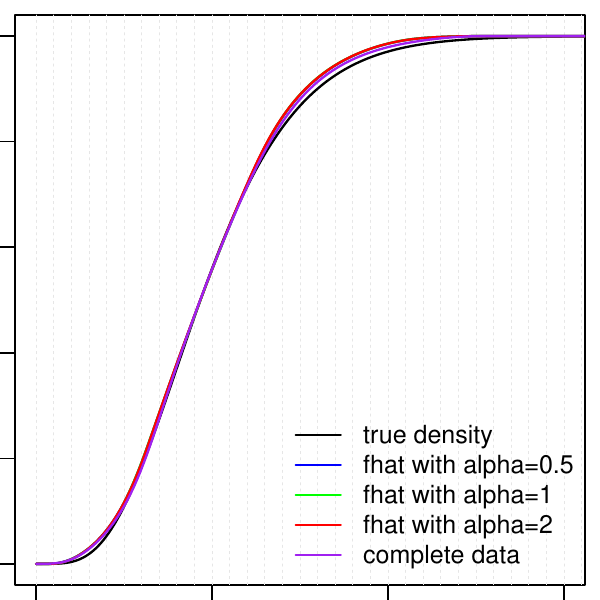}
\includegraphics[width=0.3\textwidth, height=0.22\textheight]{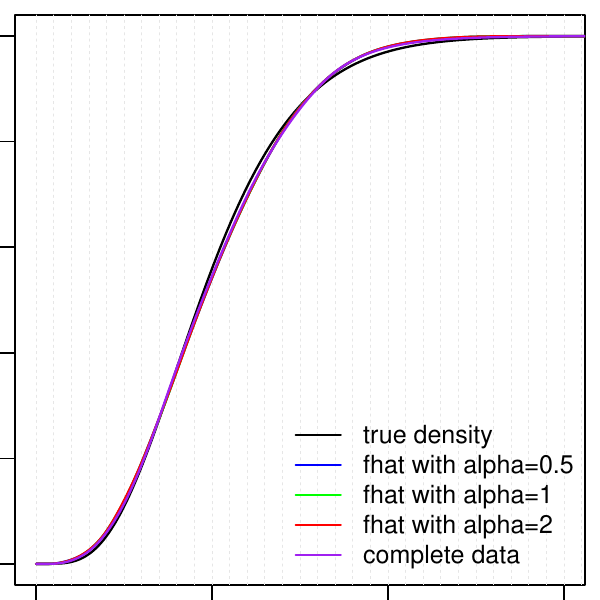}\\\caption{Example of the log-concave $\widehat f_n^{smooth}$ recovered from IID samples from a gamma distribution with shape=6 and rate=1 (top row).  The logarithm of $\widehat f_n$ is given in the middle row, and the associated cumulative distribution functions are in the bottom row.  The grouped interval width is equal to one in all examples and is shown via the light gray vertical grid in all plots.   From left to right the sample sizes are $n=100, 200, 1000$. For comparison, the purple line denotes the log-concave MLE assuming that the complete data (i.e. without grouping) was observed. In the above, the range of the $x$--axis is always the interval $[0,15]$,   The $y$--axis range is $[0,0.25]$ for the densities (top row), $[-10,2]$ for the log-densities (middle row) and $[0,1]$ for the cumulative distribution functions (bottom row).}\label{fig:example_gamma}
\end{figure}

The algorithm begins with the observed values of $p_{n,j}, j=1\ldots, k$, over the intervals given by $a_1, \ldots, a_{k+1}.$   Let $\delta = a_{j+1}-a_j,$ and recall that this is uniform.  
\begin{enumerate}
     \item  Maximize the log-likelihood for grouped data given in \eqref{line:loglik} to obtain $\widehat \mu_n$.  In the density $g_{\mu},$ choose $\sigma$ such that $\delta<2\sigma.$
   \item  Maximize the (modified) log-likelihood function
   \begin{eqnarray*}
   \frac{1}{n} \sum_{i=1}^k p_{n,j} \varphi_i -\sum_{i=1}^k e^{\varphi_i}
   \end{eqnarray*}
   over all discretely concave functions $\varphi.$  Denote the maximizer as $\widehat \varphi_n,$ and set $\widehat p_n=e^{\widehat \varphi_n}$. The resulting $\widehat p_n$ is a log-concave probability mass function.   Note that this step is easily completed using the R package \texttt{logConDiscr()}.   
    \item  
    \begin{itemize}
    \item[--] Let $Y_{n,\delta}$ denote the random variable where $P(Y_{n,\delta} = a_j) = \widehat p_{n,j}, j=1, \ldots, k.$  
     \item[--] Let $ Z_{n,\delta}$ denote a continuous random variable taking values in $[0, \delta)$ and mean given by $\widehat \mu_n - \sum_{j=1}^k p_{n,j} a_j.$   
     \item[--] Finally, define $X_{n,\delta}$ to be the convolution of $ Y_{n,\delta}$ and $ Z_{n,\delta}.$    
    \end{itemize}
    
   With these definitions, we can now define our final step.   
\begin{enumerate}
\item Generate $B$ independent samples of $X_{n,\delta}: X_{n,\delta}^b, b=1, \ldots, B.$   
\item Based on $X_{n,\delta}^b, b=1, \ldots, B,$ find the MLE of a log-concave density using \texttt{logcondens()}.   In \texttt{logcondens()}, we can choose the smoothed version of the density estimator or the original version of the estimator.   For more information on this (fully automatic) smoothing step, we refer to \citet{dumbgen2009logcondens} and \citet{chenSam2013}. 
\end{enumerate}
\end{enumerate}
We denote the resulting estimator of the density $f$ as $\widehat f_{n}$ (without smoothing), and $\widehat f_n^{smooth},$ with smoothing.  Although initially the above algorithm may seem complex, the basic idea is simple.  We obtain a log-concave MLE for the grouped data, and ``ungroup" it by finding a log-concave density closest to the convolution of the grouped distribution with a random perturbation.   The perturbation is designed in such a way that both the convolution and the log-concave density have the mean $\widehat \mu_n,$ which is our best guess at the true mean of the unknown density.   Note that although the convolution $X_{n,\delta}$ does not necessarily have a log-concave density, the MLE does exist and is unique.   Moreover, it is consistent.  We refer to \citet{Cule08} for details.  

In Step 3 of the proposed method, there is a considerable choice in the distribution of $Z_{n,\delta}$.   For convenience, we set $Z_{n,\delta}$ to be $\delta Z_n,$ where $Z_n$ is a beta random variable with parameters $\alpha-\beta$ and $\alpha+\beta.$   The value of $\alpha$ is chosen by the user while $\beta=\beta_n$ is selected so that the mean of $X_{n,\delta}$ is $\widehat \mu_n$ from Step~1.  In Figure~\ref{fig:example_gamma}, we compare $\widehat f_n$ with three different values of $\alpha= 0.5, 1, 2.$  We observe that -- at least in this example -- the choice of $\alpha$ does not appear to have a great impact on the resulting estimator.   Furthermore, comparing the proposed estimator based on grouped data to the MLE based on the true (generally unobserved) data, our proposed estimator's performance is largely driven by the quality of the data.  The main loss of information appears to be around the peak of the density. 

In Step 3(b), we choose the smoothed version of the log-concave density estimator as described in \citet{dumbgen2009logcondens}. However, a non-smoothed option is also available in our package. It is well known that the maximum likelihood estimator under the log-concave assumption preserves the empirical distribution's mean. This is also true for the smoothed version of the MLE.  Therefore, if $B$ chosen to be large, the mean of the estimator $\widehat f_n$ can be made arbitrarily close to $\widehat \mu_n.$  It is also well-known that the likelihood estimator under the log-concave assumption underestimates the variance of the empirical distribution.  The smoothed log-concave estimator is a convolution of the original log-concave MLE with a mean-zero Gaussian density with variance so that the variances of the smoothed density and the empirical distribution are equal.  The variance of the estimated density will therefore be close to (for sufficiently large $B$) \begin{eqnarray*}
\sum_{j=1}^k \widehat p_{n,j} \left(a_j -\sum_{j=1}^k p_{n,j} a_j \right)^2 + \frac{\delta^2}{4} \frac{\alpha^2-\beta^2}{\alpha^2 (2\alpha+1)}
\end{eqnarray*}
See also \citet{chenSam2013} for a more general discussion of smoothing of the log-concave density MLE.

\begin{remark}
The MALC algorithm described here is not the only approach we tested; however, in simulations, it performed best. The other methods we tried are given in the Supplementary Material A1.
\end{remark}

\subsection{Method implementation}

The method has been implemented in the \texttt{R} package \href{https://github.com/FurkanDanisman/MALC/tree/main}{\texttt{MALC}}. The main function takes as input the observed bin counts and the corresponding grid boundaries, and returns the fitted density estimator described above.  The user may choose either the smoothed or the non-smoothed version of the estimator. The package also provides functions for evaluating the fitted density, distribution function, quantile function, and the EM-based mean estimate, as well as a plotting function for the fitted estimator. The comparison methods used in Section~\ref{subsec:simstudy} are included in the same package for reproducibility.

\subsection{Consistency and compatibility}\label{sec:consistent}

We next examine the large sample behaviour of our proposed estimator, both for the cases when $n\rightarrow \infty$ as well as $\delta\rightarrow 0.$   
 
The following statement follows directly from the results of \citet{JankowskiDiscr2013}.
\begin{lemma}\label{lem:discrete_part}
Consider the probability mass function $\{p_{0,j}\}.$  Define $\widehat p_0$ as $\widehat p_0=e^{\widehat \varphi_0},$ where $\widehat \varphi_0$ is the discretely concave function which maximizes 
\begin{eqnarray*}
\sum_j \varphi_{0,j} p_{0,j} - \sum_j e^{\varphi_{0,j}} p_{0,j} +1.
\end{eqnarray*}
Then $\widehat p_0$ exists and is unique, and satisfies $\sum_j \widehat p_{0,j} a_j = \sum_j p_{0,j} a_j$.   Moreover, if $p_{0}$ is discretely log-concave, then $\widehat p_0=p_0.$  Furthermore, as $n\rightarrow \infty,$  $\widehat p_n \rightarrow \widehat p_0,$ almost surely. 
\end{lemma}

Next, define $G_{n,\delta}$ to be the cumulative distribution function of the random variable $Z_{n, \delta}.$   We assume that there exists a random variable $Z_\delta$ with cumulative distribution function $G_\delta,$ such that  
\begin{eqnarray}\label{line:Zcond}
\int_0^\delta \vert \widetilde G_{n,\delta}-\widetilde G_{\delta} \vert(x) dx \rightarrow 0
\end{eqnarray}
as $n\rightarrow \infty.$  Note that this is convergence in Mallow's distance.   From \citet[Lemma 8.3]{bickel84}, this implies that  $\widehat \mu_n -\sum_{j=1}^k p_{n,j}a_j = E[Z_{n,\delta}] \rightarrow E[Z_\delta].$   Therefore, $E[Z_\delta]=\widehat \mu_0 -\sum_j a_j p_{0,j},$ from Theorem~\ref{thm:mean_reco}.  Note also that our specific choice of $Z_{n,\delta}$ in the previous section satisfies condition \eqref{line:Zcond}.     Indeed, the distribution of $G_\delta$ is $\delta Z,$ where $Z$ is a beta random variable with $\alpha$ as originally chosen by the user and
\begin{eqnarray*}
\beta &=& \alpha \left(\frac{\delta}{\widehat \mu_0 - \sum_j a_j p_{0,j}}-1\right).
\end{eqnarray*}

Finally, let $Q_{n,\delta}$ denote the distribution of the convolution of the distributions $\widehat p_n$ and $G_{n,
\delta}$ (this is also the distribution of $X_{n,\delta}$ as defined in the previous section).  Similarly, let $Q_{\delta}$ denote the convolution of $\widehat p_0$ and $G_{
\delta}$.

\begin{theorem}\label{thm:projection}
Let $\widehat f_0$ denote the closest log-concave density to the distribution $Q_\delta$ in the sense of Kullback-Liebler divergence.   That is, the density $\widehat f_0$ is the unique log-concave density which minimizes 
\begin{eqnarray*}
\int \varphi dQ_\delta - \int  e^\varphi dQ_\delta +1,
\end{eqnarray*}
over all concave functions $\varphi.$   Denoting the minimizer as $\widehat \varphi_0,$ we have that $\widehat f_0 = e^{\widehat \varphi_0}.$
With $\delta$ fixed, as $n\rightarrow \infty$, the density estimator $\widehat f_n$ converges to $\widehat f_0$.   That is, 
\begin{eqnarray*}
\limsup_{n} \int_{-\infty}^{\infty} |\widehat f_n-\widehat f_0|(x)dx  &= & 0.
\end{eqnarray*}
\end{theorem}
This result relies heavily on \citet[Theorem 2.15]{DSS2011}, where log-concave projections are shown to be continuous in Mallows' distance.  {Recall that the problem we set out to solve is not indentifiable for a fixed $\delta.$  Theorem~\ref{thm:projection} says that our estimator converges to $\widehat f_0$ in this case.   Although $\widehat f_0$ in general is different from $f_0,$ it is the \emph{closest} density to $f_0$ in the sense defined above.}   As a corollary, we also obtain consistency as $\delta$ decreases.  {That is, }

\begin{corollary}\label{cor:cons}
Suppose that the generating density $f_0$ is log-concave.  Then, as $n\rightarrow \infty$ and $\delta\rightarrow 0,$ the density estimator $\widehat f_n$ is consistent.   That is, 
\begin{eqnarray*}
\limsup_{\delta} \limsup_n \int_{-\infty}^{\infty} |f^*_n-f_0|(x)dx  &= & 0,
\end{eqnarray*}
almost surely, for $f^*_n=\widehat f_n$ or $\widehat f_n^{smooth}.$
\end{corollary}

\begin{remark}
We have thus shown that our estimator is consistent in $n$ and $\delta.$   Ideally, we would also be able to quantify the difference between $\widehat f_0$ and $f_0$ in the well-specified setting and as a function of $\delta.$  Alas, the proof does not provide such a quantification.  One might also ask if the density $\widehat f_0$ is \emph{compatible} when $f_0$ is log-concave.   We say that $\widehat f_0$ is compatible with $f_0$ if $\int_{a_j}^{a_{j+1}}\widehat f_0(x)dx = \int_{a_j}^{a_{j+1}} f_0(x)dx$ for all $j.$  Simulations imply that this does not hold in general. This is most likely a byproduct of our choice of  $Z_{n,\delta},$ which does not depend on the interval $[a_j, a_{j+1})$.  We feel that the ease of the proposed method trumps any attempts to vary $Z_{n,\delta}.$  Furthermore, the performance of our method is favourable as seen in Section \ref{sec:results}. In Supplementary Material A1, we discuss an alternative estimator that was developed to be compatible.  In practice, however, that estimator performed worse than the MALC. 
\end{remark}

\section{Results}
\label{sec:results}


In this section, we present a comprehensive simulation study across multiple data-generating distributions. For each distribution, we evaluate the performance of our method over a range of grid widths and sample sizes and compare it with three kernel-based methods introduced in Section \ref{sec:intro}. The first competing method is the one proposed by \cite{BlowerKelsall2002}, which we refer to as BK2002 in our analyses. The second method, which we call BinnedNP, is implemented in the R package \textit{binnednp} \citep{binnednp2019} proposed by \cite{BarreiroUresEtAl2019}. The third method, KernSmooth, is implemented in the R package of the same name \citep{KernSmoothPackage}, which is based on the book by \cite{WandJones1995}. We note that of these three methods, only BK2002 works on both multivariate and univariate data, while BK2002 and BinnedNP work for non-uniform grids, and KernSmooth does not.  

Since all three competing approaches rely on kernel density estimation, their performance depends heavily on bandwidth selection. The specific bandwidth-selection procedures and additional details for each method are provided in the Supplementary Material C, for ease of reading. For the BK2002 method, no publicly available software implementation is available; therefore, we translated the proposed methodology into an \texttt{R} implementation. For reproducibility, all three kernel-based methods are included in the \texttt{MALC} package and can be called using \texttt{BK2002()}, \texttt{BinnedNP()}, and \texttt{KernSmooth()}.

\subsection{Simulation study}
\label{subsec:simstudy}

We evaluated the performance of the proposed and compared methodologies under both log-concave (normal, gamma, beta, logistic, Laplace, chi-square, Weibull) and non-log-concave (Student’s $t$, log-normal, Pareto) data-generating distributions. We are interested in how performance scales with different sample sizes and standardized grid widths (defined as the grid width relative to the true standard deviation). Therefore, our simulation study consists of two main scenarios: (1) varying sample sizes ($n \in \{10^2, 10^3, 10^4, 10^5, 10^6\}$) with a fixed standardized grid width ($\delta/\sigma = 0.5$); and (2) varying standardized grid widths ($\delta/\sigma \in \{0.5, 1.0, 1.5, 2.0, 2.5\}$) with a fixed sample size ($n = 10^3$). The full parameters and simulation variables are detailed in Table~1 of the Supplementary Material. We generated $100$ independent replications for each combination and compared the $L_2$ norm between the true density and its estimator to evaluate performance.

\subsubsection{Results}
\label{subsec:simresults}

\begin{figure}[!htb]
    \centering
    \begin{tabular}{c | c c c c}
        & \small MALC & \small BK2002 & \small BinnedNP & \small KernSmooth \\[1ex]
        \hline\rule{0pt}{3ex} 
        
        \raisebox{0.08\textwidth}{\rotatebox{90}{\small normal}} &
        \subfloat{\includegraphics[width=0.19\textwidth,height=0.13\textheight,keepaspectratio]{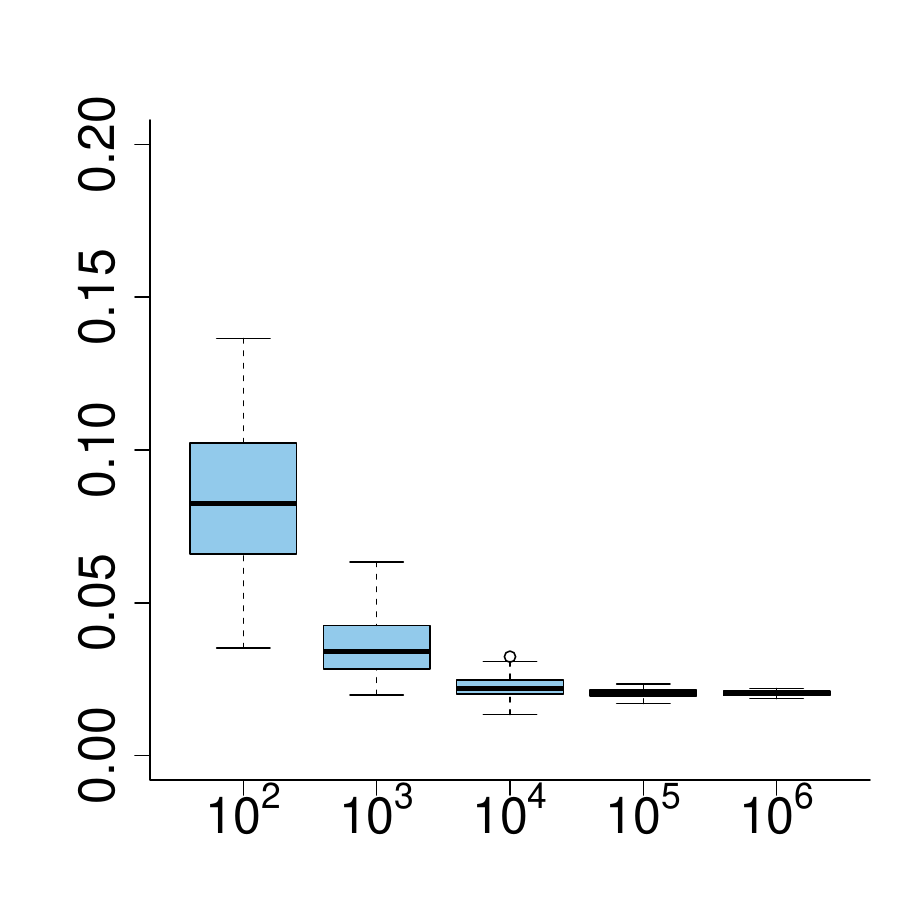}} &
        \subfloat{\includegraphics[width=0.19\textwidth,height=0.13\textheight,keepaspectratio]{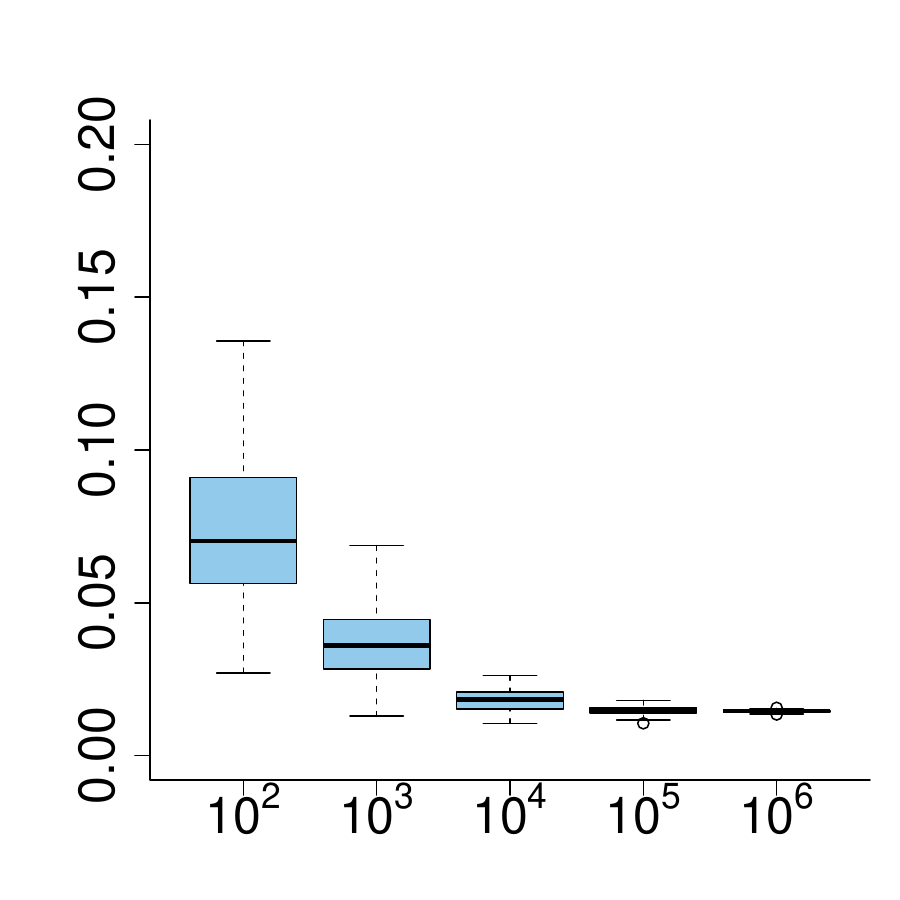}} &
        \subfloat{\includegraphics[width=0.19\textwidth,height=0.13\textheight,keepaspectratio]{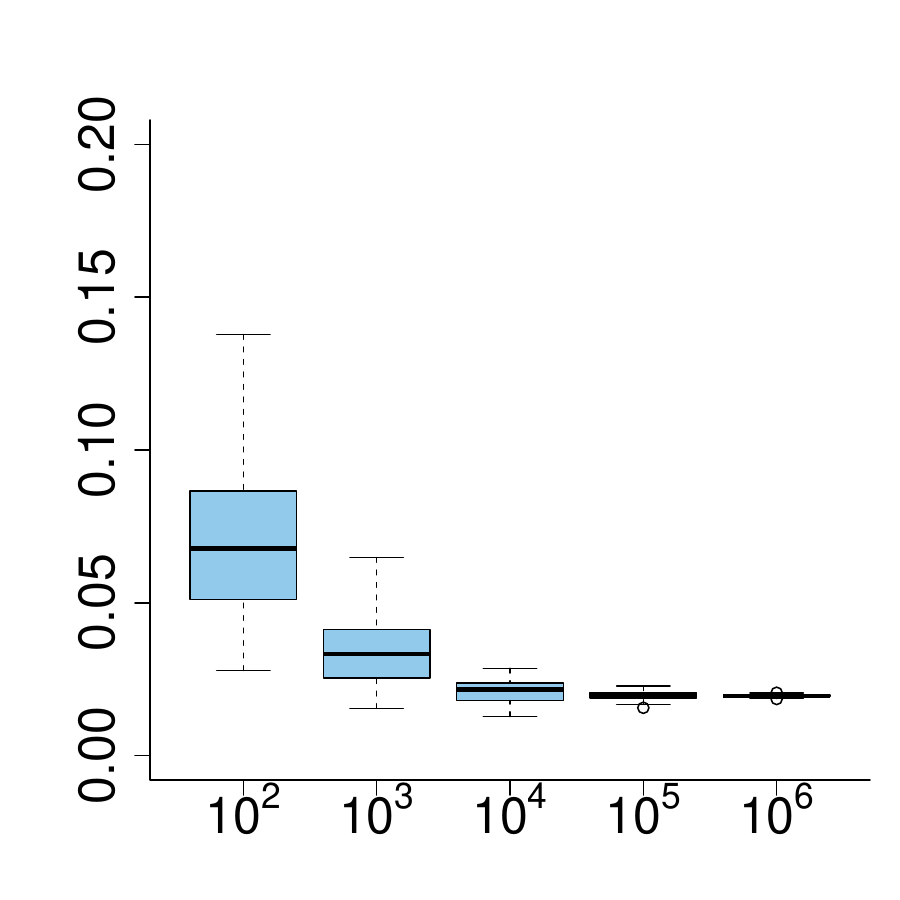}} &
        \subfloat{\includegraphics[width=0.19\textwidth,height=0.13\textheight,keepaspectratio]{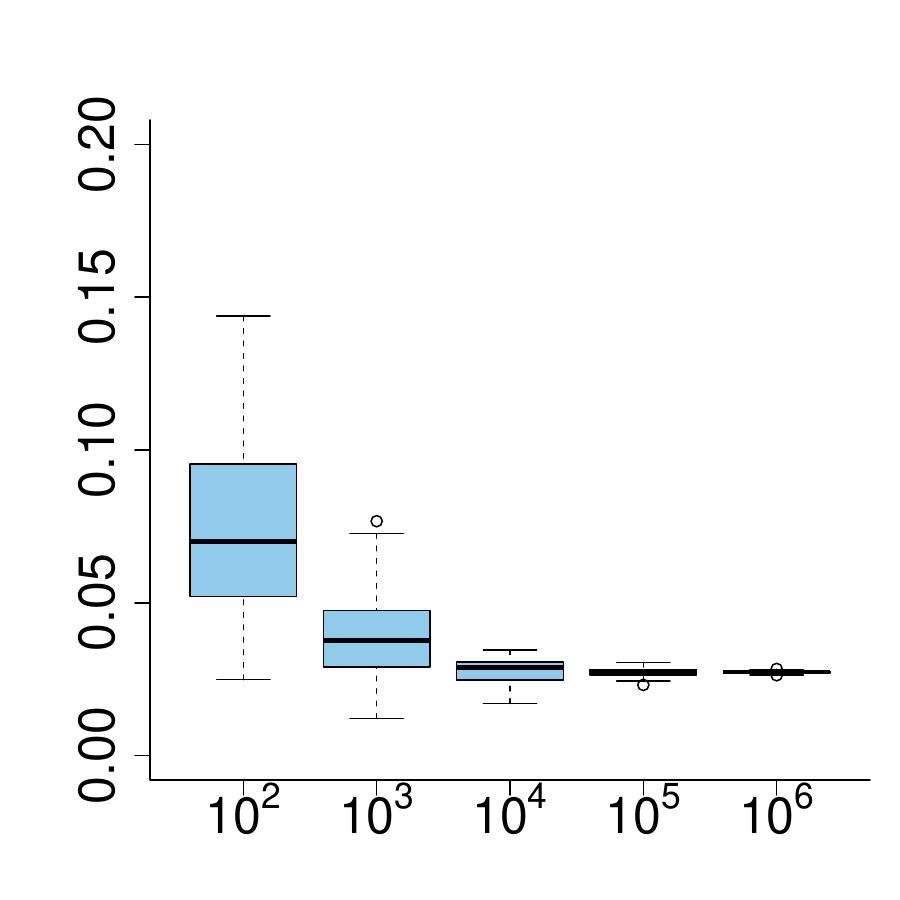}} \\[1ex]

        \raisebox{0.08\textwidth}{\rotatebox{90}{\small Beta}} &
        \subfloat{\includegraphics[width=0.19\textwidth,height=0.13\textheight,keepaspectratio]{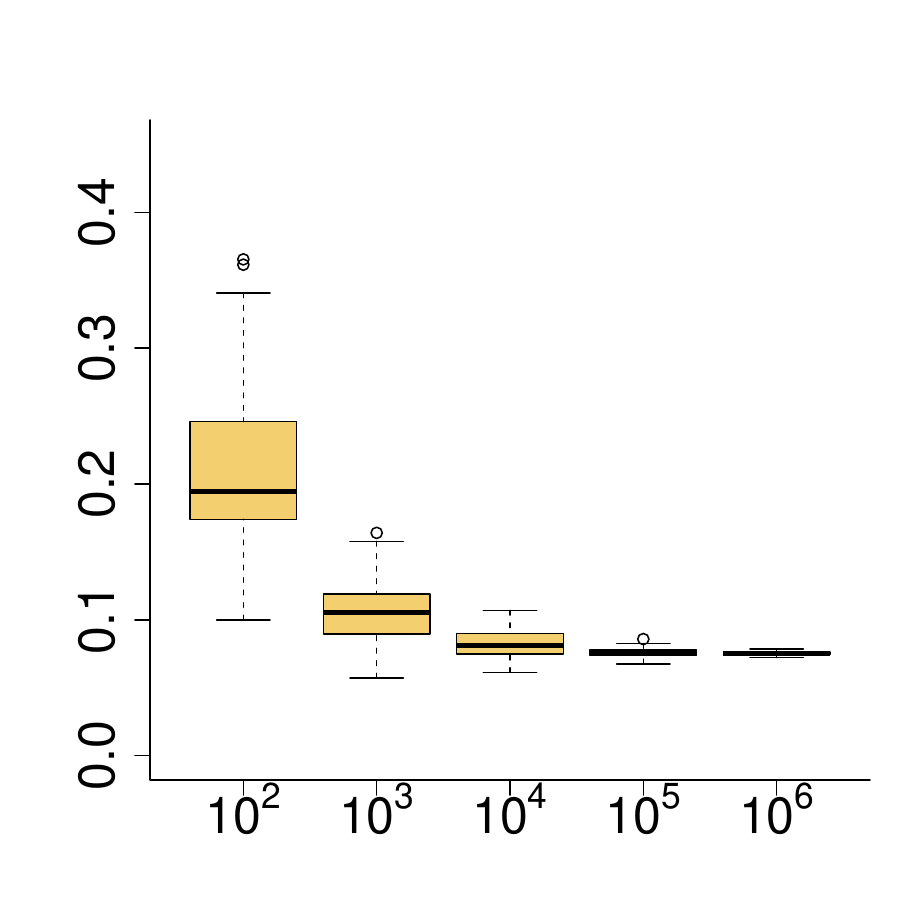}} &
        \subfloat{\includegraphics[width=0.19\textwidth,height=0.13\textheight,keepaspectratio]{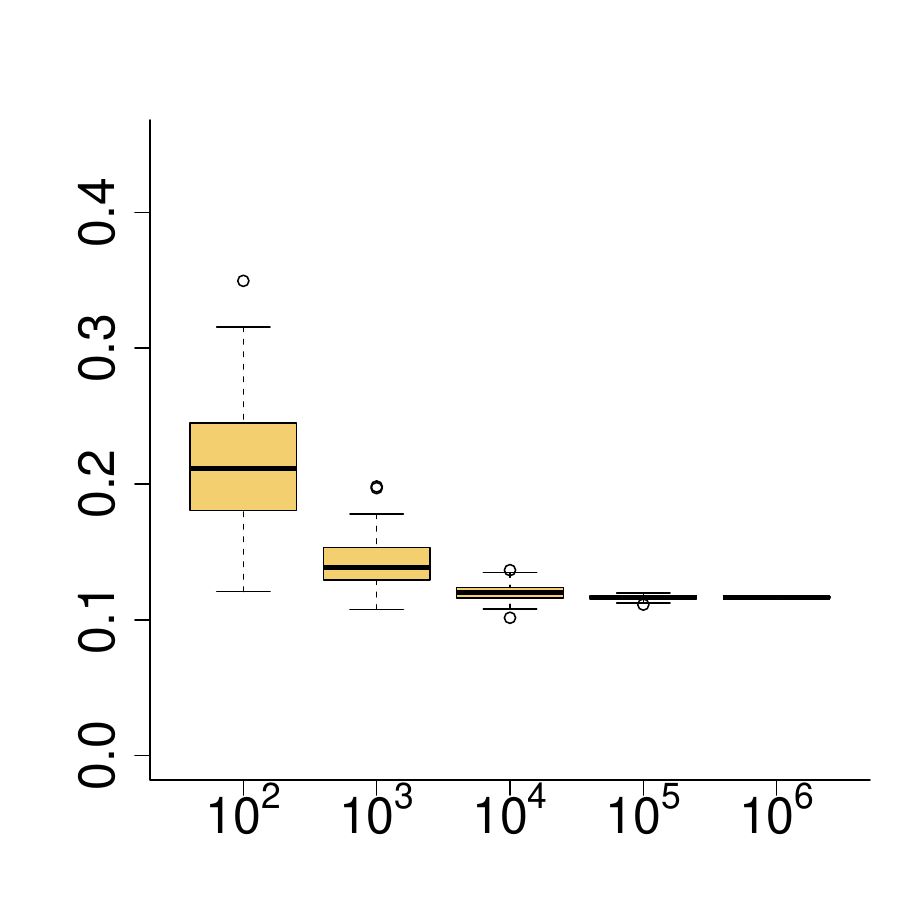}} &
        \subfloat{\includegraphics[width=0.19\textwidth,height=0.13\textheight,keepaspectratio]{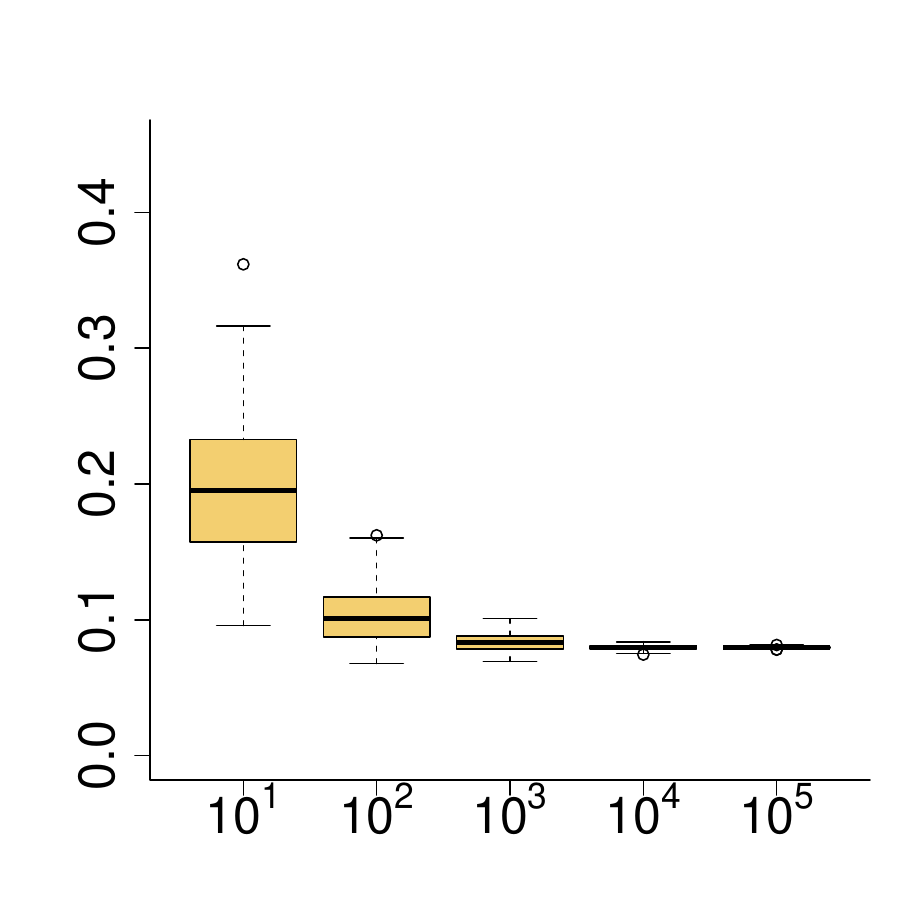}} &
        \subfloat{\includegraphics[width=0.19\textwidth,height=0.13\textheight,keepaspectratio]{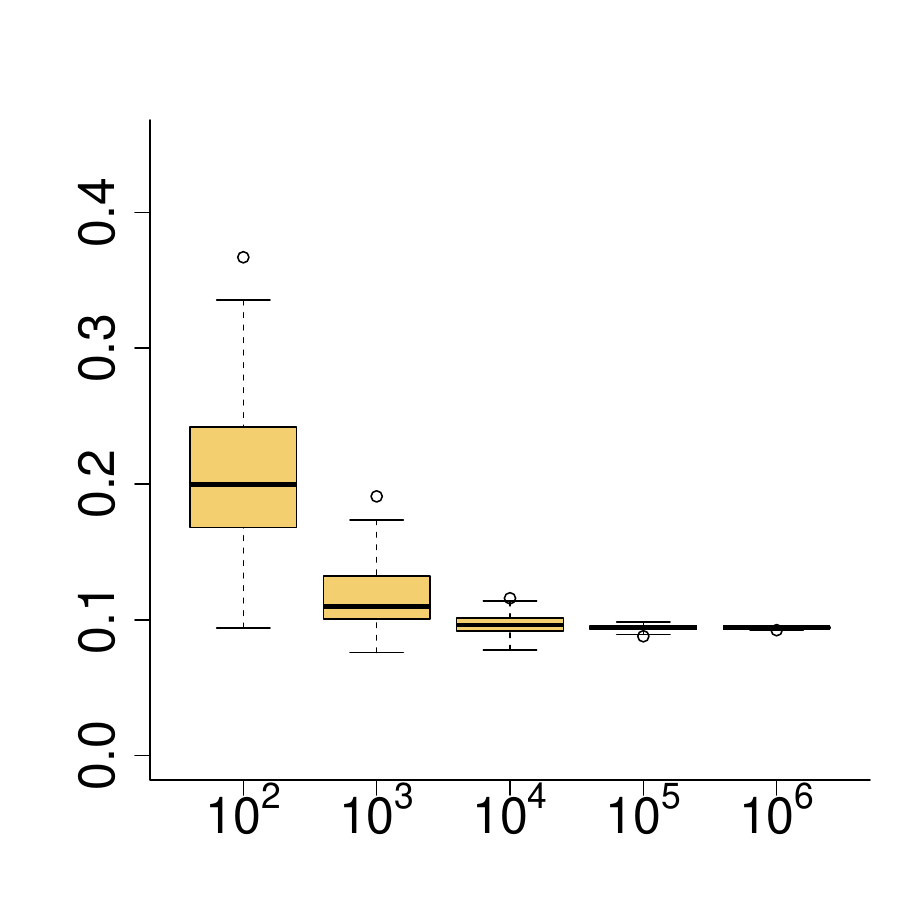}} \\[1ex]

        \raisebox{0.08\textwidth}{\rotatebox{90}{\small Gamma}} &
        \subfloat{\includegraphics[width=0.19\textwidth,height=0.13\textheight,keepaspectratio]{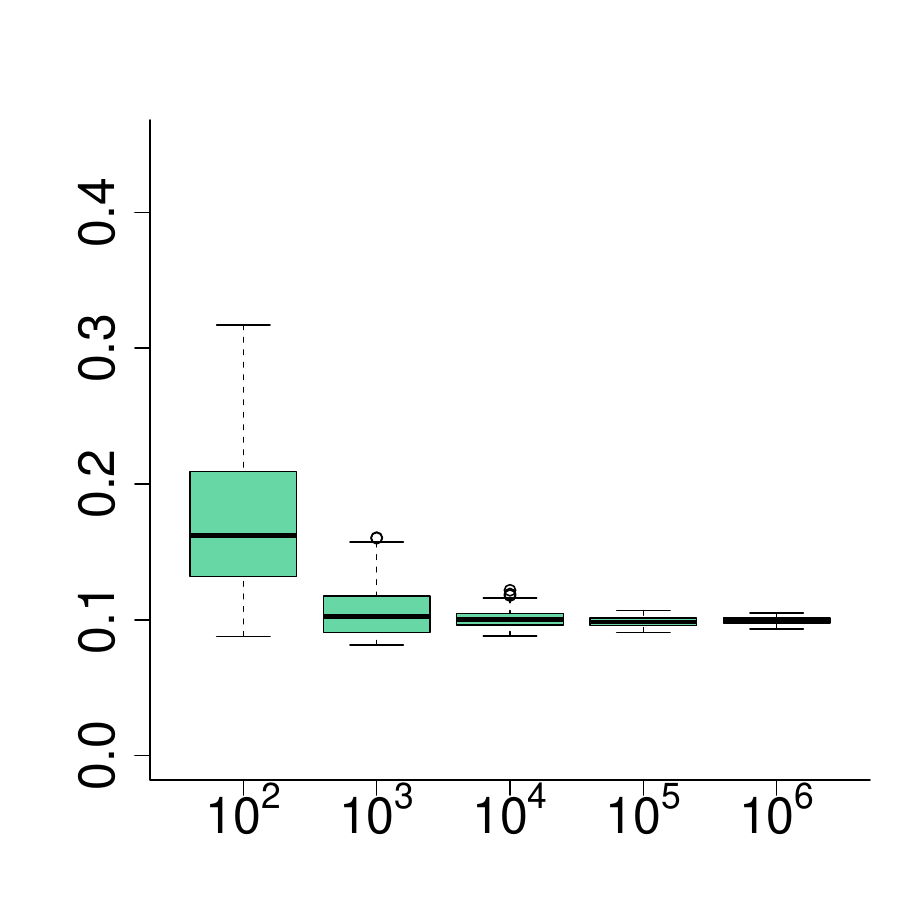}} &
        \subfloat{\includegraphics[width=0.19\textwidth,height=0.13\textheight,keepaspectratio]{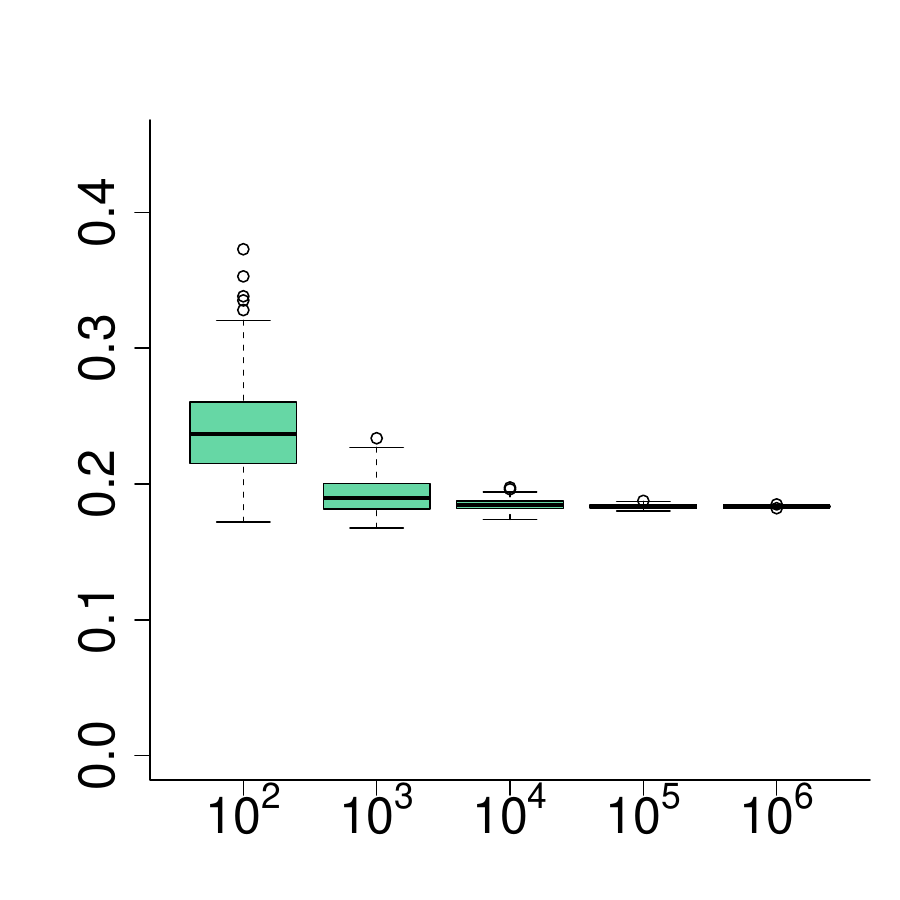}} &
        \subfloat{\includegraphics[width=0.19\textwidth,height=0.13\textheight,keepaspectratio]{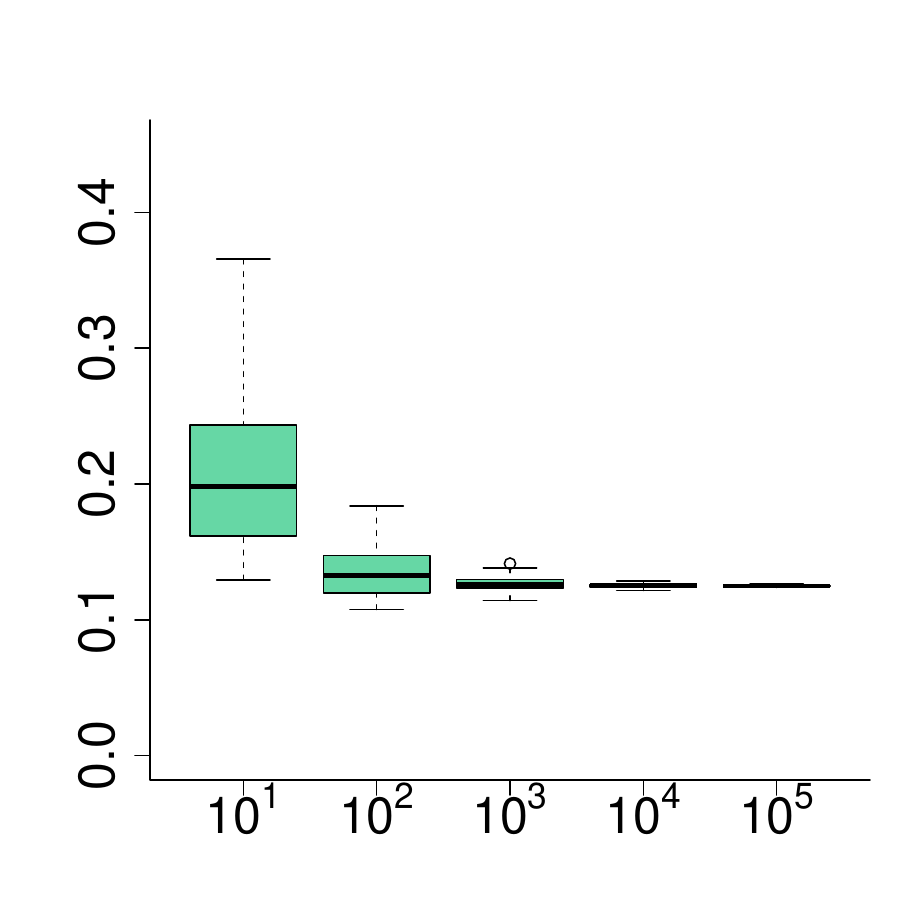}} &
        \subfloat{\includegraphics[width=0.19\textwidth,height=0.13\textheight,keepaspectratio]{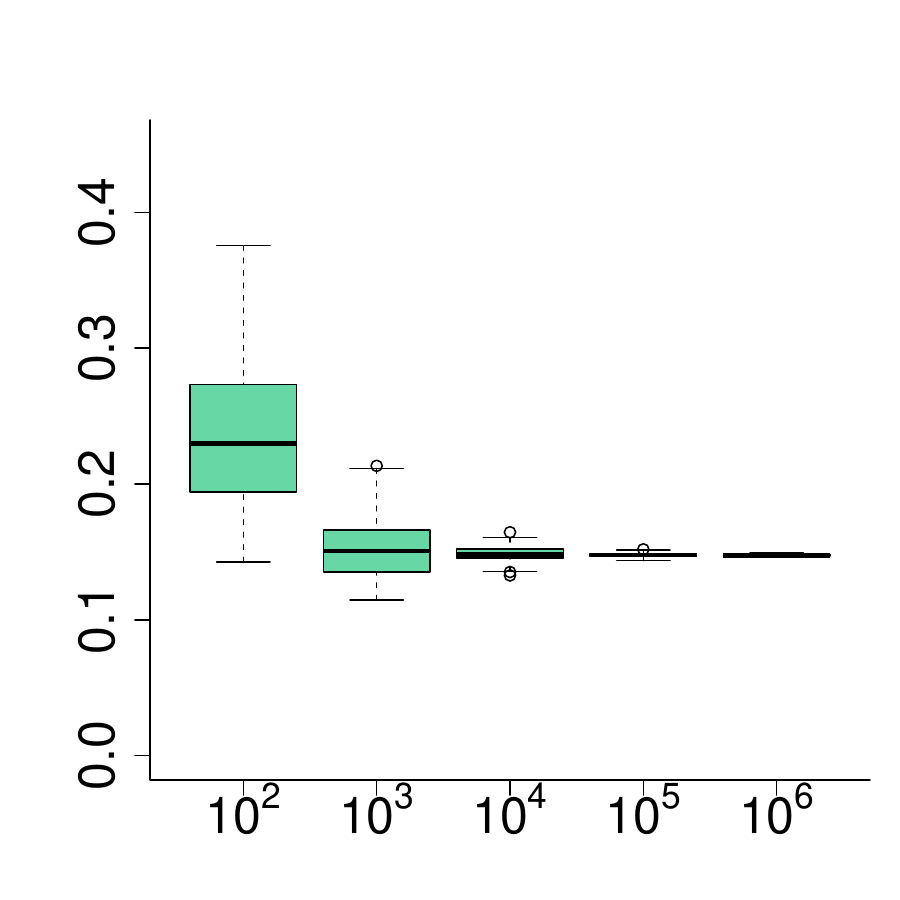}} \\[1ex]

        \raisebox{0.08\textwidth}{\rotatebox{90}{\small logistic}} &
        \subfloat{\includegraphics[width=0.19\textwidth,height=0.13\textheight,keepaspectratio]{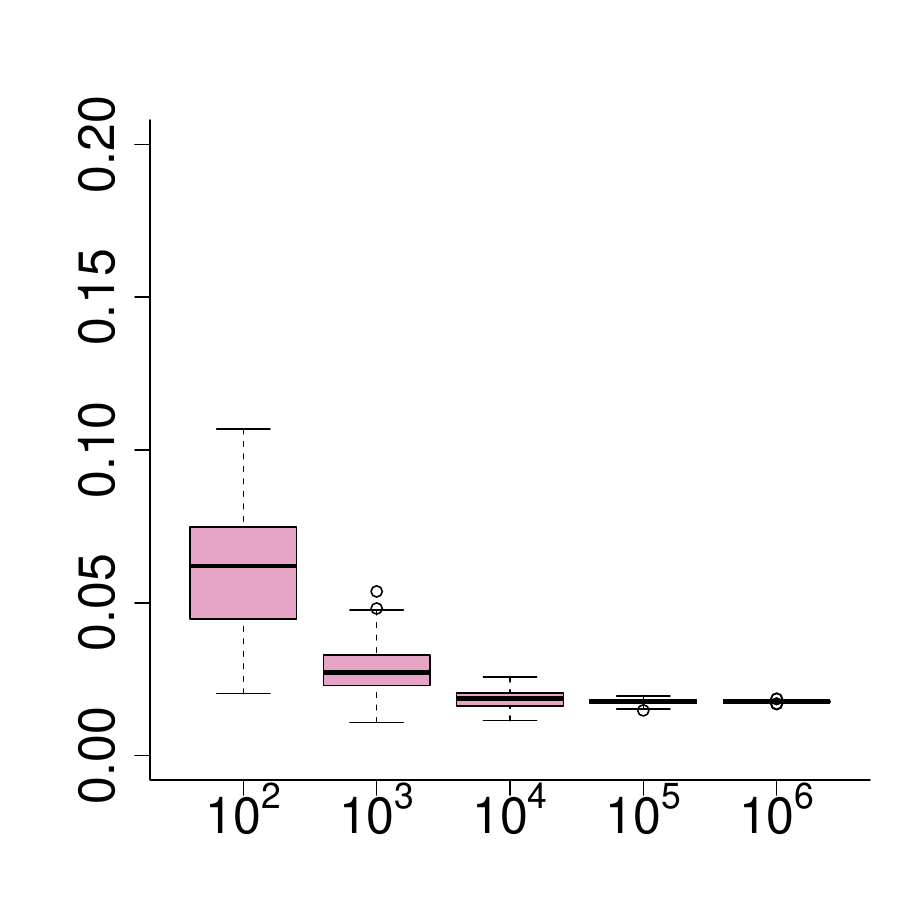}} &
        \subfloat{\includegraphics[width=0.19\textwidth,height=0.13\textheight,keepaspectratio]{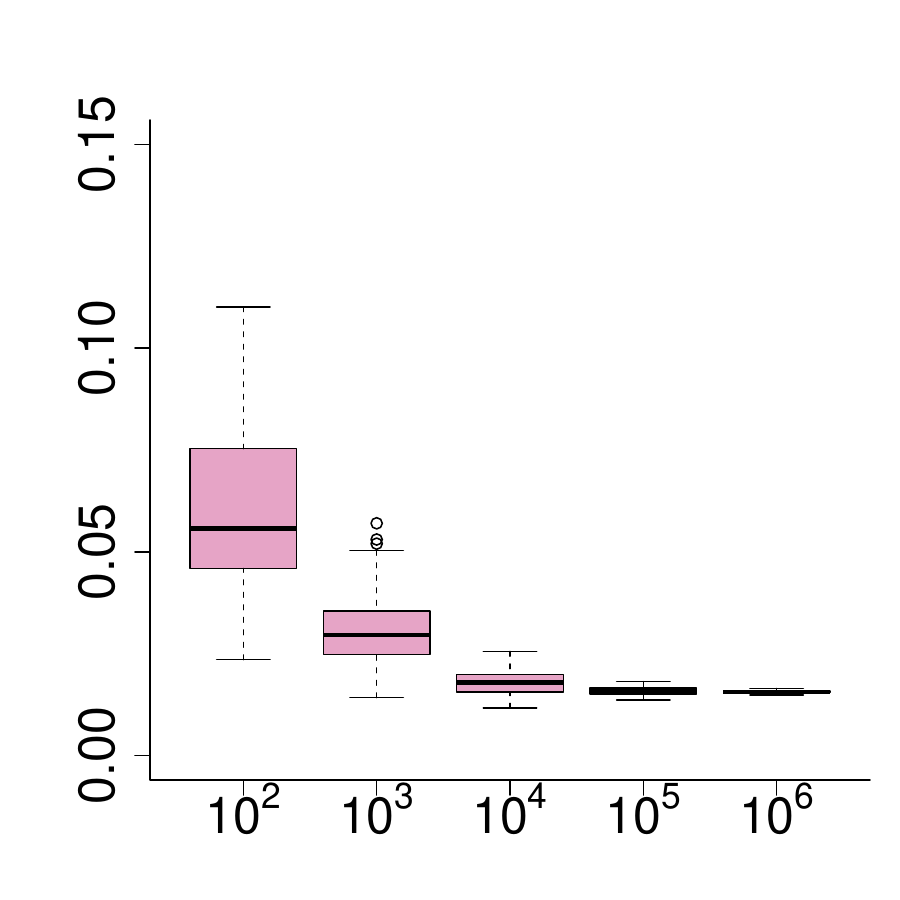}} &
        \subfloat{\includegraphics[width=0.19\textwidth,height=0.13\textheight,keepaspectratio]{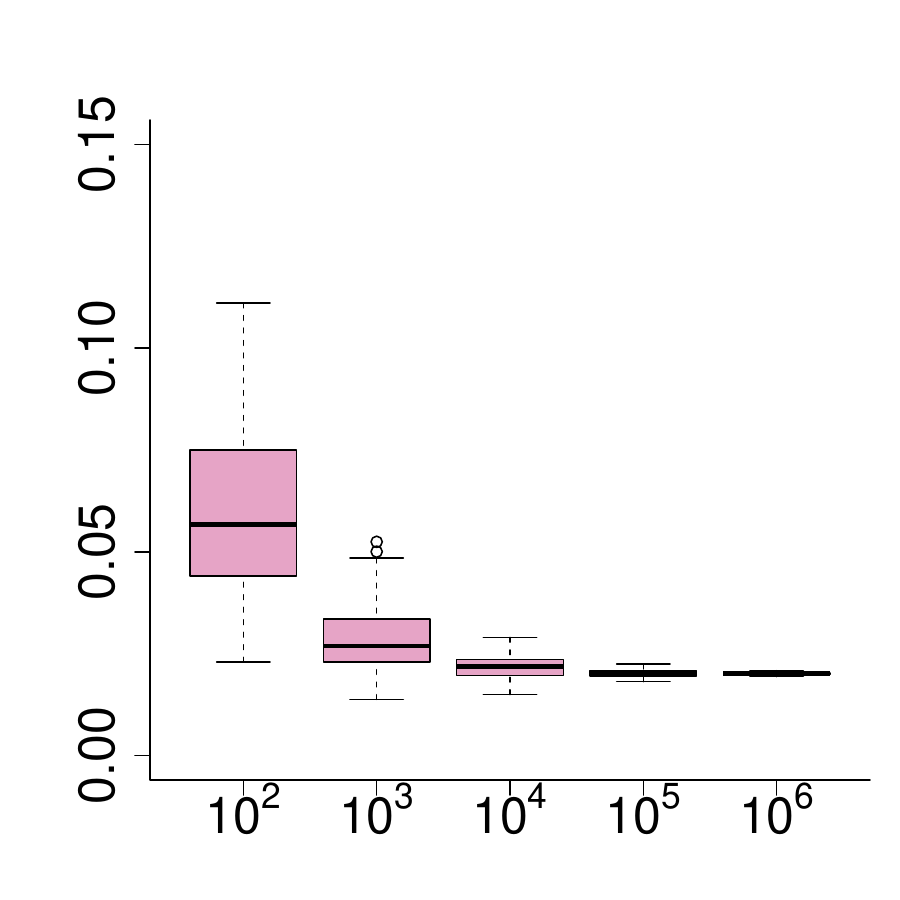}} &
        \subfloat{\includegraphics[width=0.19\textwidth,height=0.13\textheight,keepaspectratio]{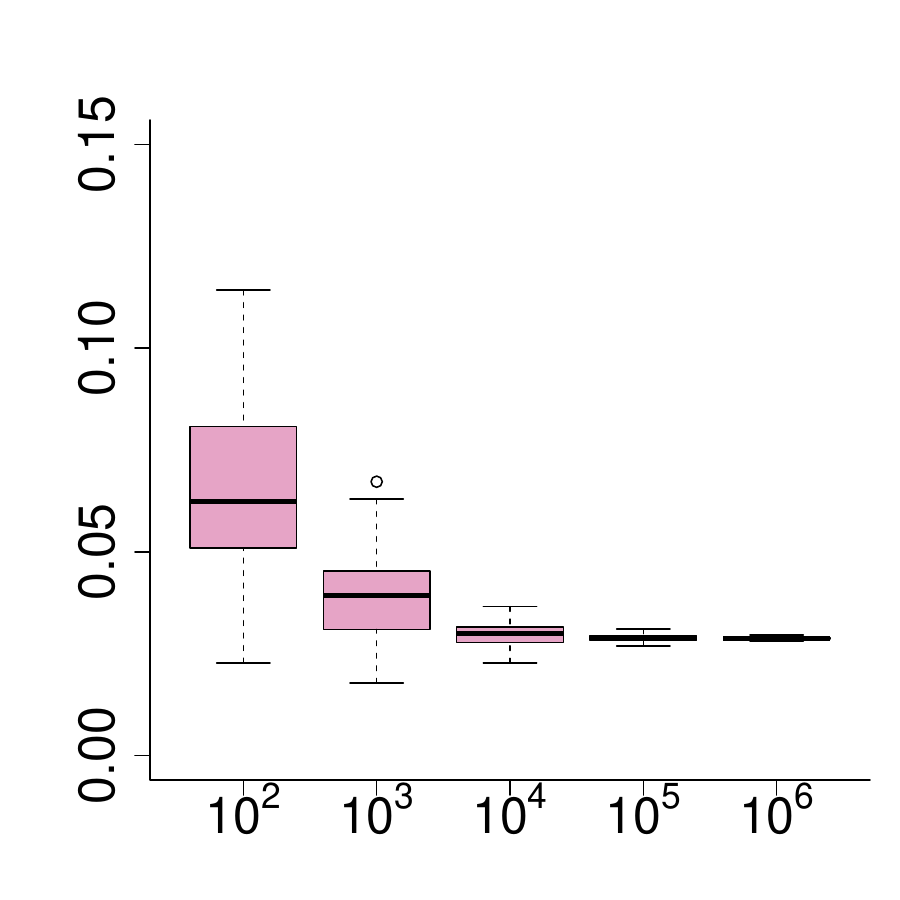}} \\[1ex]

        \raisebox{0.08\textwidth}{\rotatebox{90}{\small Student's $t$}} &
        \subfloat{\includegraphics[width=0.19\textwidth,height=0.13\textheight,keepaspectratio]{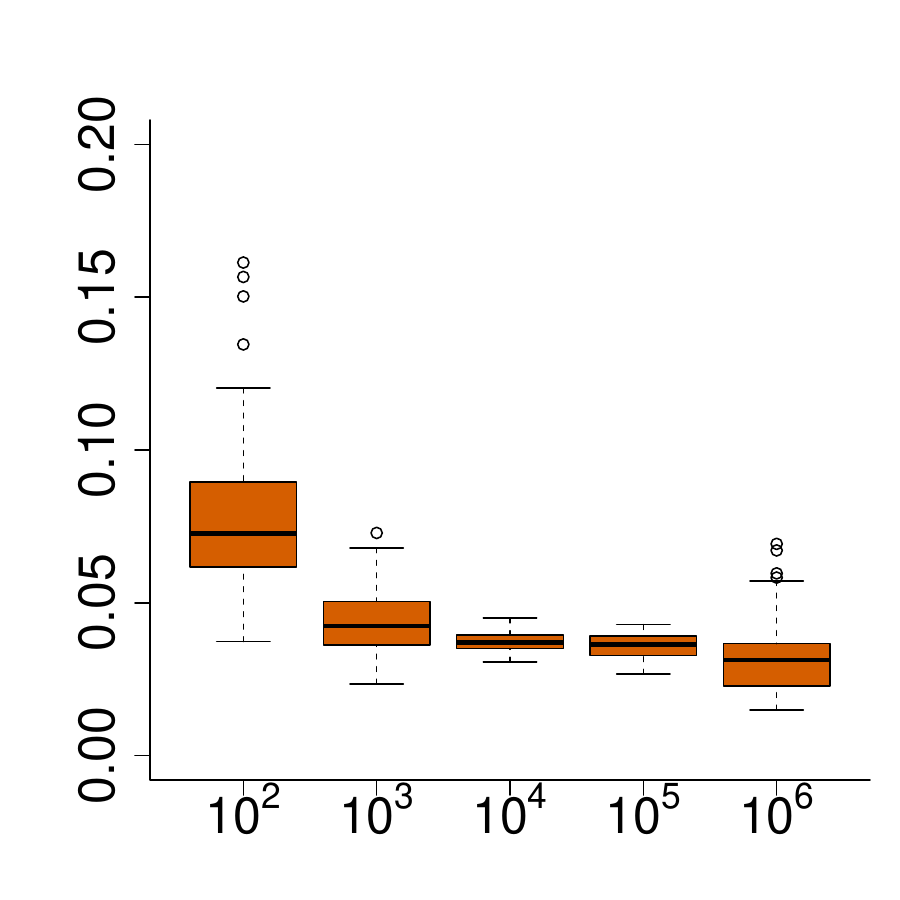}} &
        \subfloat{\includegraphics[width=0.19\textwidth,height=0.13\textheight,keepaspectratio]{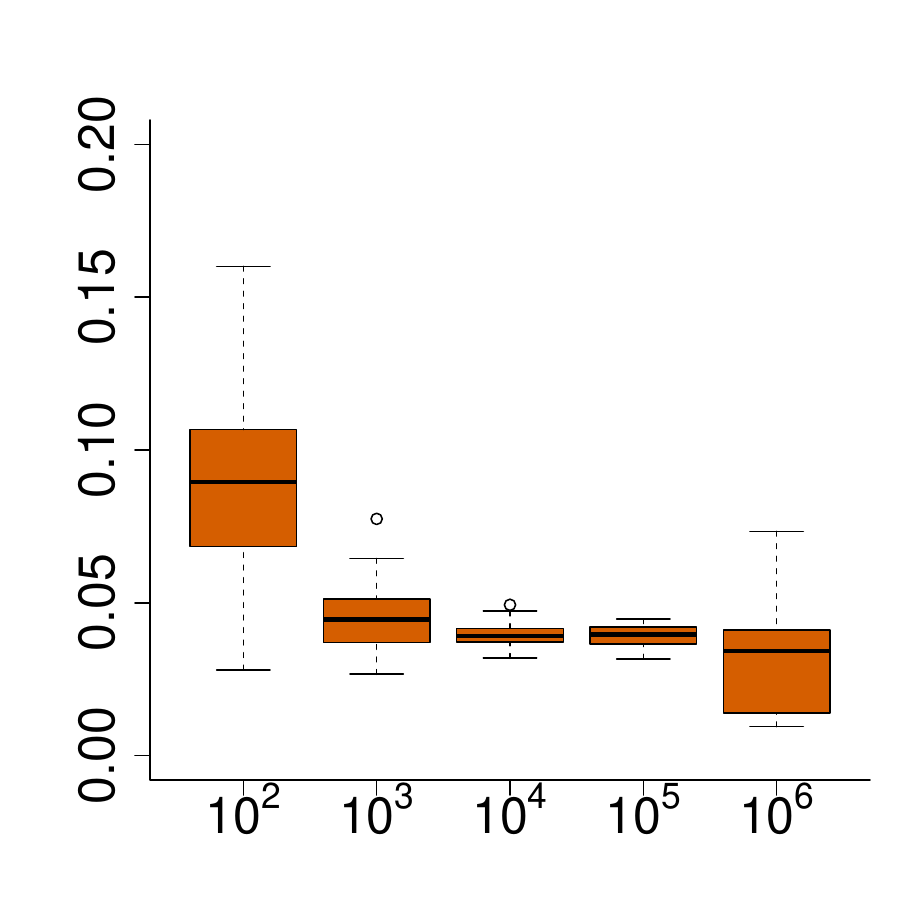}} &
        \subfloat{\includegraphics[width=0.19\textwidth,height=0.13\textheight,keepaspectratio]{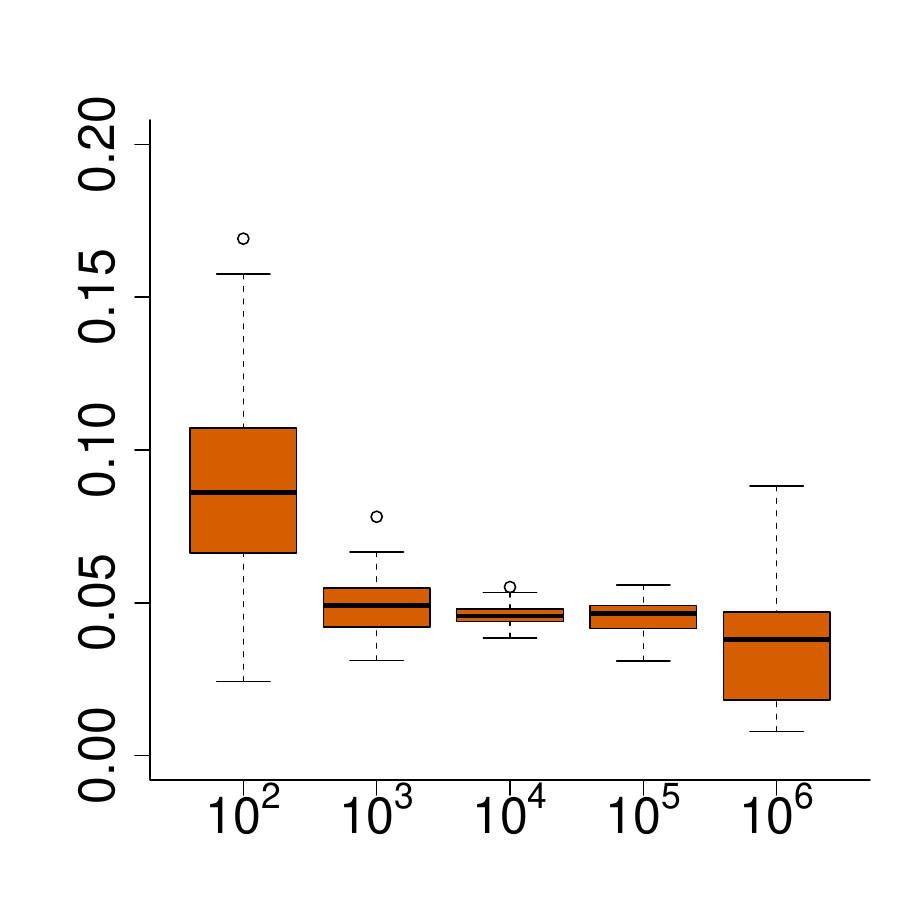}} &
        \subfloat{\includegraphics[width=0.19\textwidth,height=0.13\textheight,keepaspectratio]{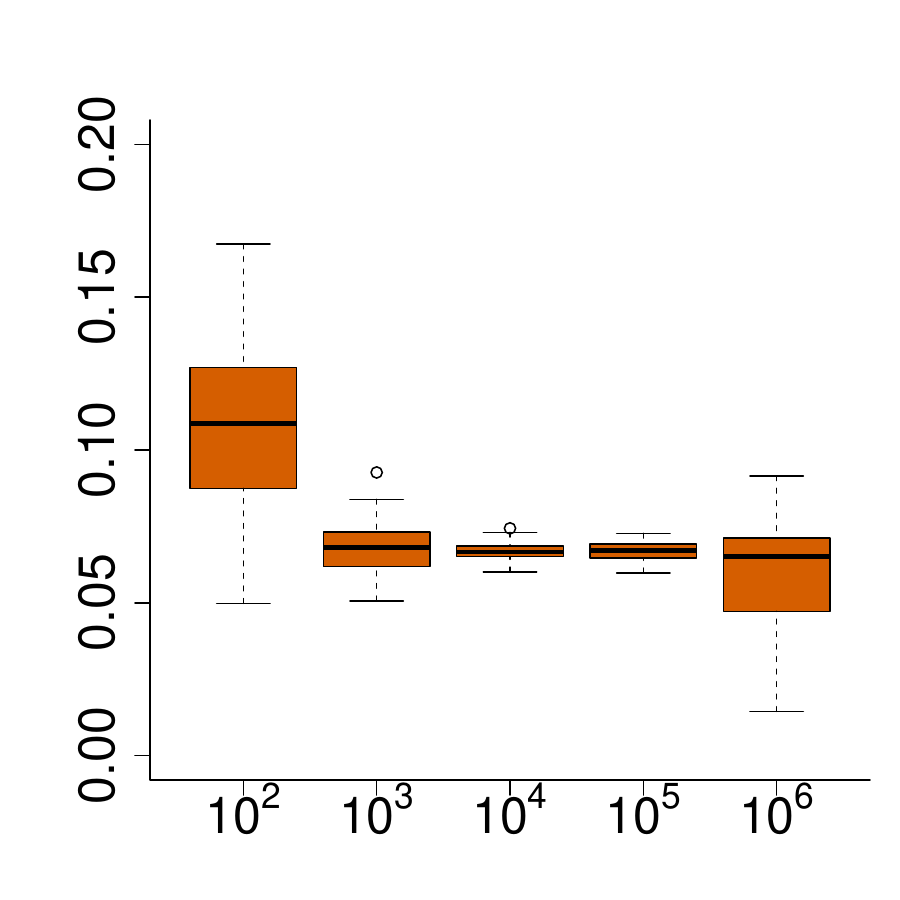}} \\
    \end{tabular}
    
    \caption{Simulation results under sample size variation ($n \in \{10^2, 10^3, 10^4, 10^5, 10^6\}$) with a fixed standardized grid width ($\delta/\sigma = 0.5$). The plots display the $L_2$ norm between the true and estimated densities for the normal, beta, gamma, logistic, and Student's $t$ distributions, using the proposed MALC and three kernel-based methods.}
    \label{fig:pdfn1}
\end{figure}

\begin{figure}[!htb]
    \centering
    \begin{tabular}{c | c c c c}
        & \small MALC & \small BK2002 & \small BinnedNP & \small KernSmooth \\[1ex]
        \hline\rule{0pt}{3ex}
        
        \raisebox{0.08\textwidth}{\rotatebox{90}{\small Laplace}} &
        \subfloat{\includegraphics[width=0.19\textwidth,height=0.13\textheight,keepaspectratio]{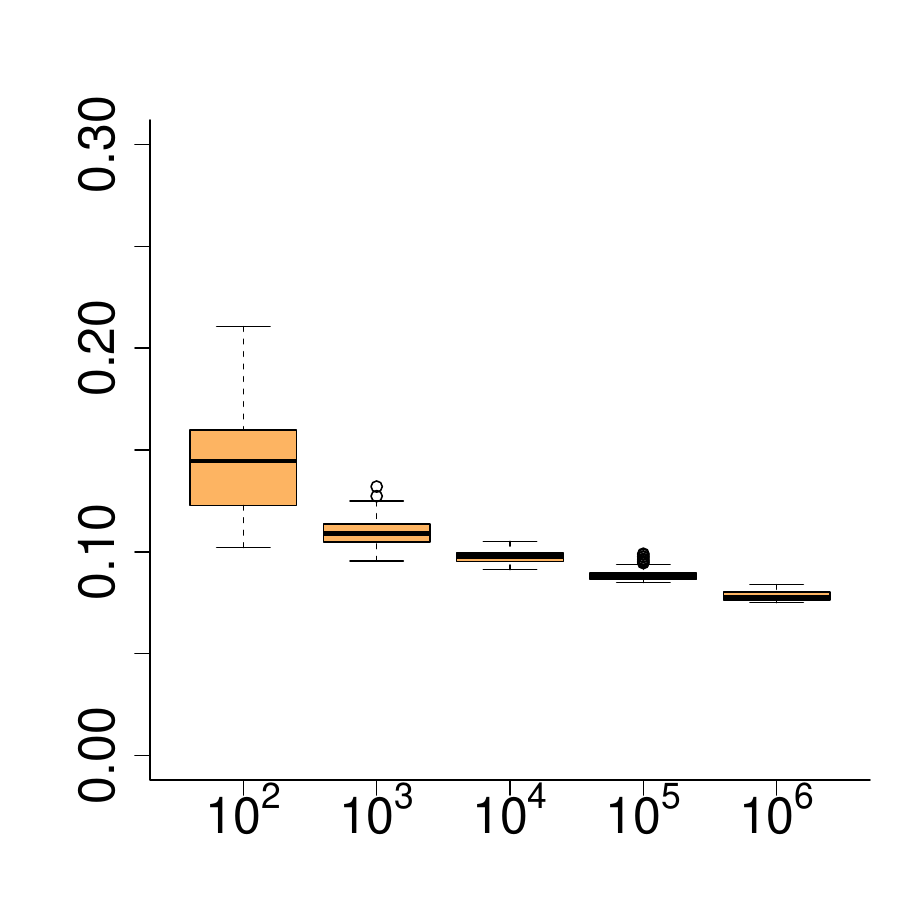}} &
        \subfloat{\includegraphics[width=0.19\textwidth,height=0.13\textheight,keepaspectratio]{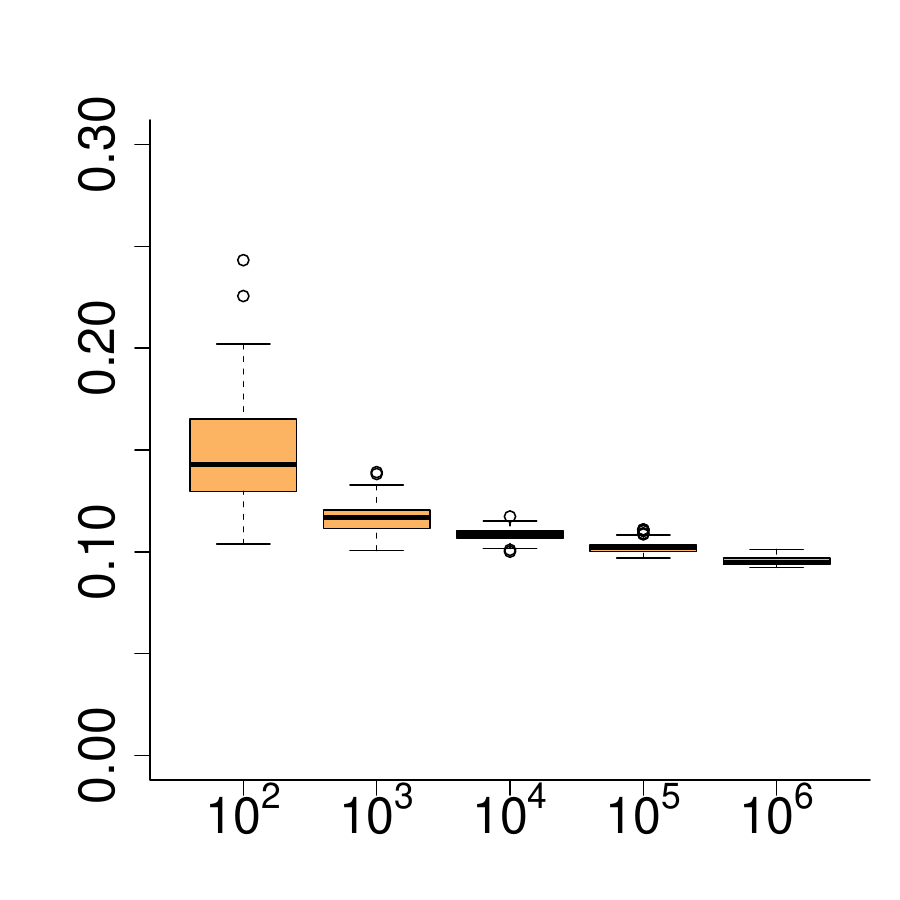}} &
        \subfloat{\includegraphics[width=0.19\textwidth,height=0.13\textheight,keepaspectratio]{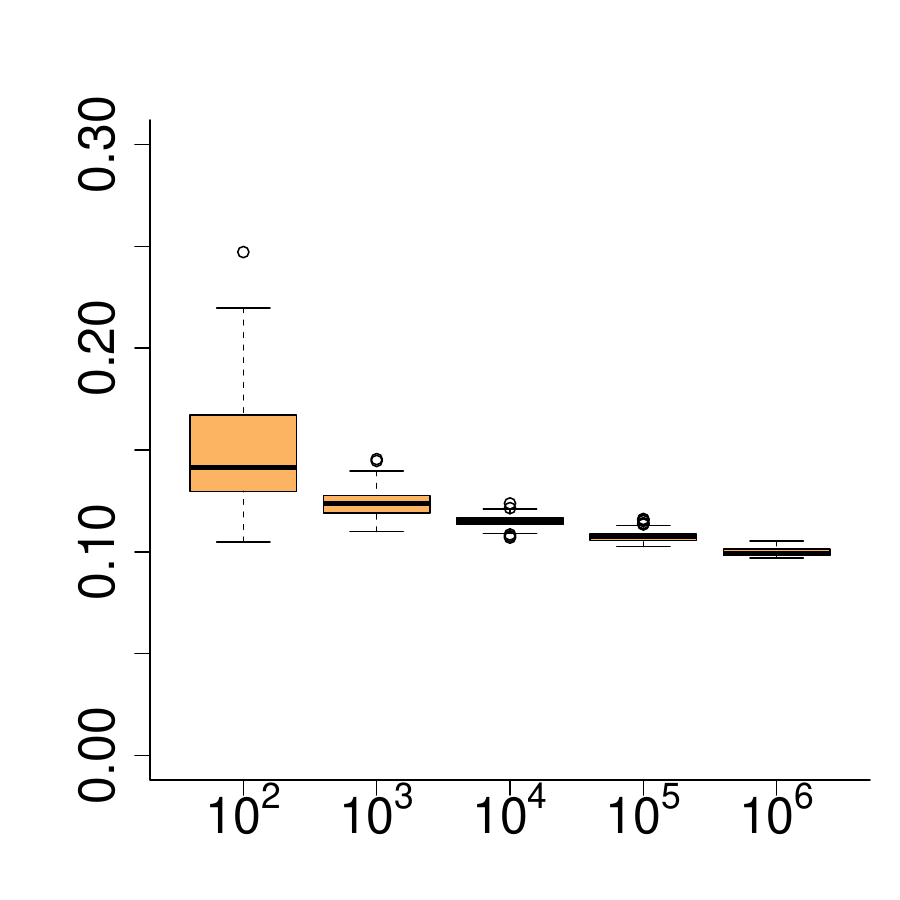}} &
        \subfloat{\includegraphics[width=0.19\textwidth,height=0.13\textheight,keepaspectratio]{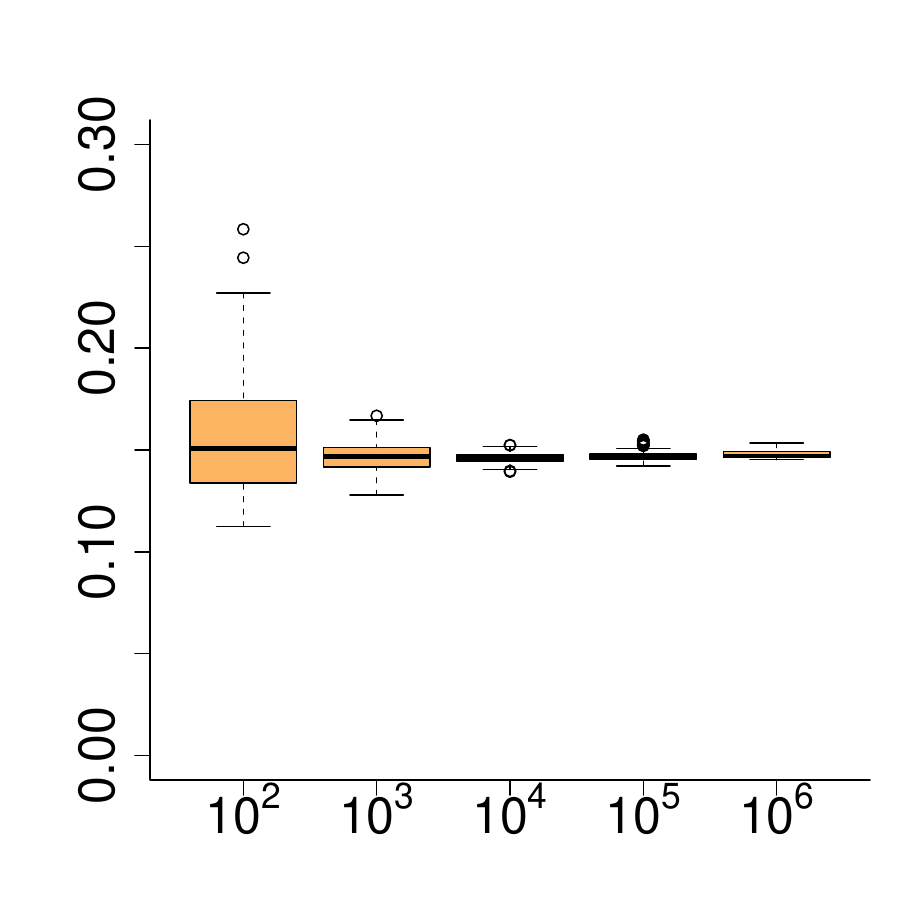}} \\[1ex]

        \raisebox{0.08\textwidth}{\rotatebox{90}{\small chi-square}} &
        \subfloat{\includegraphics[width=0.19\textwidth,height=0.13\textheight,keepaspectratio]{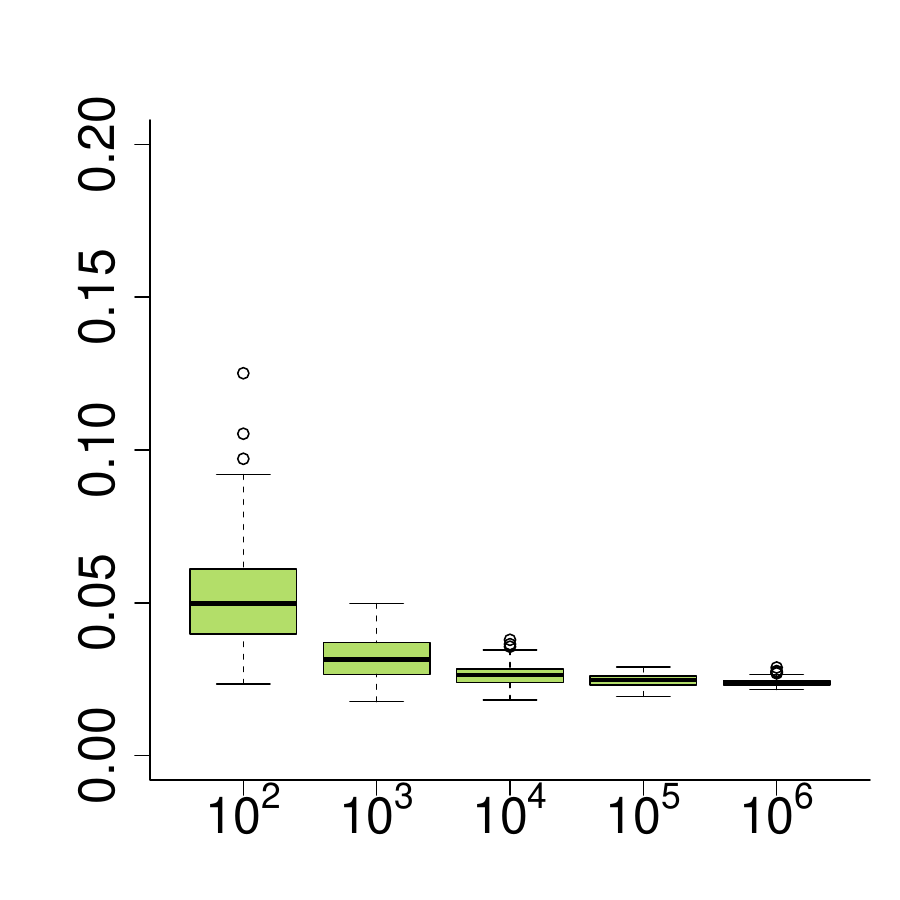}} &
        \subfloat{\includegraphics[width=0.19\textwidth,height=0.13\textheight,keepaspectratio]{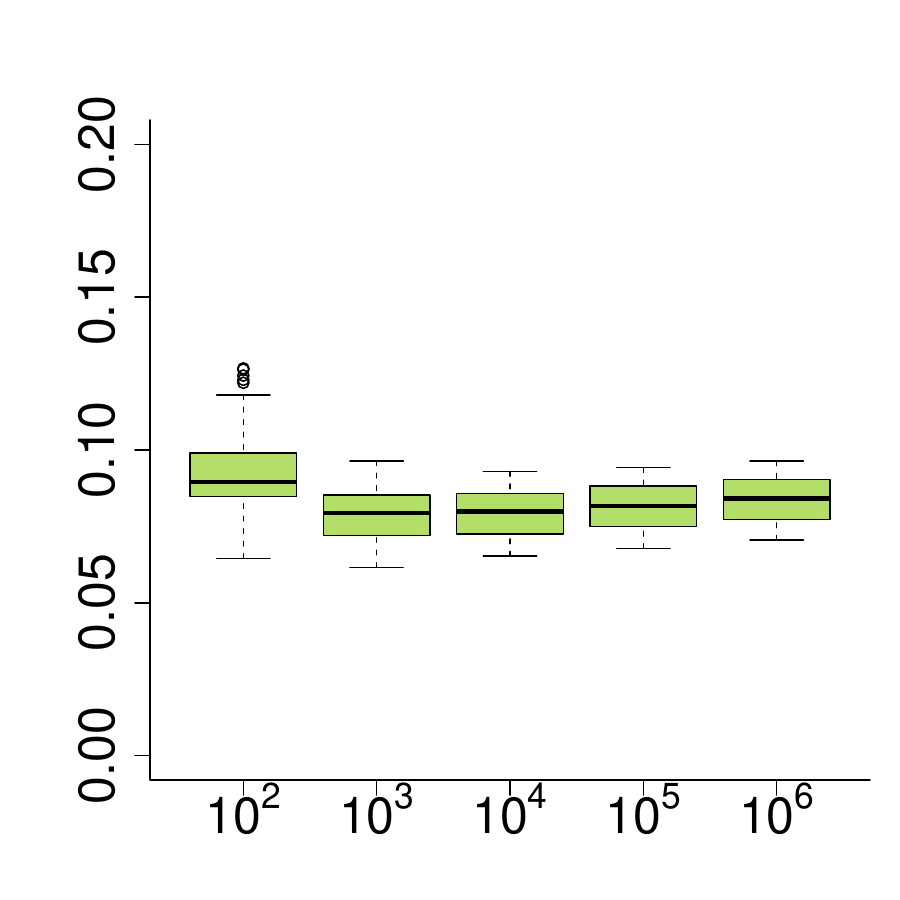}} &
        \subfloat{\includegraphics[width=0.19\textwidth,height=0.13\textheight,keepaspectratio]{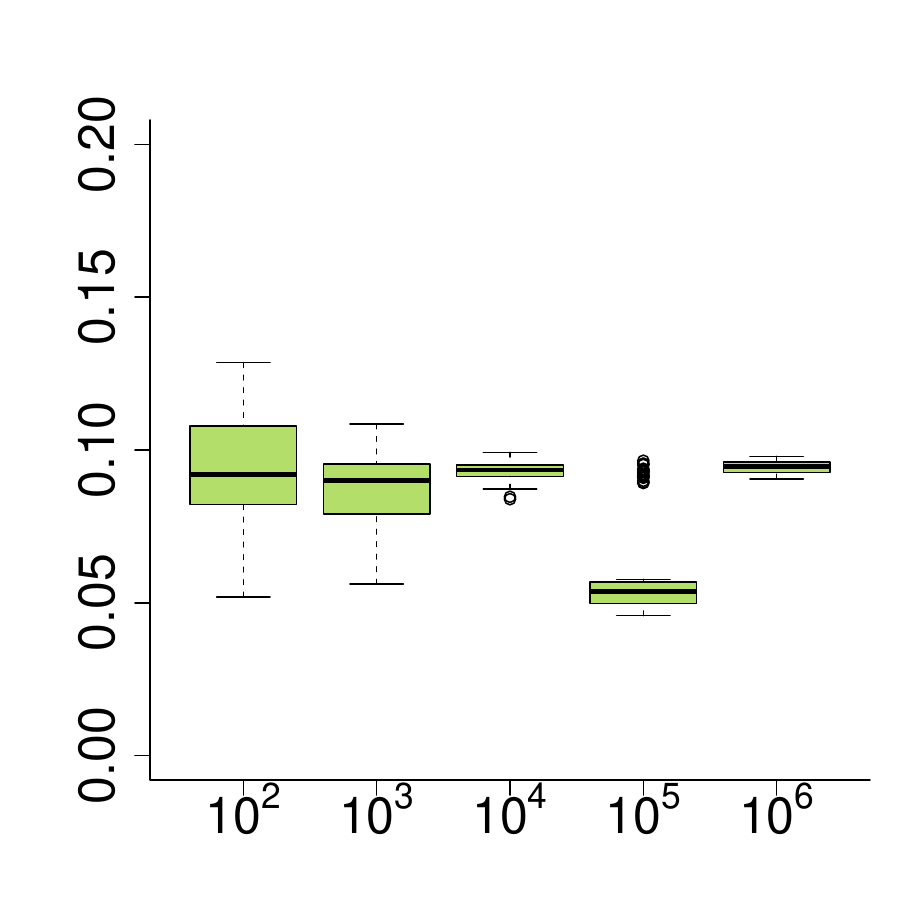}} &
        \subfloat{\includegraphics[width=0.19\textwidth,height=0.13\textheight,keepaspectratio]{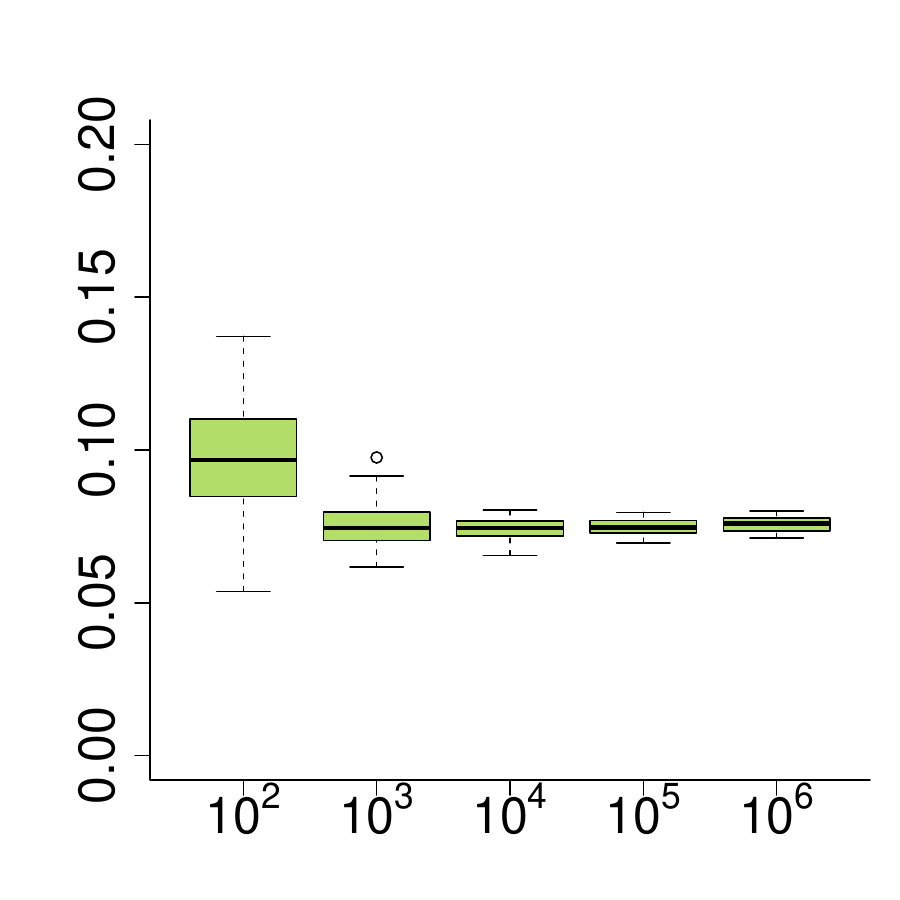}} \\[1ex]

        \raisebox{0.08\textwidth}{\rotatebox{90}{\small log-normal}} &
        \subfloat{\includegraphics[width=0.19\textwidth,height=0.13\textheight,keepaspectratio]{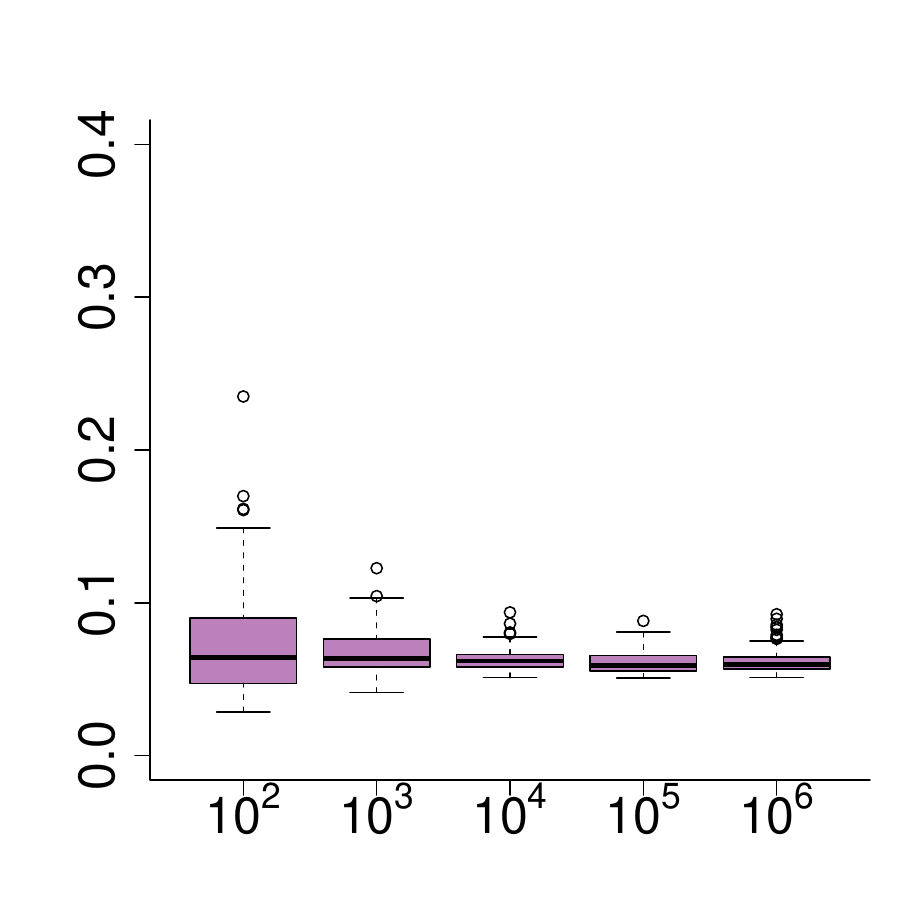}} &
        \subfloat{\includegraphics[width=0.19\textwidth,height=0.13\textheight,keepaspectratio]{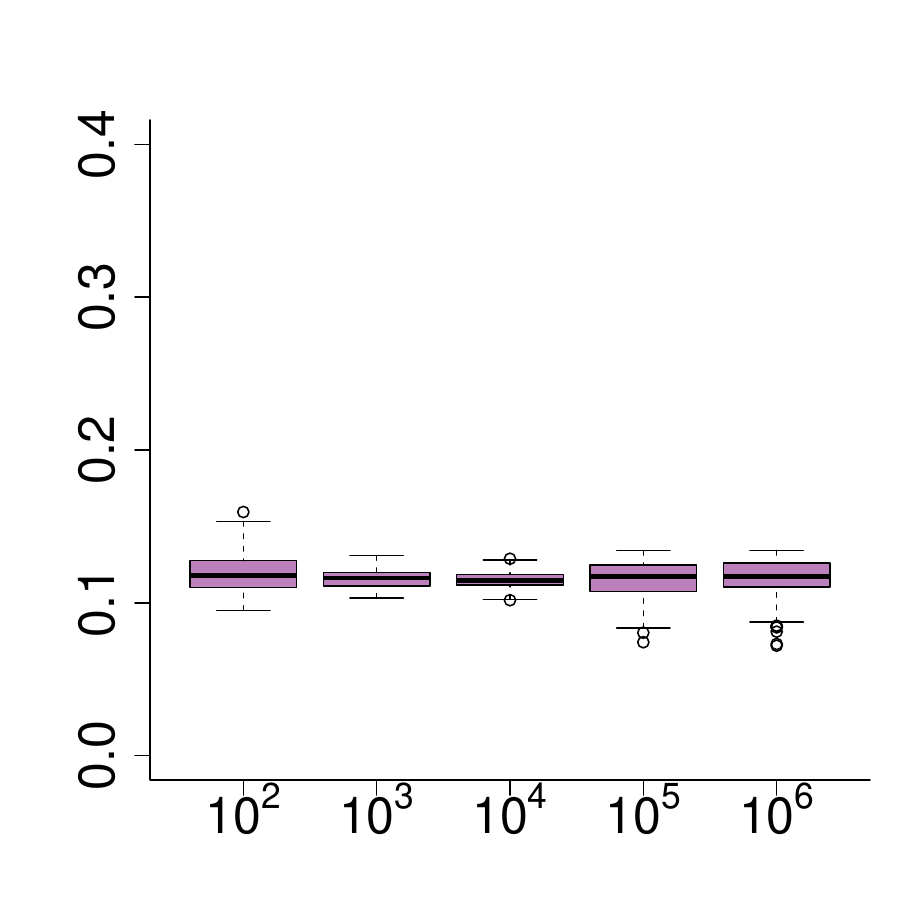}} &
        \subfloat{\includegraphics[width=0.19\textwidth,height=0.13\textheight,keepaspectratio]{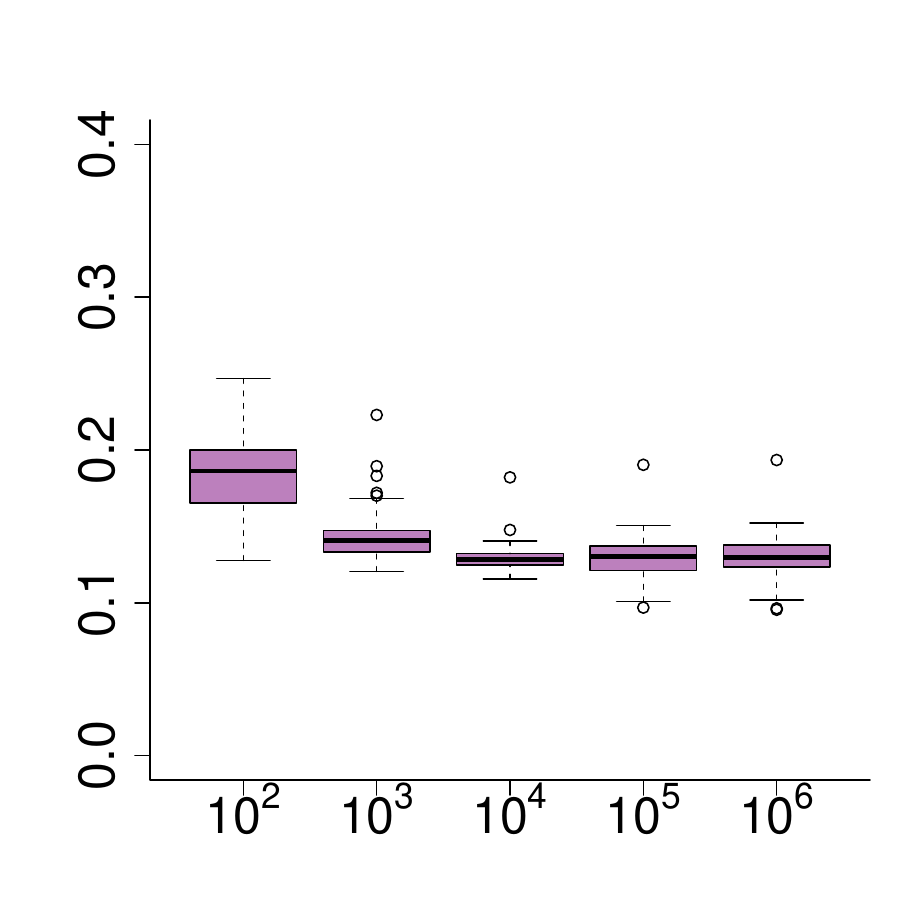}} &
        \subfloat{\includegraphics[width=0.19\textwidth,height=0.13\textheight,keepaspectratio]{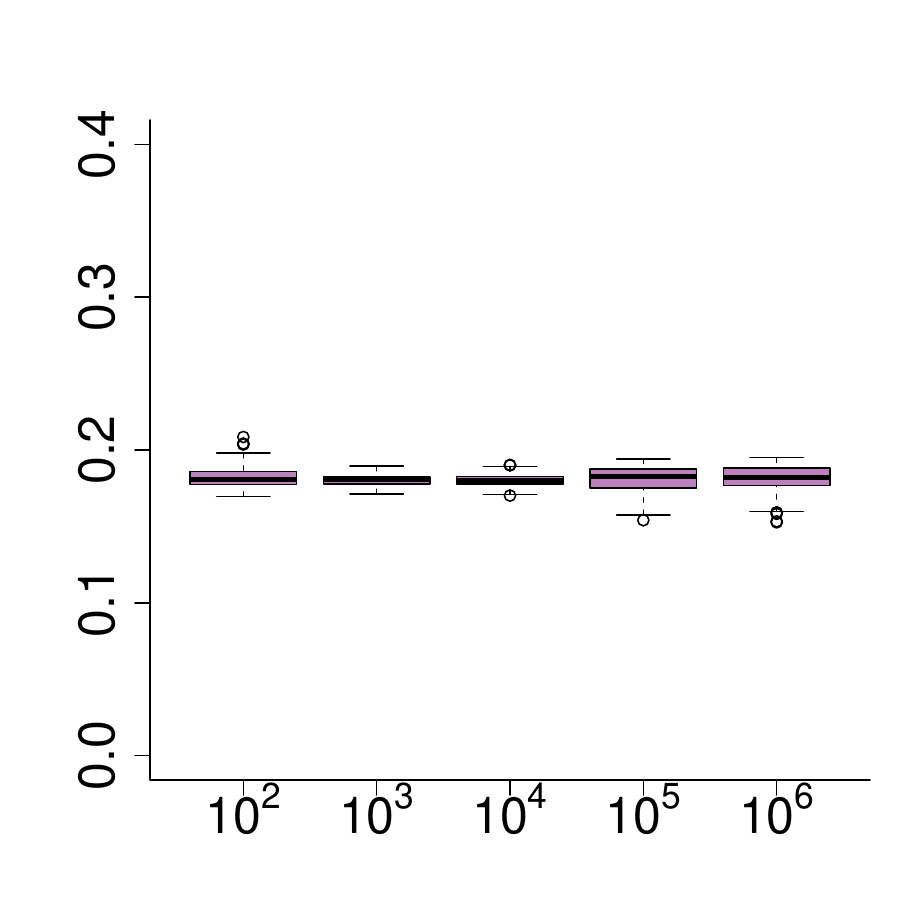}} \\[1ex]

        \raisebox{0.08\textwidth}{\rotatebox{90}{\small Weibull}} &
        \subfloat{\includegraphics[width=0.19\textwidth,height=0.13\textheight,keepaspectratio]{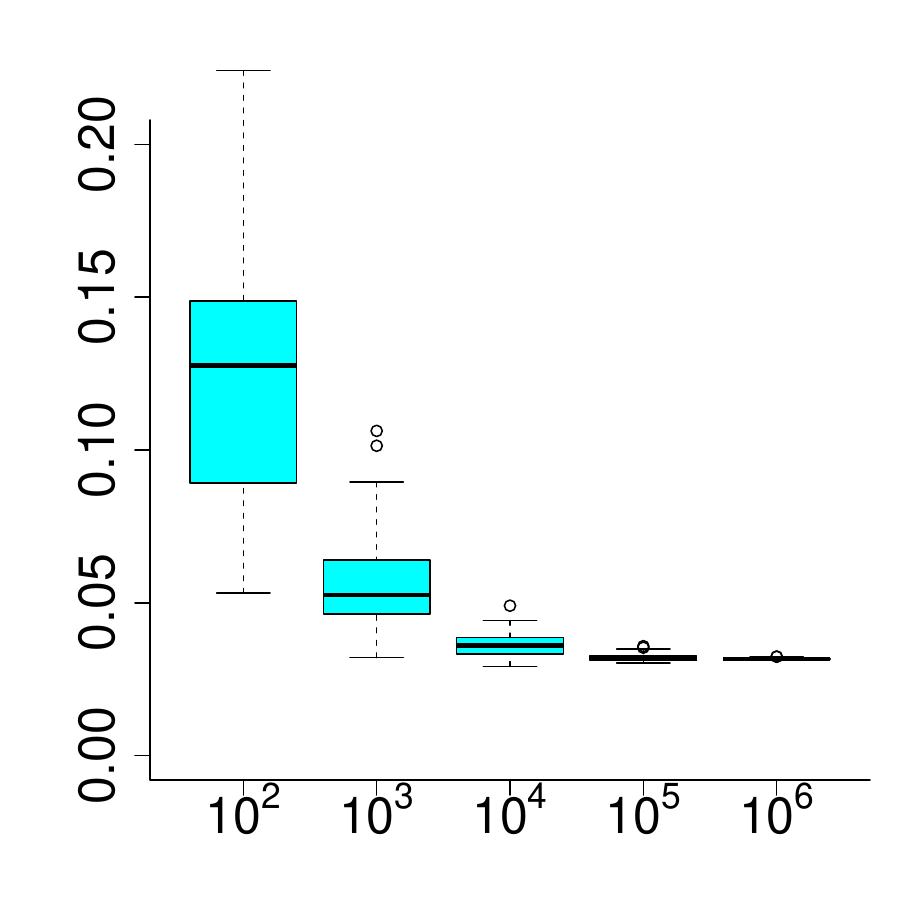}} &
        \subfloat{\includegraphics[width=0.19\textwidth,height=0.13\textheight,keepaspectratio]{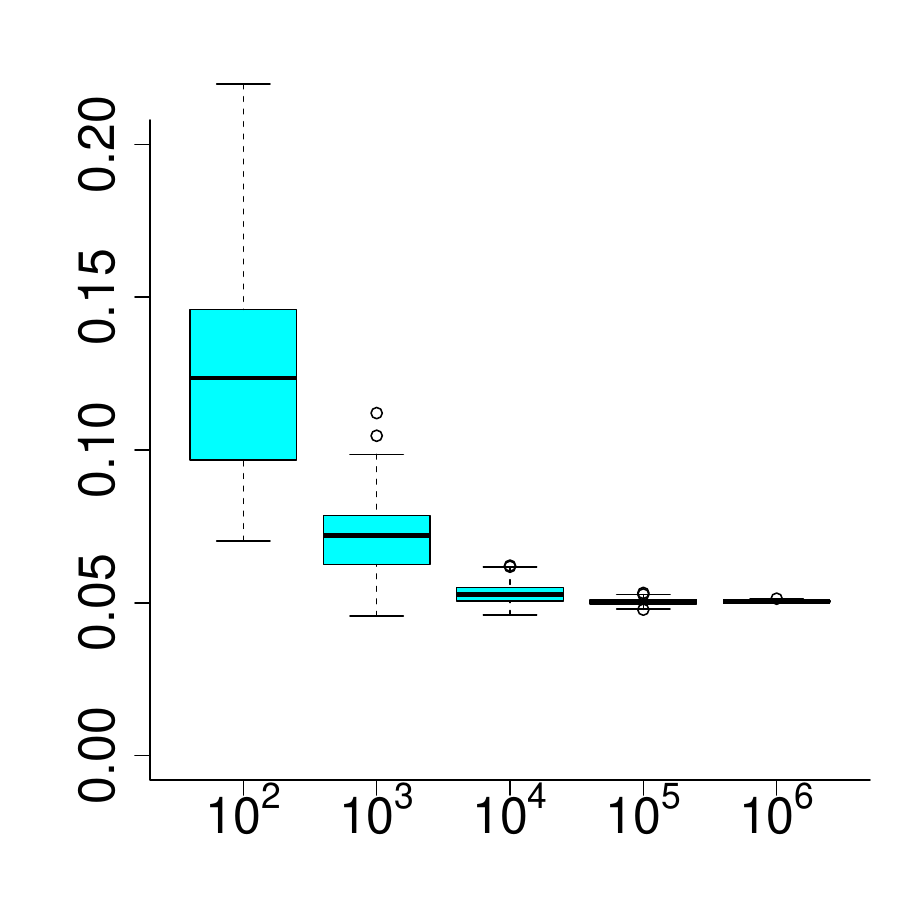}} &
        \subfloat{\includegraphics[width=0.19\textwidth,height=0.13\textheight,keepaspectratio]{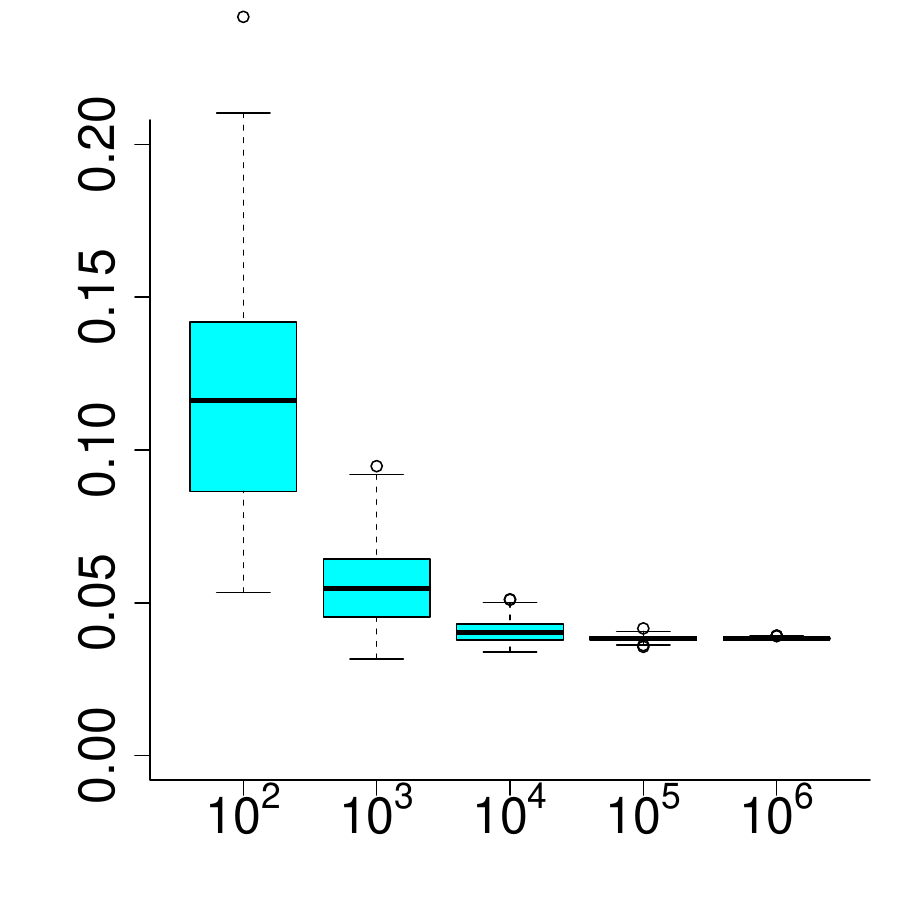}} &
        \subfloat{\includegraphics[width=0.19\textwidth,height=0.13\textheight,keepaspectratio]{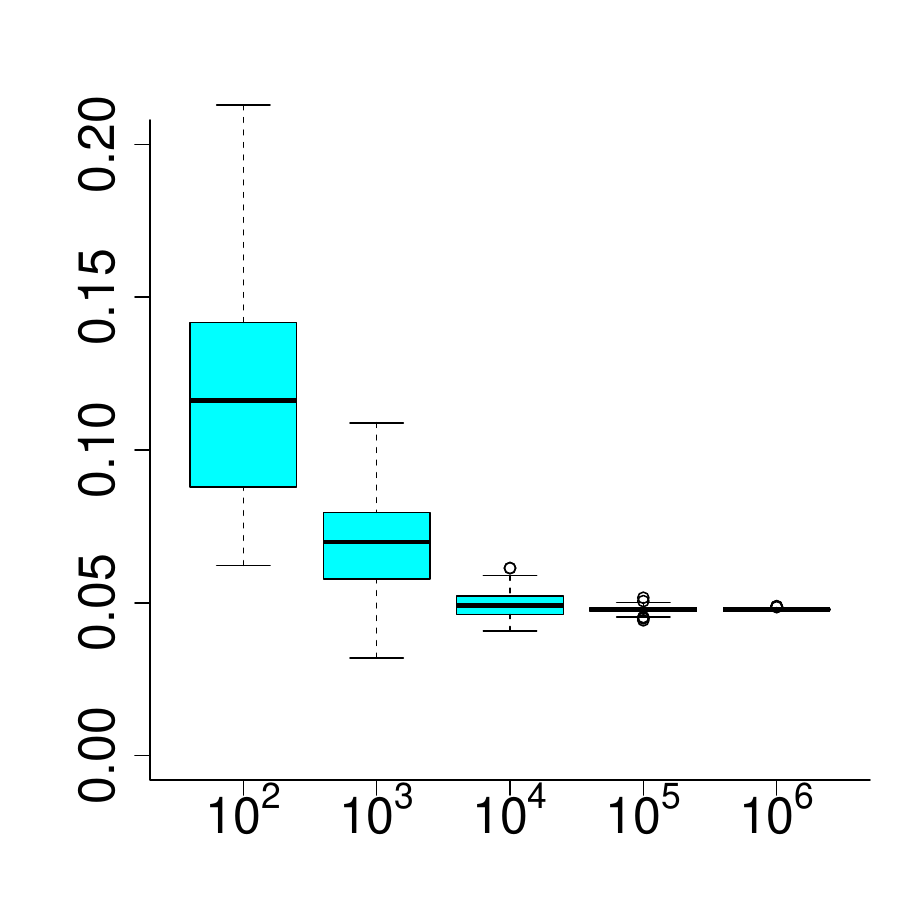}} \\[1ex]

        \raisebox{0.08\textwidth}{\rotatebox{90}{\small Pareto}} &
        \subfloat{\includegraphics[width=0.19\textwidth,height=0.13\textheight,keepaspectratio]{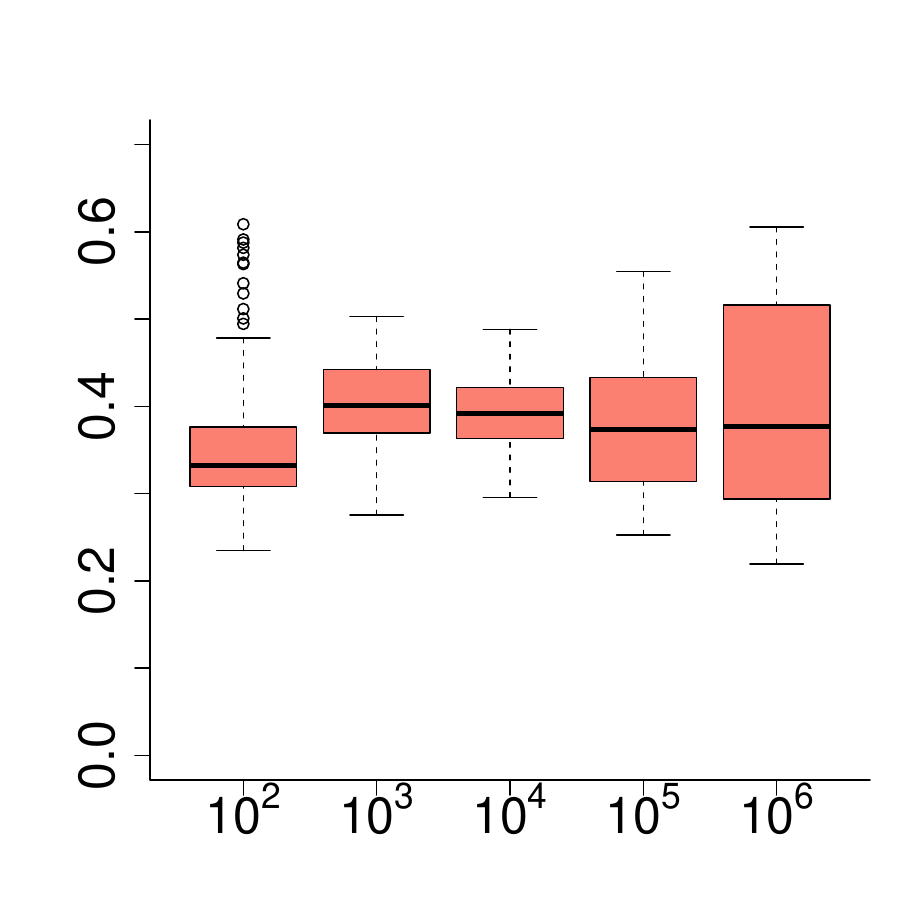}} &
        \subfloat{\includegraphics[width=0.19\textwidth,height=0.13\textheight,keepaspectratio]{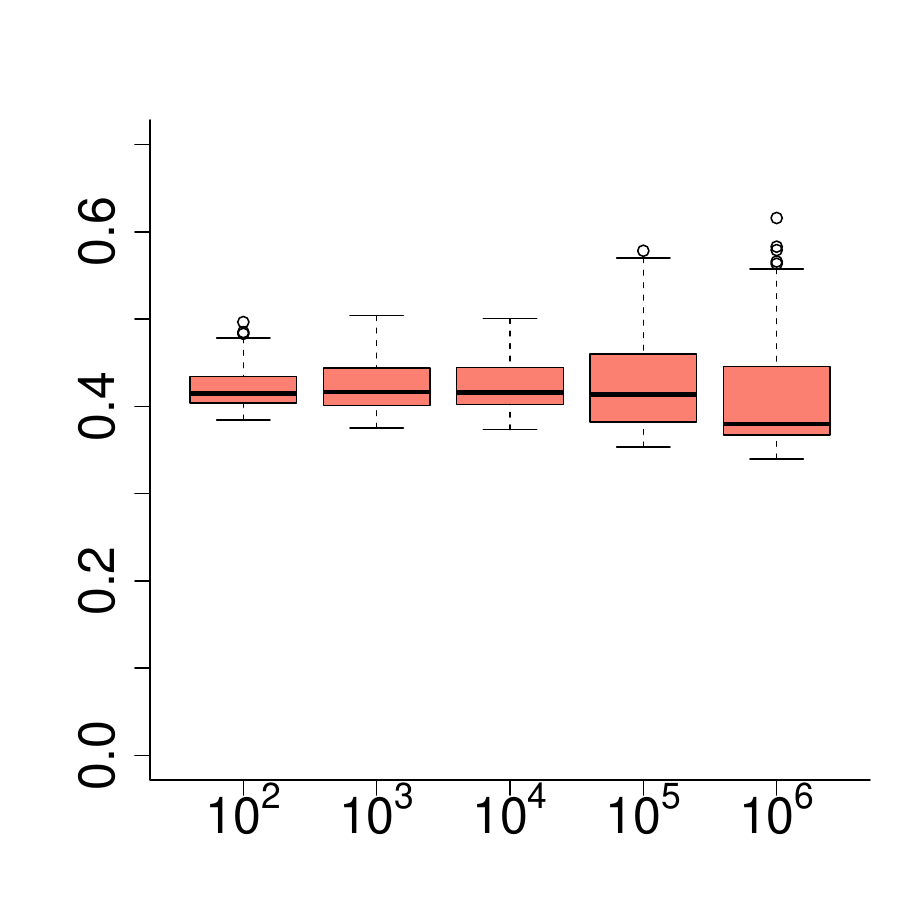}} &
        \subfloat{\includegraphics[width=0.19\textwidth,height=0.13\textheight,keepaspectratio]{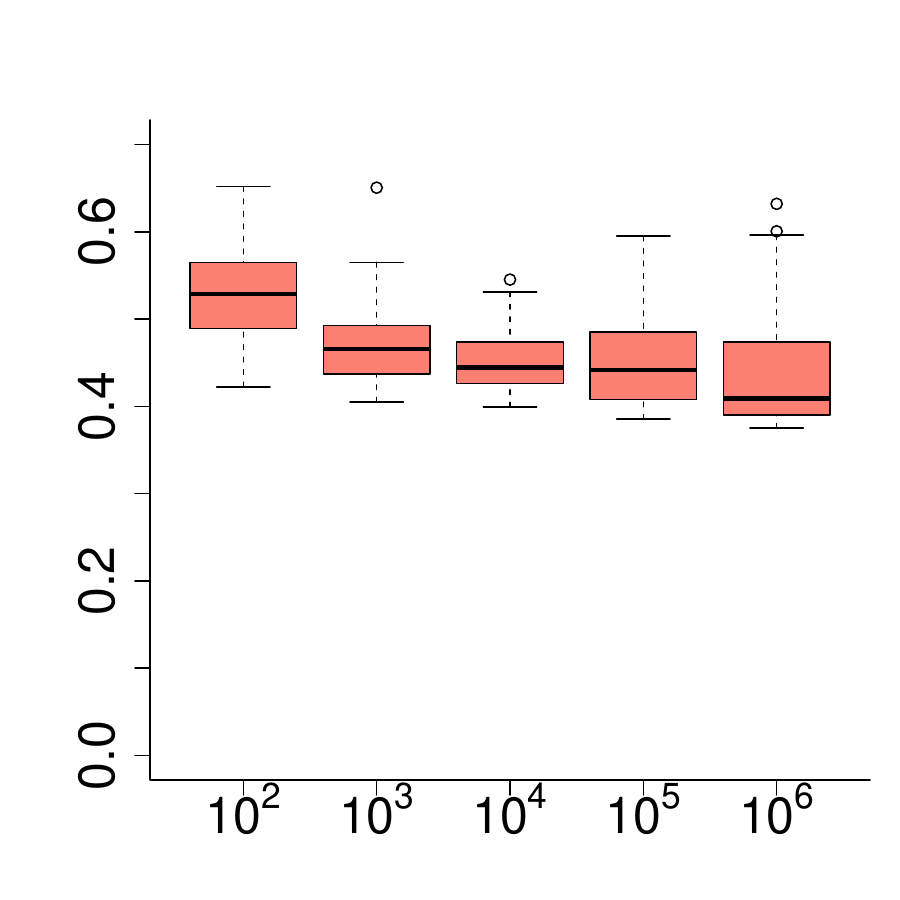}} &
        \subfloat{\includegraphics[width=0.19\textwidth,height=0.13\textheight,keepaspectratio]{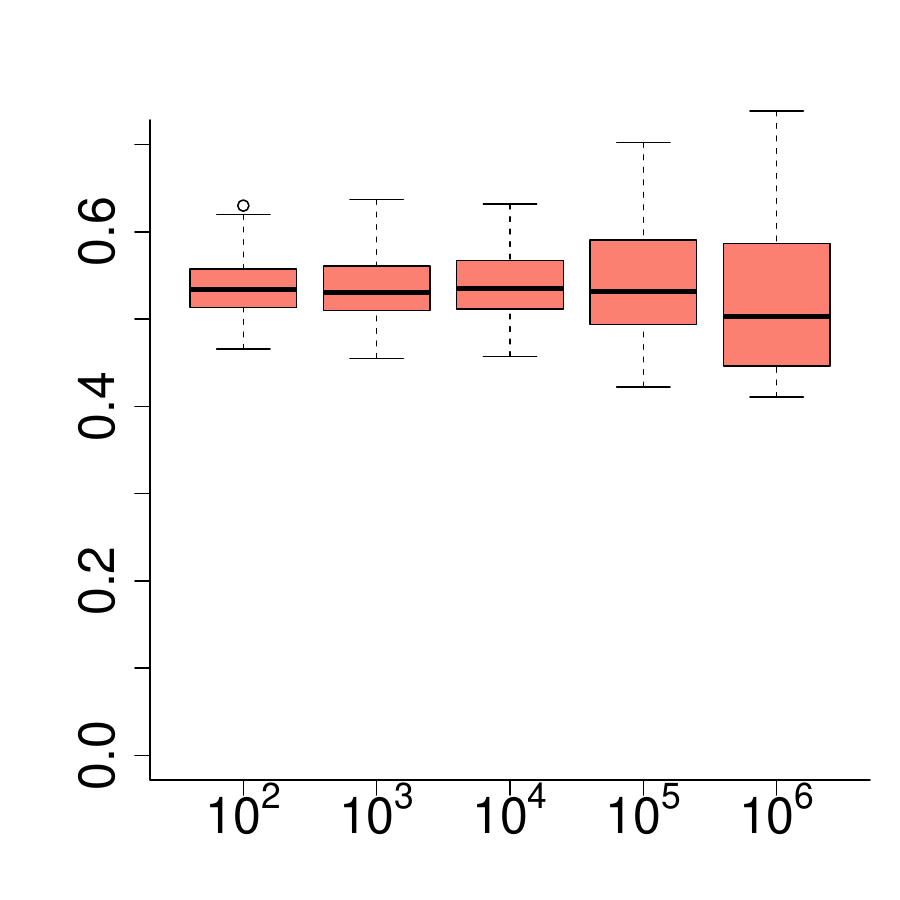}} \\
    \end{tabular}
    
    \caption{Simulation results under sample size variation ($n \in \{10^2, 10^3, 10^4, 10^5, 10^6\}$) with a fixed standardized grid width ($\delta/\sigma = 0.5$). The plots show the $L_2$ norm between the true and estimated densities for the Laplace, chi-square, log-normal, Weibull, and Pareto distributions, using the proposed MALC and three kernel-based methods.}
    \label{fig:pdfn2}
\end{figure}

Figures~\ref{fig:pdfn1} and~\ref{fig:pdfn2} report the results of our simulation study when the sample size varies.   We observe that MALC achieves the lowest errors, with a clear separation from the competing methods, for the log-concave gamma and chi-square distributions, as well as for the non-log-concave log-normal distribution. For the remaining distributions, except for the normal and logistic distributions, MALC also attains the lowest $L_2$ error, although the differences are smaller. The corresponding numerical results are provided in Tables 2-5 of the Supplementary Material D. Importantly, this performance is achieved without requiring user-specified inputs or bandwidth selection, unlike the kernel-based approaches.

\captionsetup[subfigure]{justification=centering}

\captionsetup[subfigure]{justification=centering}

\begin{figure}[p]
    \centering
    \begin{tabular}{c c c c}
        & \small $n=100$ & \small $n=200$ & \small $n=1000$ \\[0.5ex]
        \rowlab{normal} &
        \includegraphics[width=\densityfigw, height=\densityfigh]{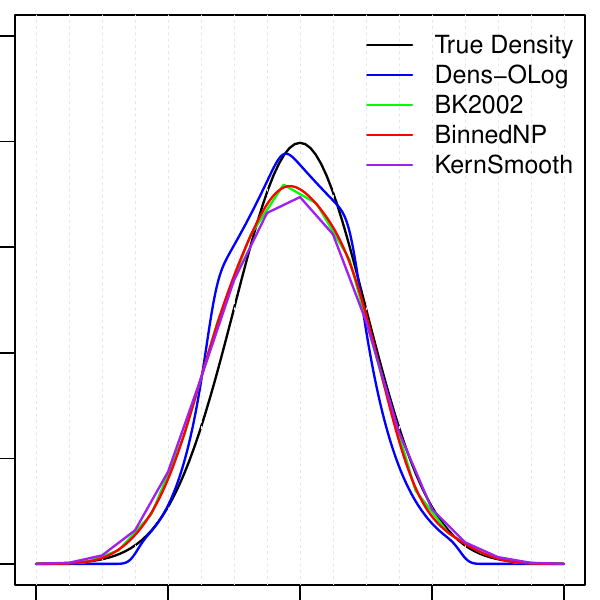} &
        \includegraphics[width=\densityfigw, height=\densityfigh]{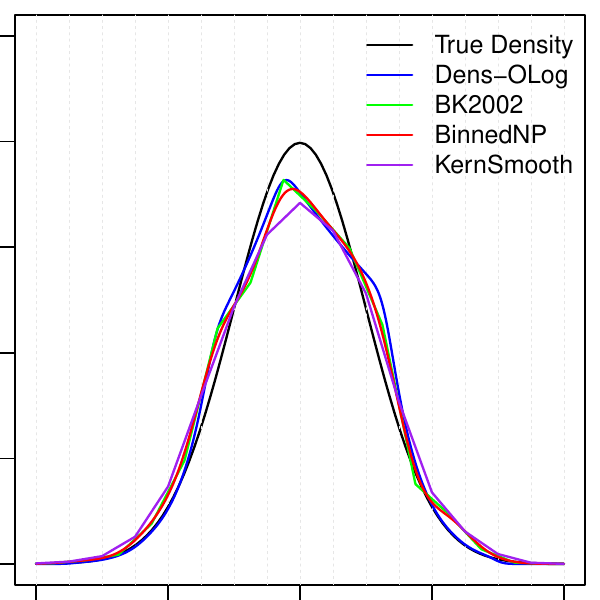} &
        \includegraphics[width=\densityfigw, height=\densityfigh]{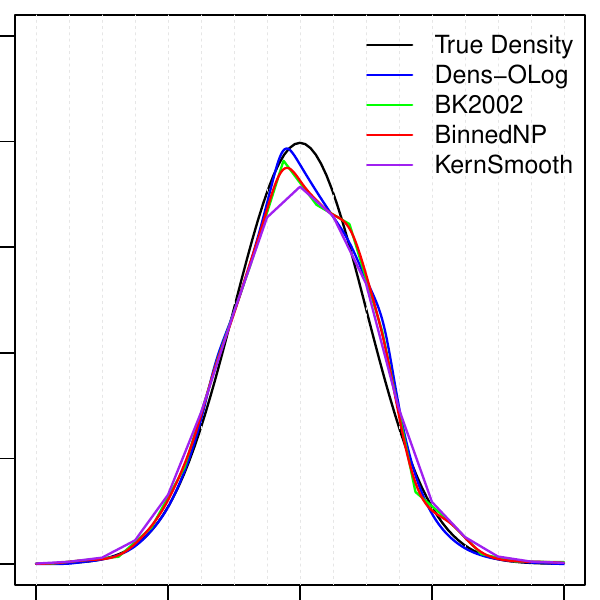} \\
        \rowlab{Gamma} &
        \includegraphics[width=\densityfigw, height=\densityfigh]{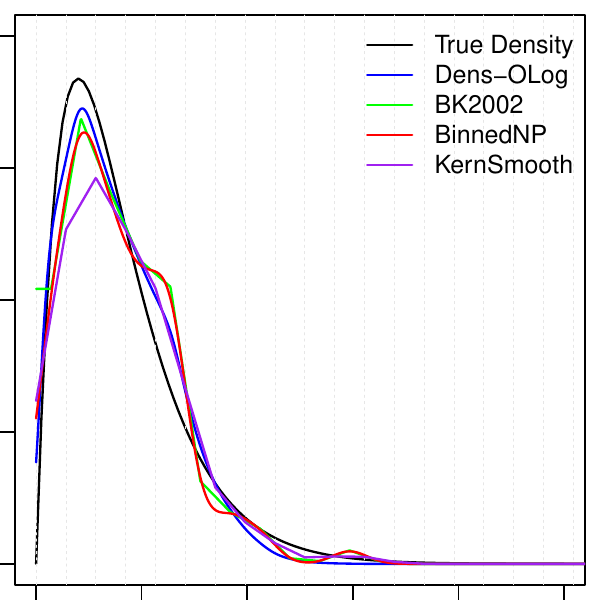} &
        \includegraphics[width=\densityfigw, height=\densityfigh]{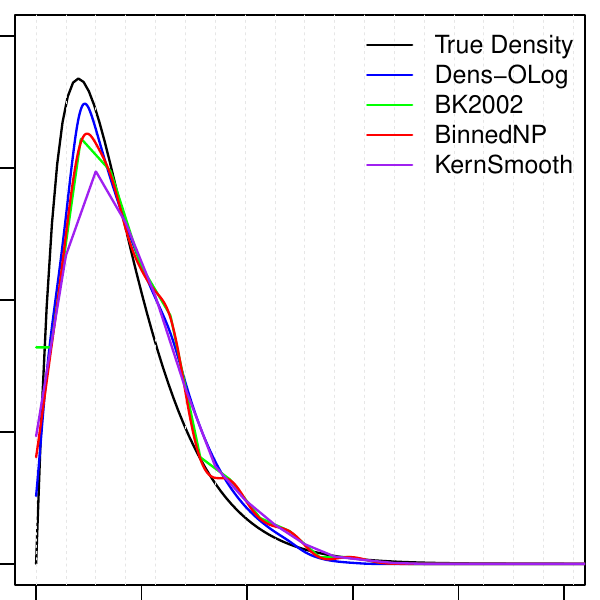} &
        \includegraphics[width=\densityfigw, height=\densityfigh]{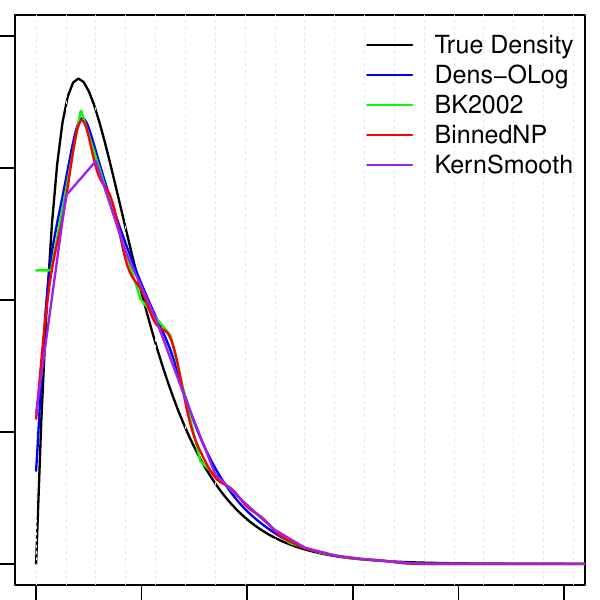} \\
        \rowlab{chi-square} &
        \includegraphics[width=\densityfigw, height=\densityfigh]{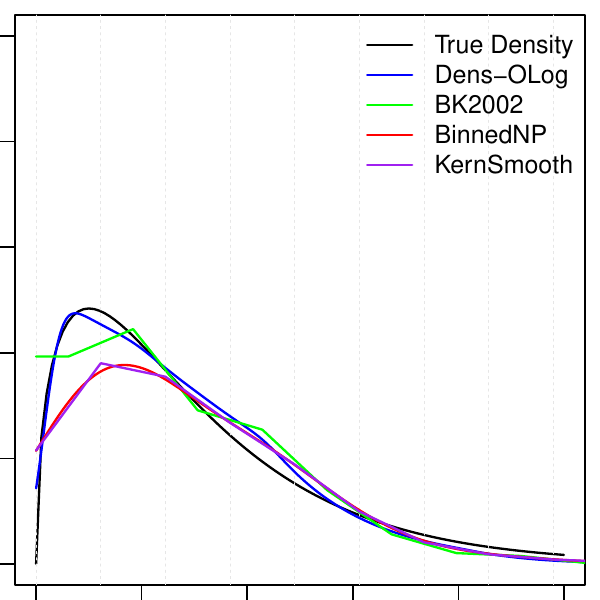} &
        \includegraphics[width=\densityfigw, height=\densityfigh]{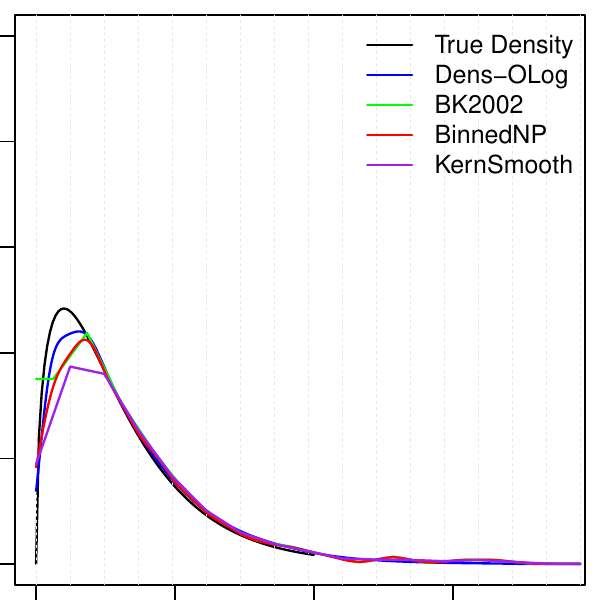} &
        \includegraphics[width=\densityfigw, height=\densityfigh]{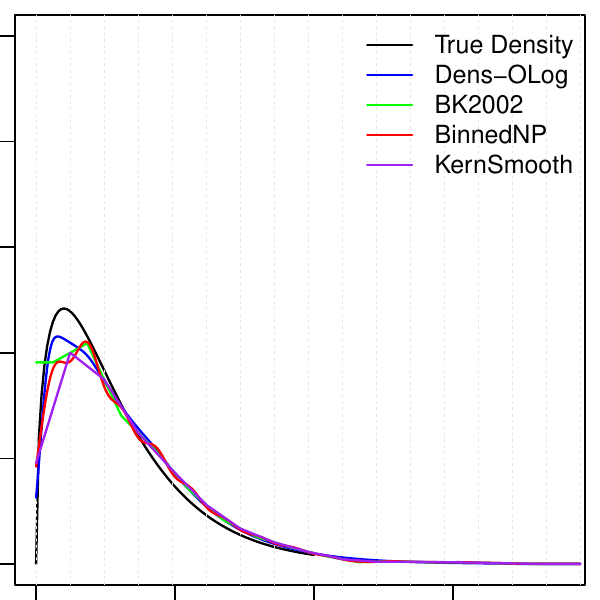} \\
        \rowlab{Pareto} &
        \includegraphics[width=\densityfigw, height=\densityfigh]{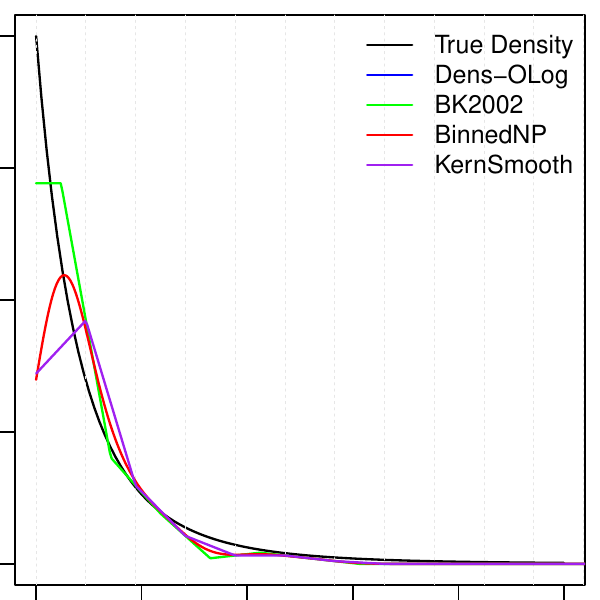} &
        \includegraphics[width=\densityfigw, height=\densityfigh]{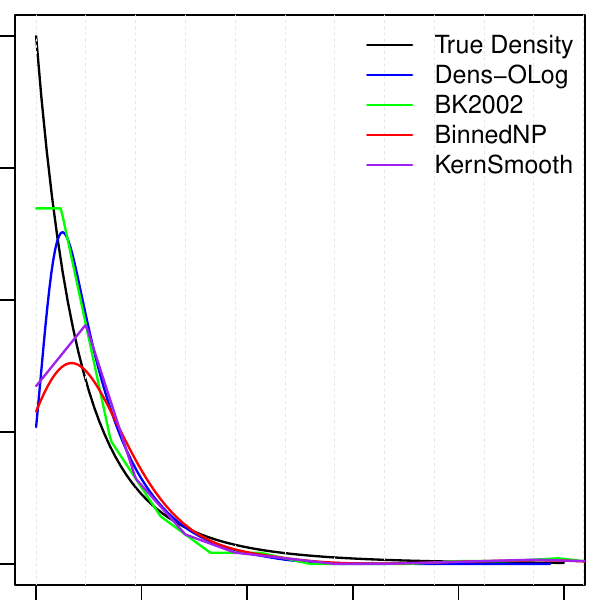} &
        \includegraphics[width=\densityfigw, height=\densityfigh]{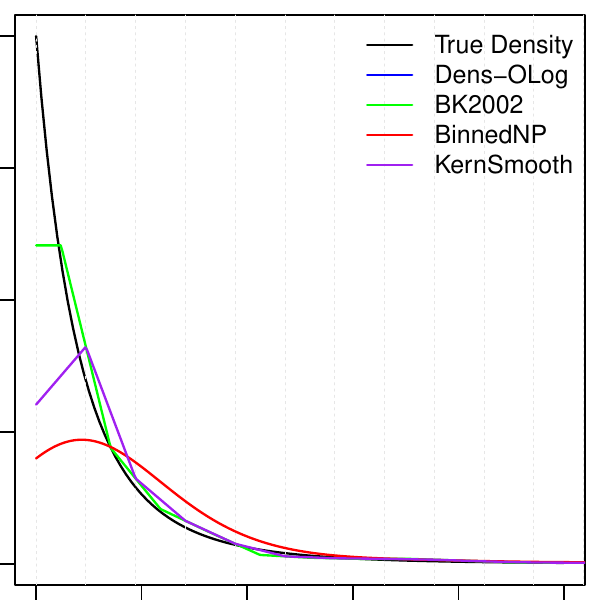}
    \end{tabular}
\caption{Comparison of MALC density estimates against true underlying densities. Rows (top to bottom) correspond to the normal, gamma, chi-square, and Pareto distributions. Columns (left to right) correspond to varying sample sizes $n \in \{100, 200, 1000\}$ with the standardized grid width fixed at $\delta/\sigma = 0.5$.}\label{fig:comparison_densities}
\end{figure}

To provide some intuition for the simulation results, in Figure~\ref{fig:comparison_densities}, we plot examples of density estimates obtained for the normal, gamma, chi-square, and Pareto distributions at sample sizes $n \in \{100, 200, 1000\}$ with the standardized grid width fixed at $\delta/\sigma = 0.5$. At smaller sample sizes, MALC more closely follows the true density across the evaluated distributions. Although the simulation results indicate that this advantage becomes less pronounced as $n$ increases to $10^4$, $10^5$, and $10^6$, the small-sample advantage remains practically relevant because grouped data are often observed with limited sample sizes in real-world settings.

\begin{figure}[!htb]
    \centering
    \begin{tabular}{c | c c c c}
        & \small MALC & \small BK2002 & \small BinnedNP & \small KernSmooth \\[1ex]
        \hline\rule{0pt}{3ex}
        
        \raisebox{0.08\textwidth}{\rotatebox{90}{\small normal}} &
        \subfloat{\includegraphics[width=0.19\textwidth,height=0.13\textheight,keepaspectratio]{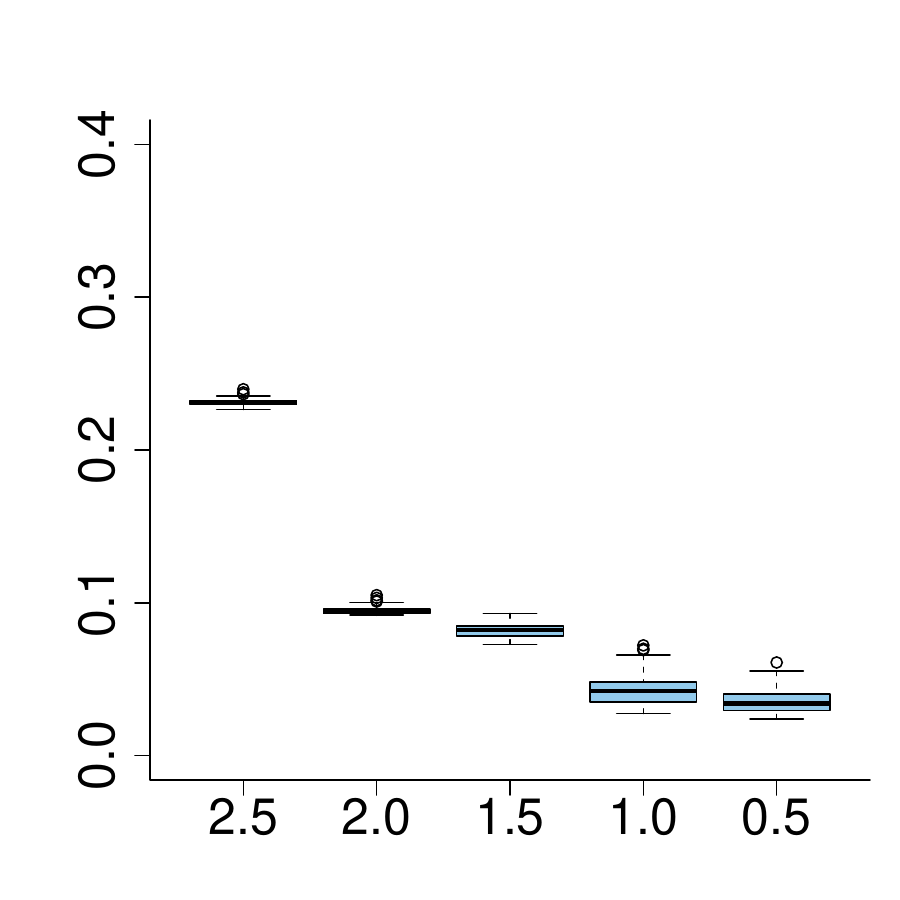}} &
        \subfloat{\includegraphics[width=0.19\textwidth,height=0.13\textheight,keepaspectratio]{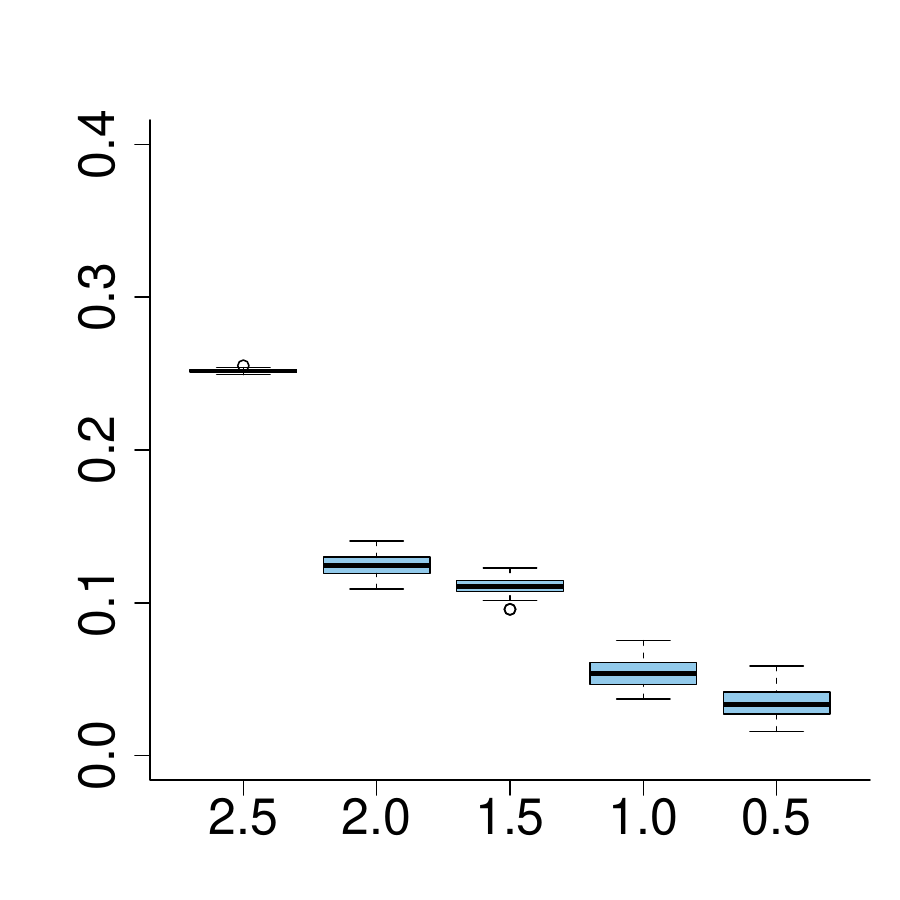}} &
        \subfloat{\includegraphics[width=0.19\textwidth,height=0.13\textheight,keepaspectratio]{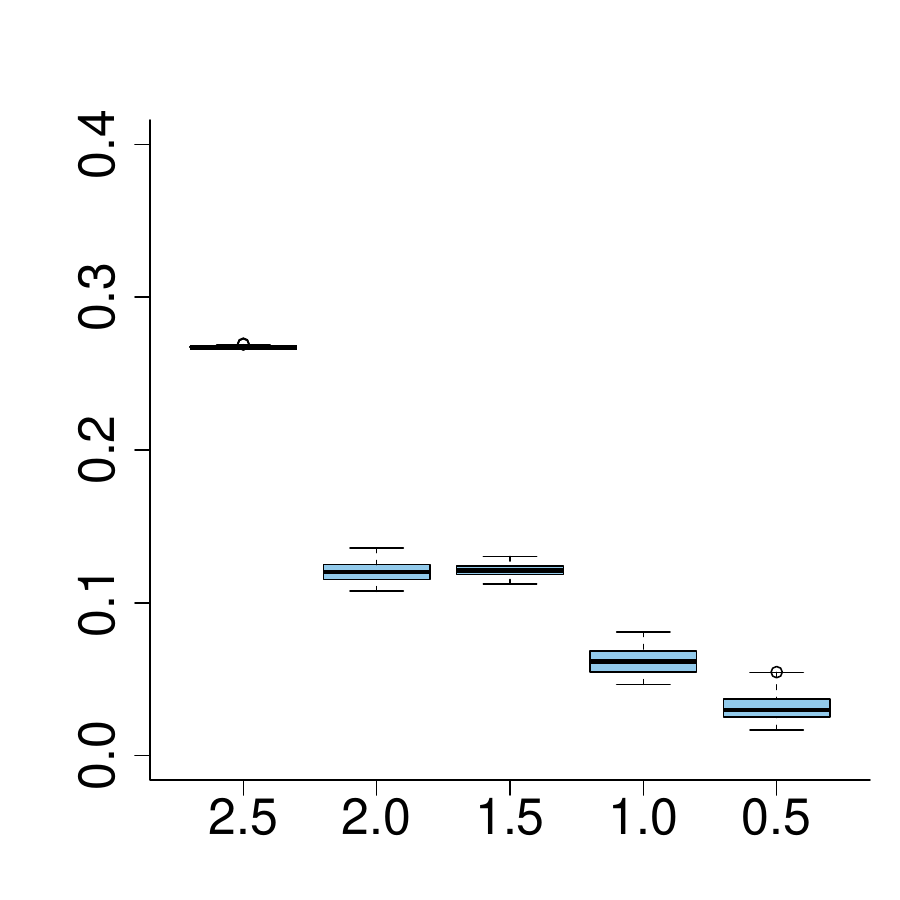}} &
        \subfloat{\includegraphics[width=0.19\textwidth,height=0.13\textheight,keepaspectratio]{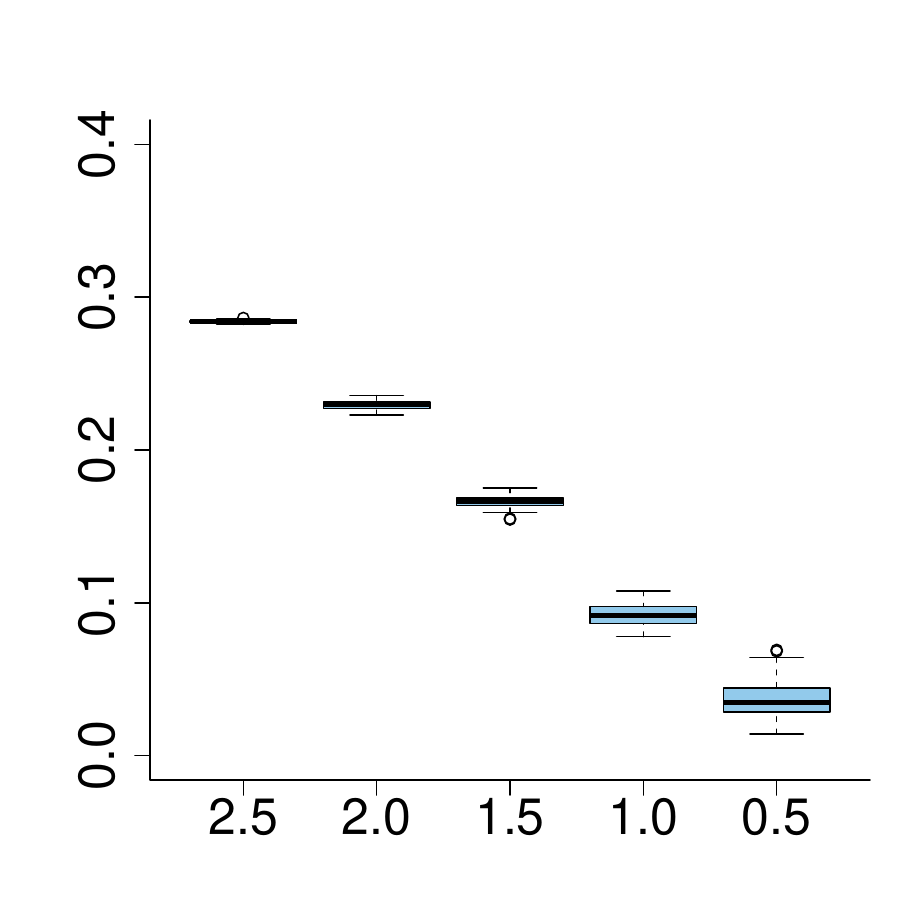}} \\[1ex]

        \raisebox{0.08\textwidth}{\rotatebox{90}{\small Beta}} &
        \subfloat{\includegraphics[width=0.19\textwidth,height=0.13\textheight,keepaspectratio]{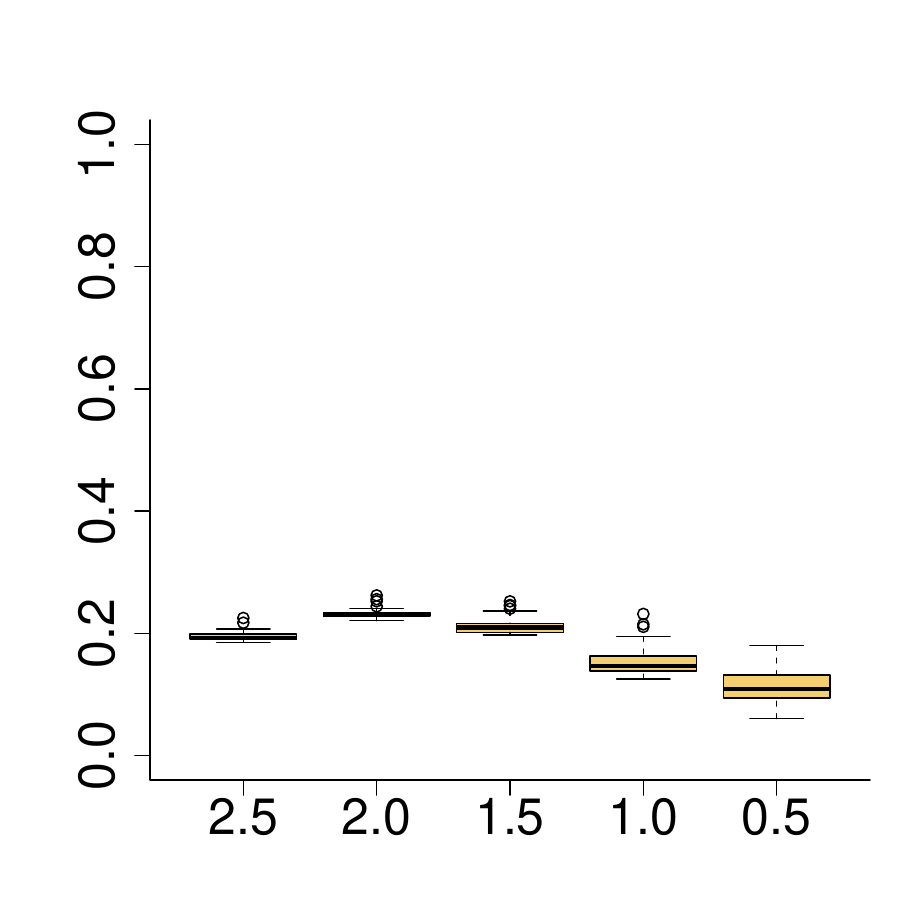}} &
        \subfloat{\includegraphics[width=0.19\textwidth,height=0.13\textheight,keepaspectratio]{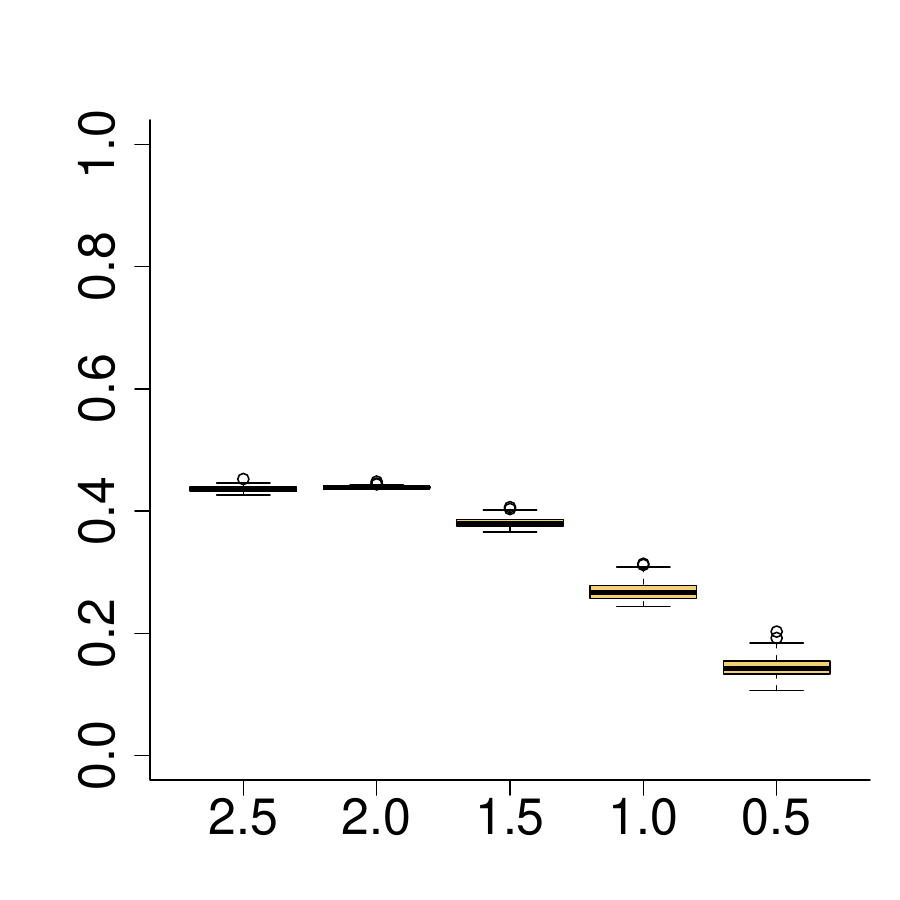}} &
        \subfloat{\includegraphics[width=0.19\textwidth,height=0.13\textheight,keepaspectratio]{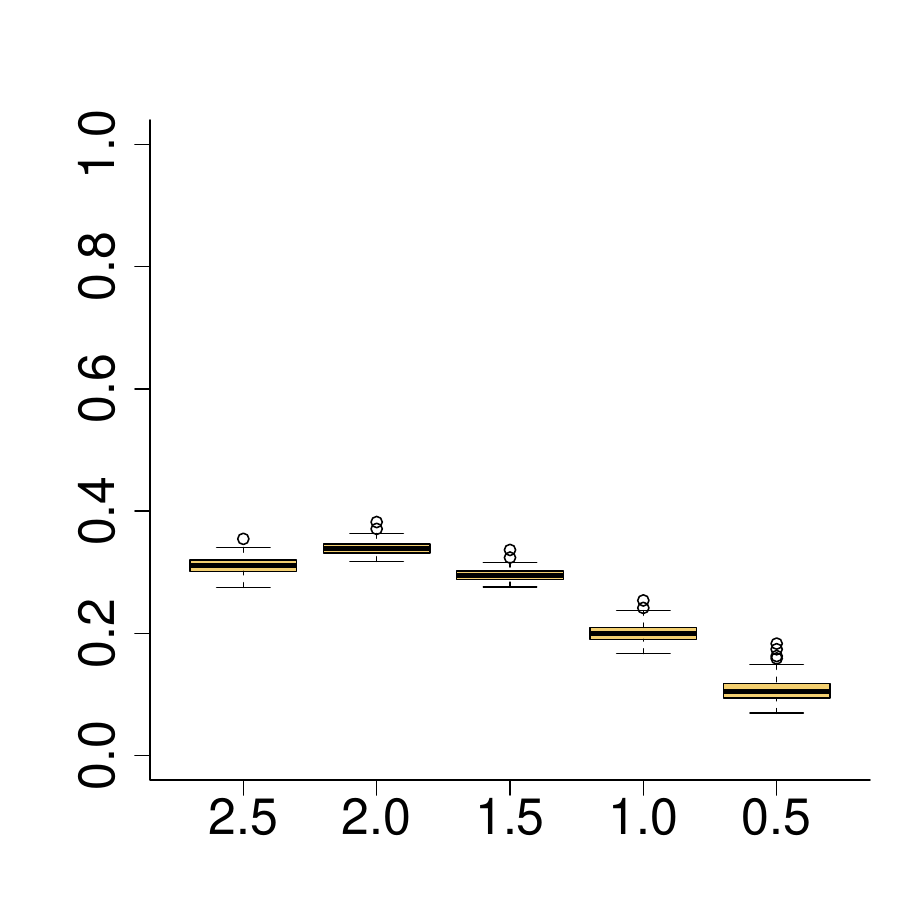}} &
        \subfloat{\includegraphics[width=0.19\textwidth,height=0.13\textheight,keepaspectratio]{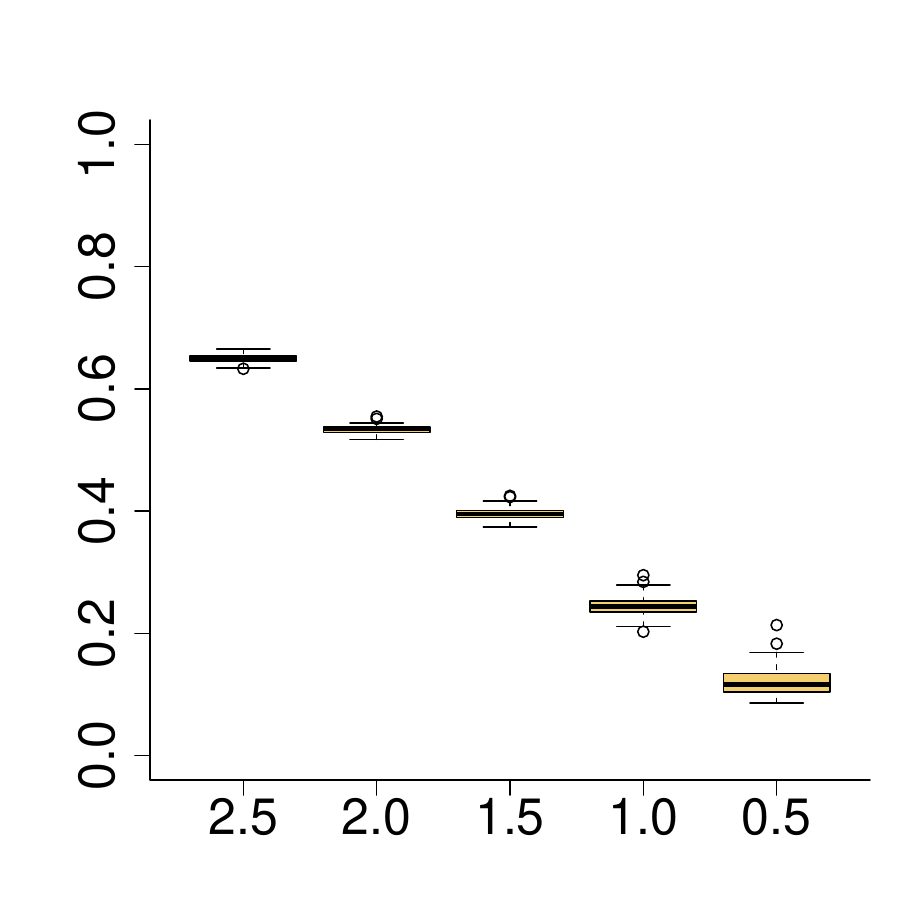}} \\[1ex]

        \raisebox{0.08\textwidth}{\rotatebox{90}{\small Gamma}} &
        \subfloat{\includegraphics[width=0.19\textwidth,height=0.13\textheight,keepaspectratio]{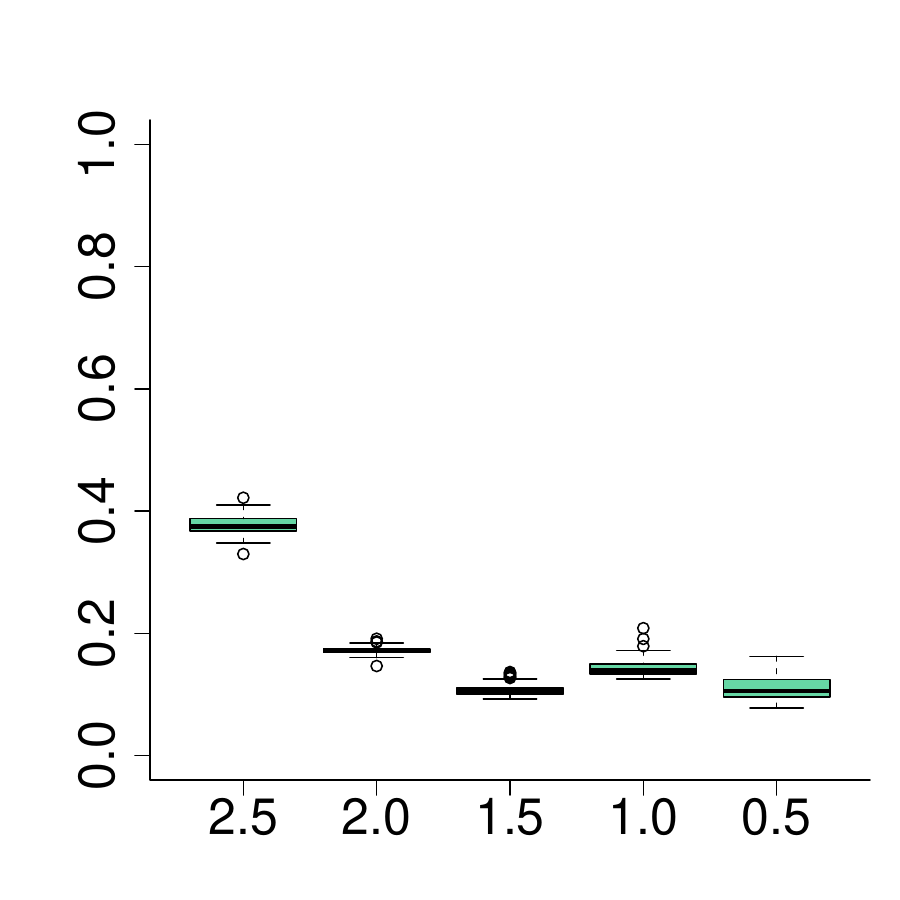}} &
        \subfloat{\includegraphics[width=0.19\textwidth,height=0.13\textheight,keepaspectratio]{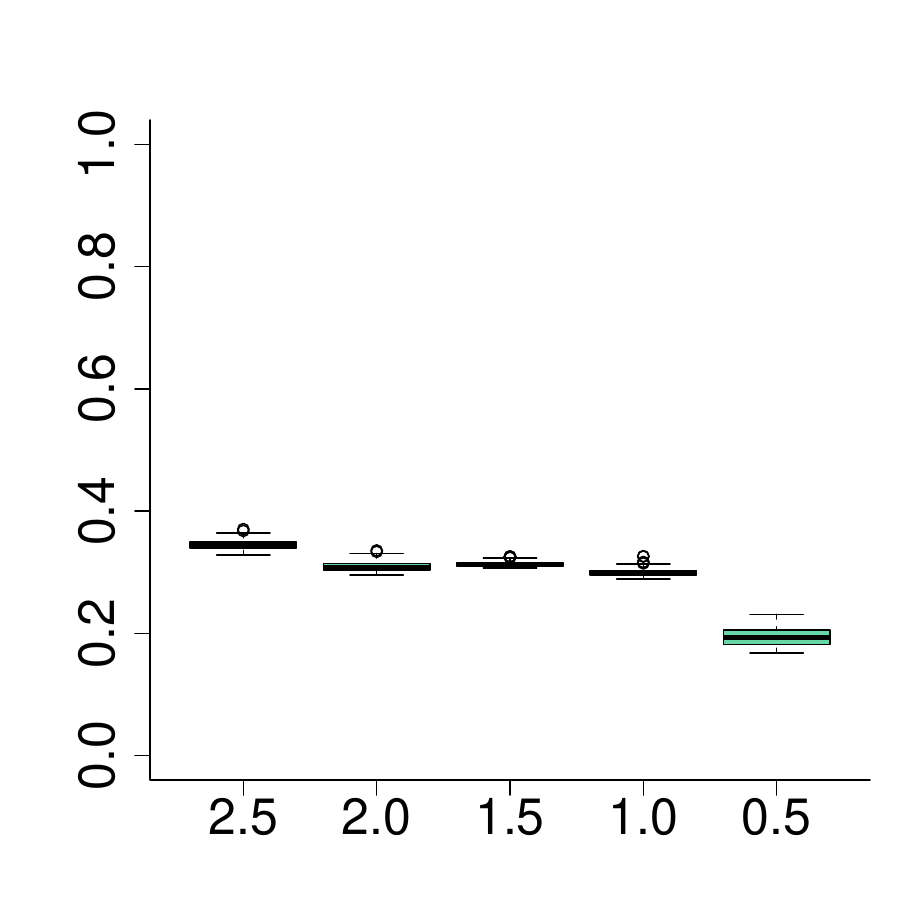}} &
        \subfloat{\includegraphics[width=0.19\textwidth,height=0.13\textheight,keepaspectratio]{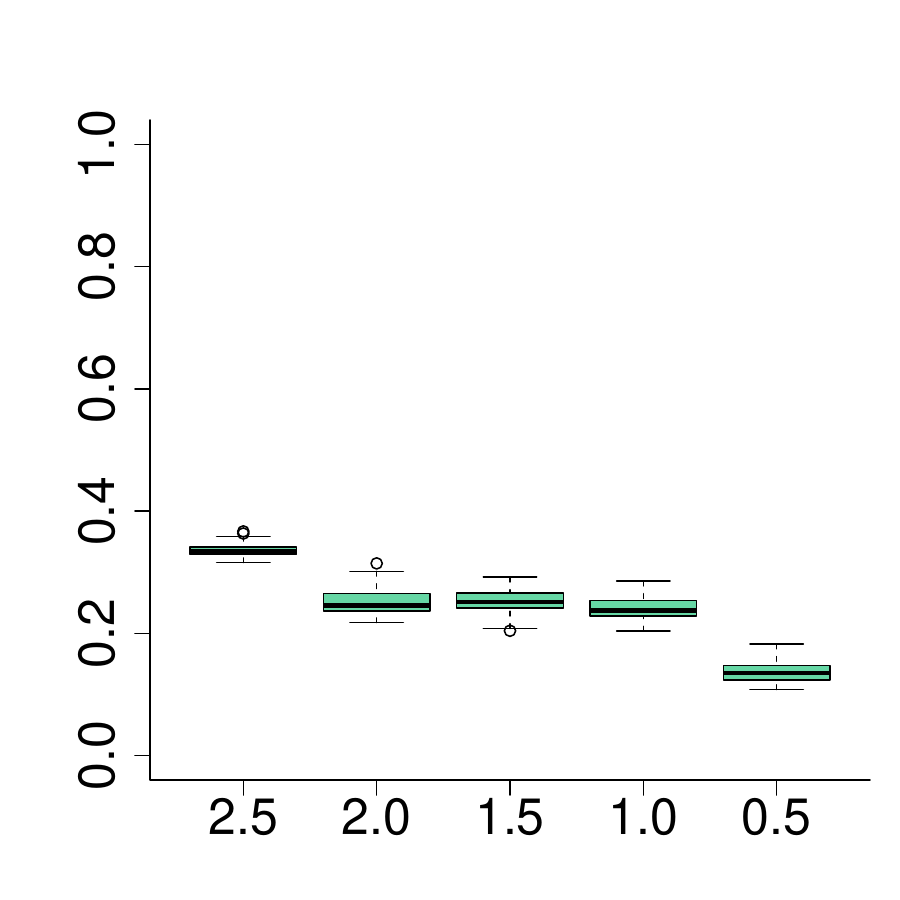}} &
        \subfloat{\includegraphics[width=0.19\textwidth,height=0.13\textheight,keepaspectratio]{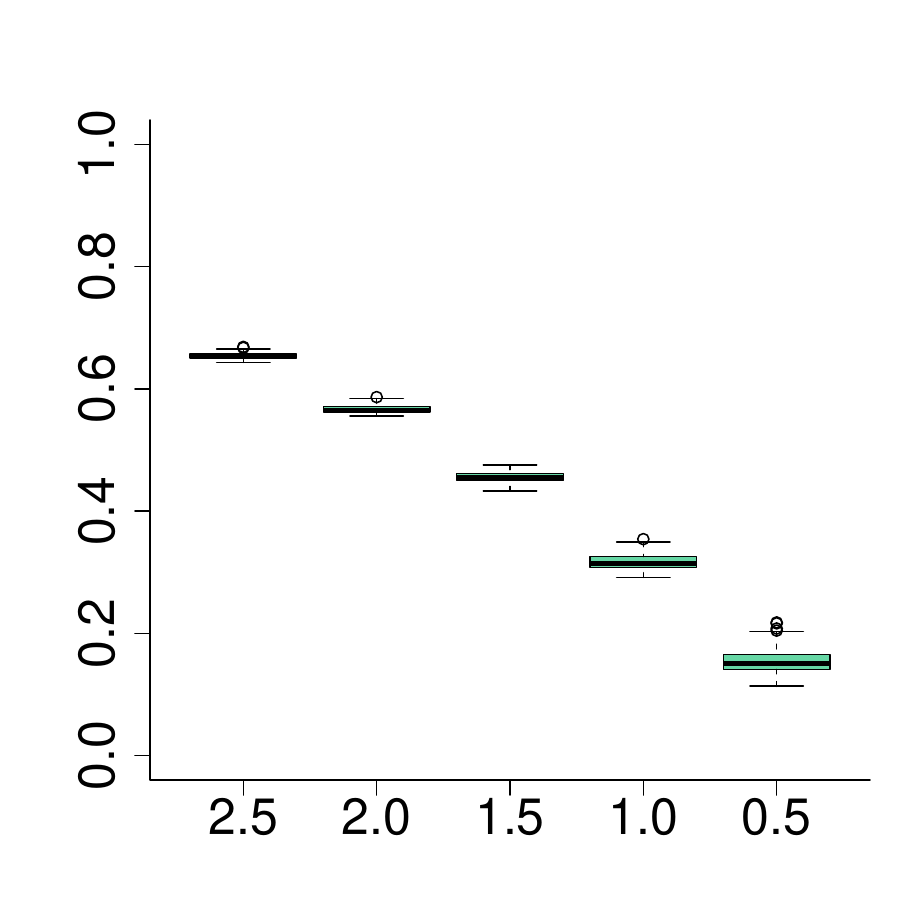}} \\[1ex]

        \raisebox{0.08\textwidth}{\rotatebox{90}{\small logistic}} &
        \subfloat{\includegraphics[width=0.19\textwidth,height=0.13\textheight,keepaspectratio]{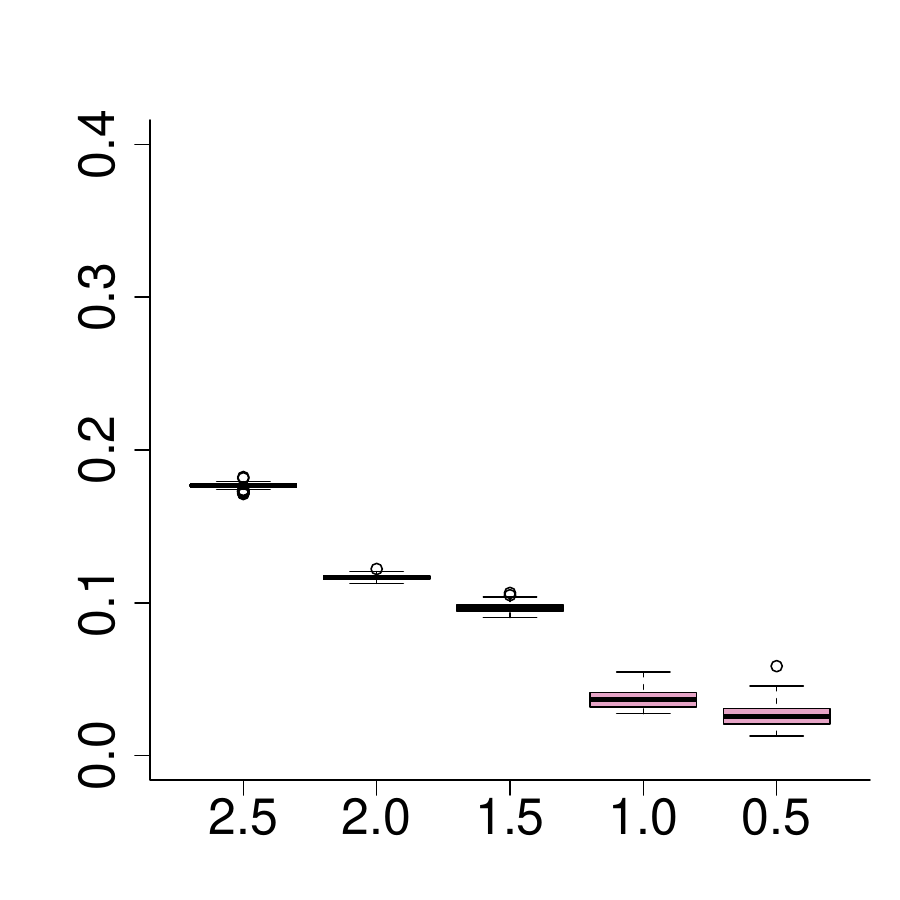}} &
        \subfloat{\includegraphics[width=0.19\textwidth,height=0.13\textheight,keepaspectratio]{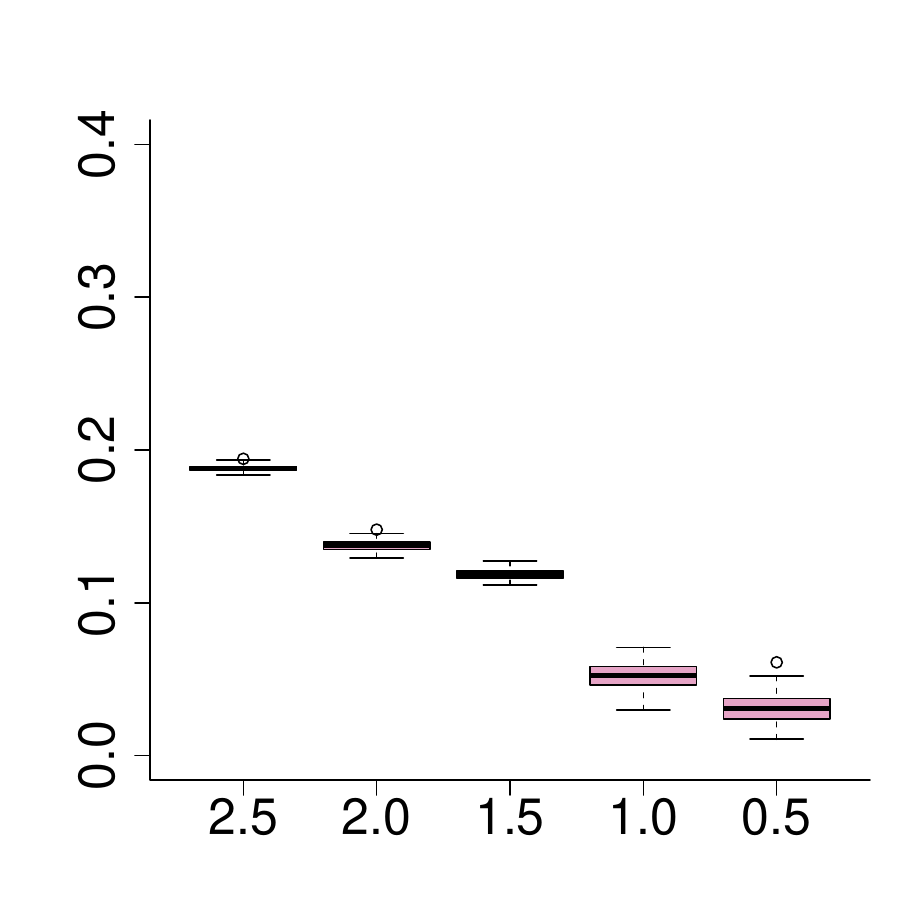}} &
        \subfloat{\includegraphics[width=0.19\textwidth,height=0.13\textheight,keepaspectratio]{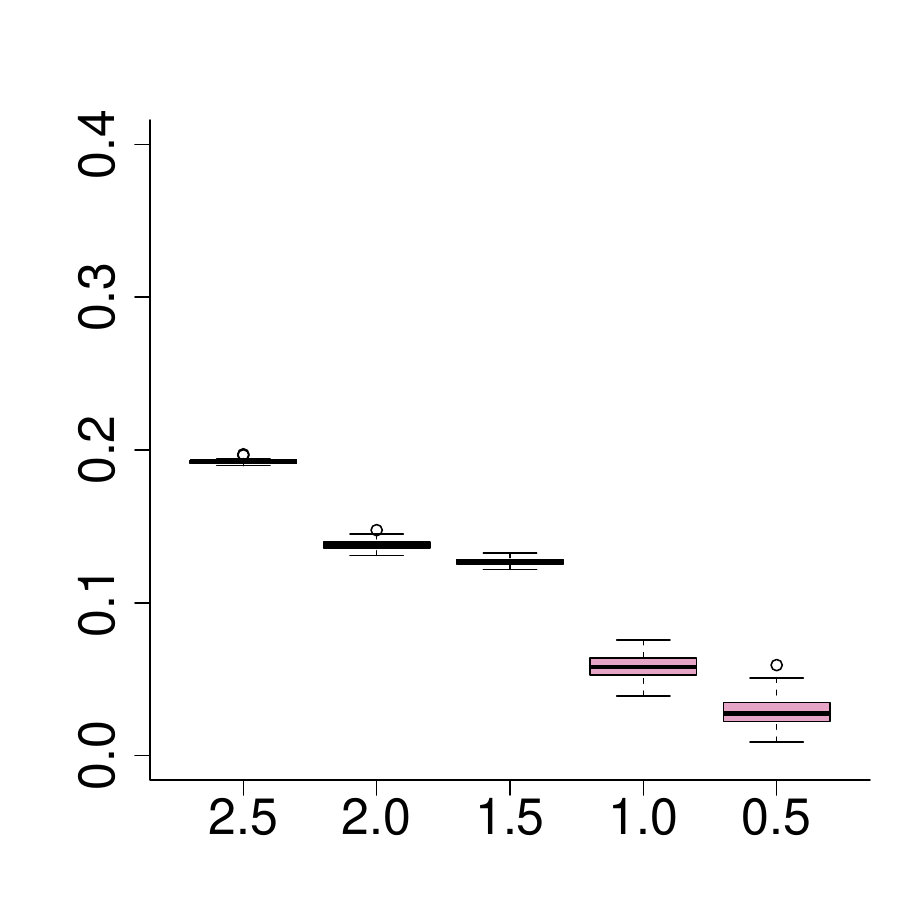}} &
        \subfloat{\includegraphics[width=0.19\textwidth,height=0.13\textheight,keepaspectratio]{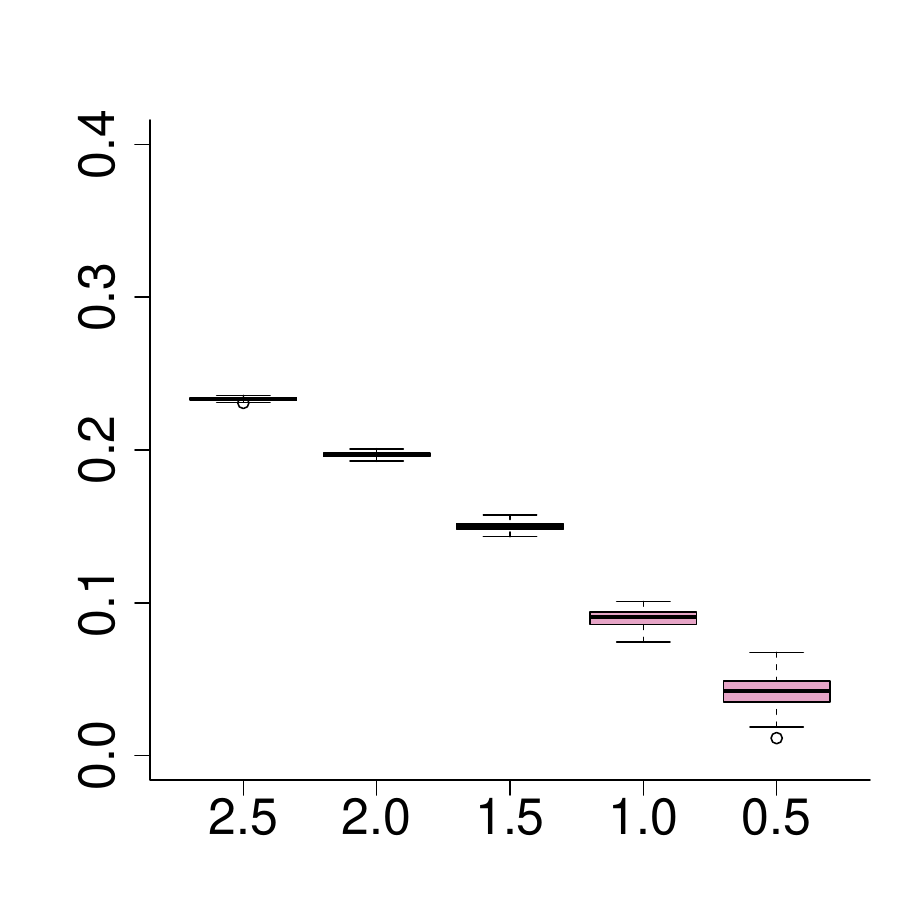}} \\[1ex]

        \raisebox{0.08\textwidth}{\rotatebox{90}{\small Student's $t$}} &
        \subfloat{\includegraphics[width=0.19\textwidth,height=0.13\textheight,keepaspectratio]{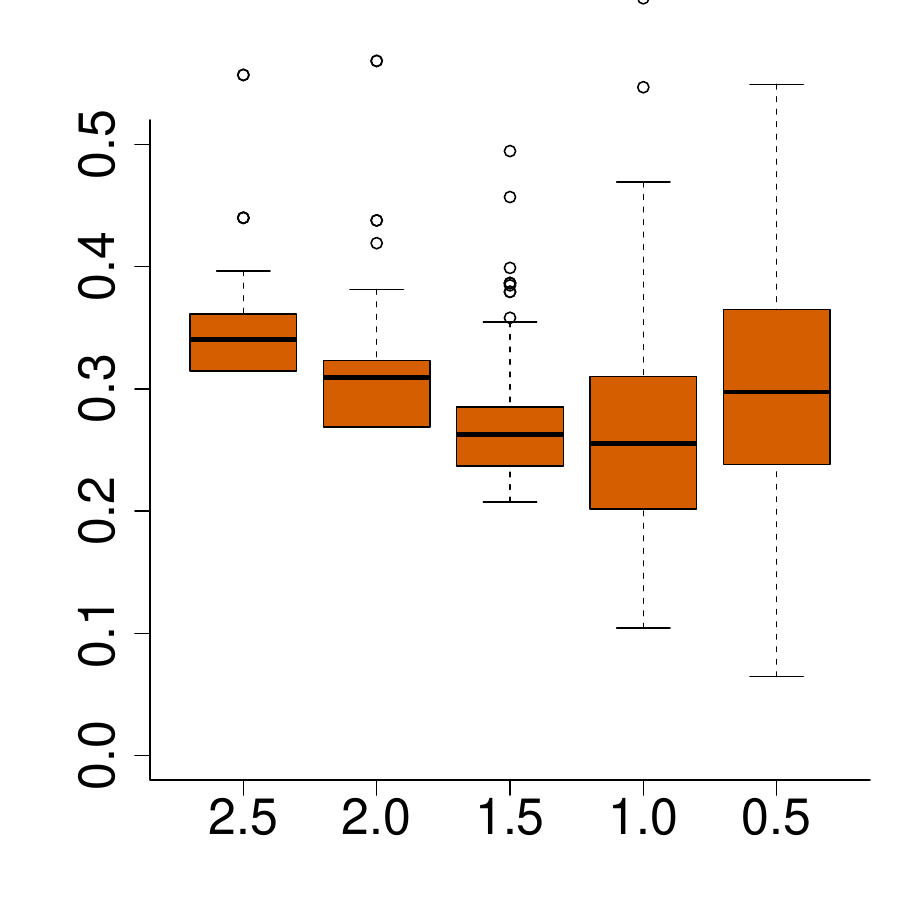}} &
        \subfloat{\includegraphics[width=0.19\textwidth,height=0.13\textheight,keepaspectratio]{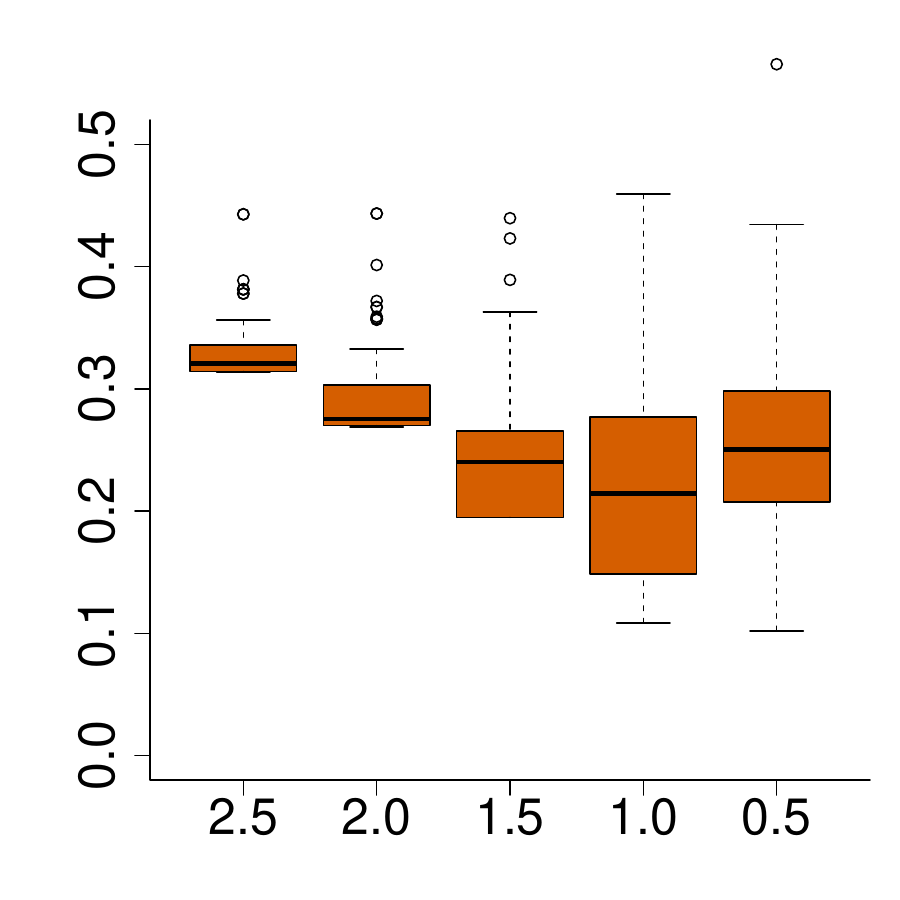}} &
        \subfloat{\includegraphics[width=0.19\textwidth,height=0.13\textheight,keepaspectratio]{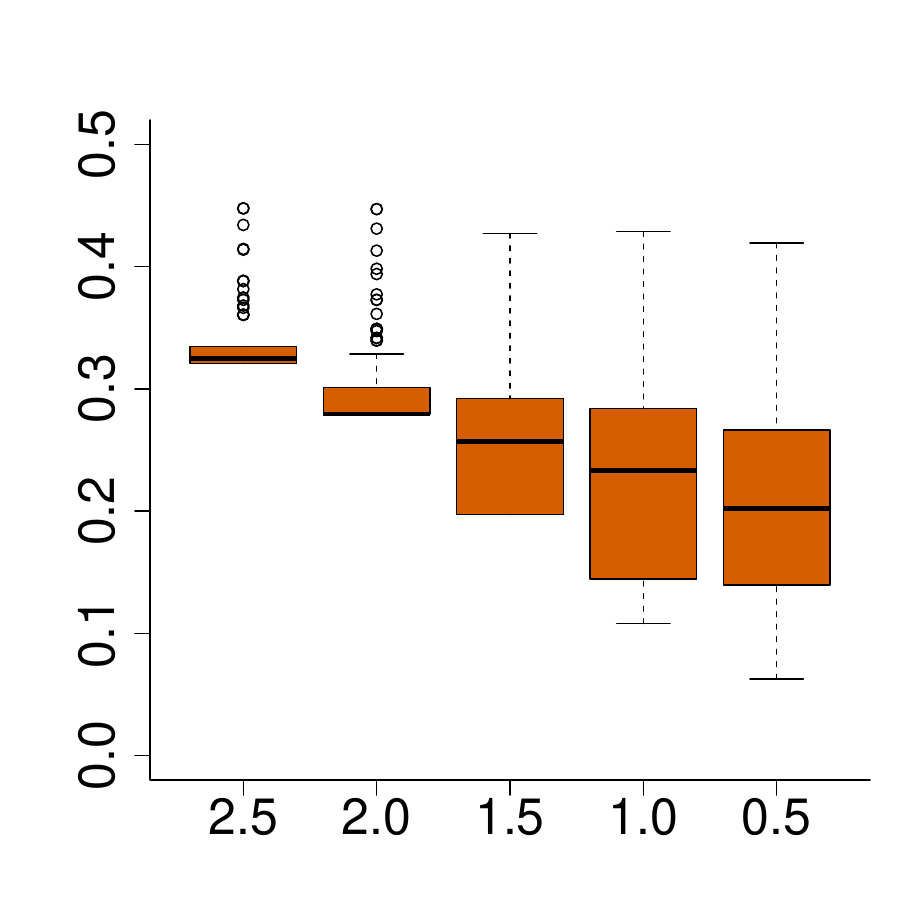}} &
        \subfloat{\includegraphics[width=0.19\textwidth,height=0.13\textheight,keepaspectratio]{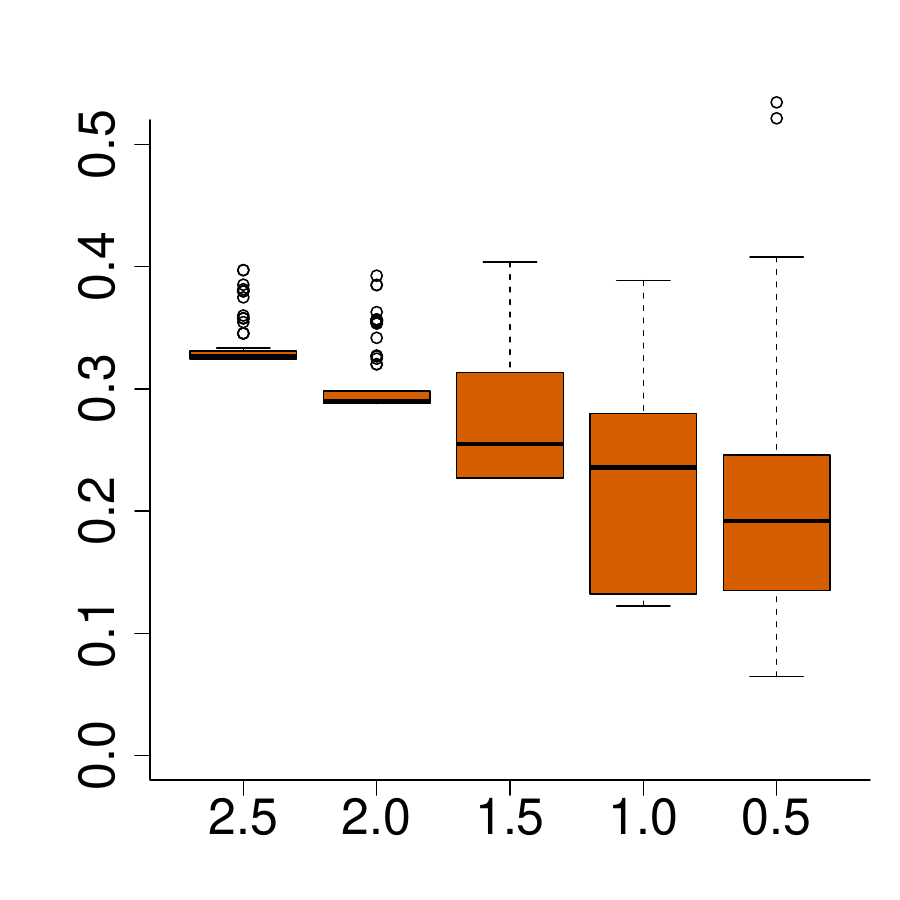}} \\
    \end{tabular}
    
    \caption{Simulation results under grid width variation ($\delta/\sigma \in \{0.5, 1.0, 1.5, 2.0, 2.5\}$) with a fixed sample size ($n = 10^3$). The plots display the $L_2$ norm between the true and estimated densities for the normal, beta, gamma, logistic, and Student's $t$ distributions, using the proposed MALC and three kernel-based methods.}
    \label{fig:pdfgw1}
\end{figure}

\begin{figure}[!htb]
    \centering
    \begin{tabular}{c | c c c c}
        & \small MALC & \small BK2002 & \small BinnedNP & \small KernSmooth \\[1ex]
        \hline\rule{0pt}{3ex}
        
        \raisebox{0.08\textwidth}{\rotatebox{90}{\small Laplace}} &
        \subfloat{\includegraphics[width=0.19\textwidth,height=0.13\textheight,keepaspectratio]{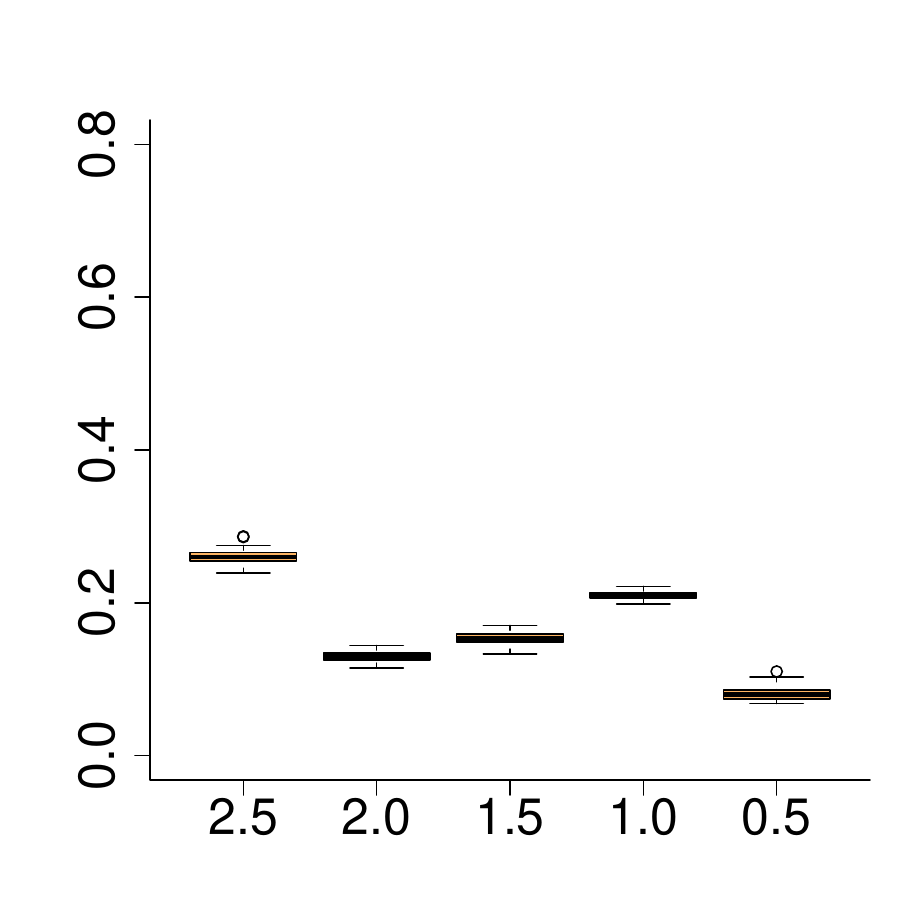}} &
        \subfloat{\includegraphics[width=0.19\textwidth,height=0.13\textheight,keepaspectratio]{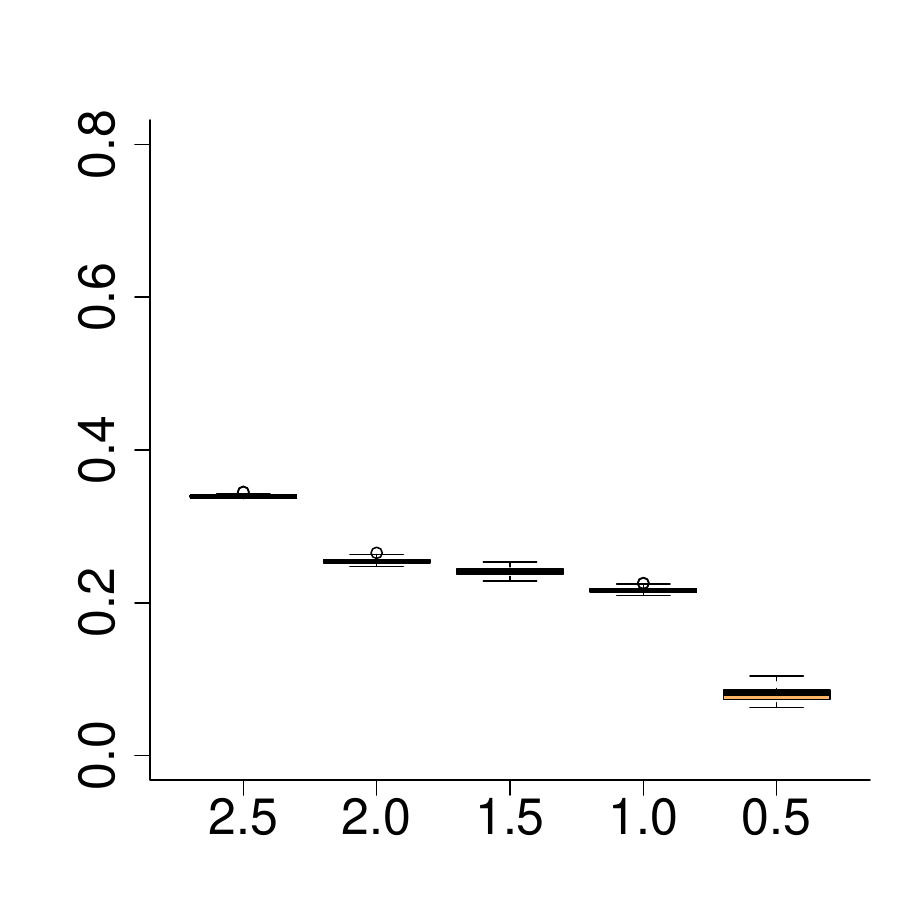}} &
        \subfloat{\includegraphics[width=0.19\textwidth,height=0.13\textheight,keepaspectratio]{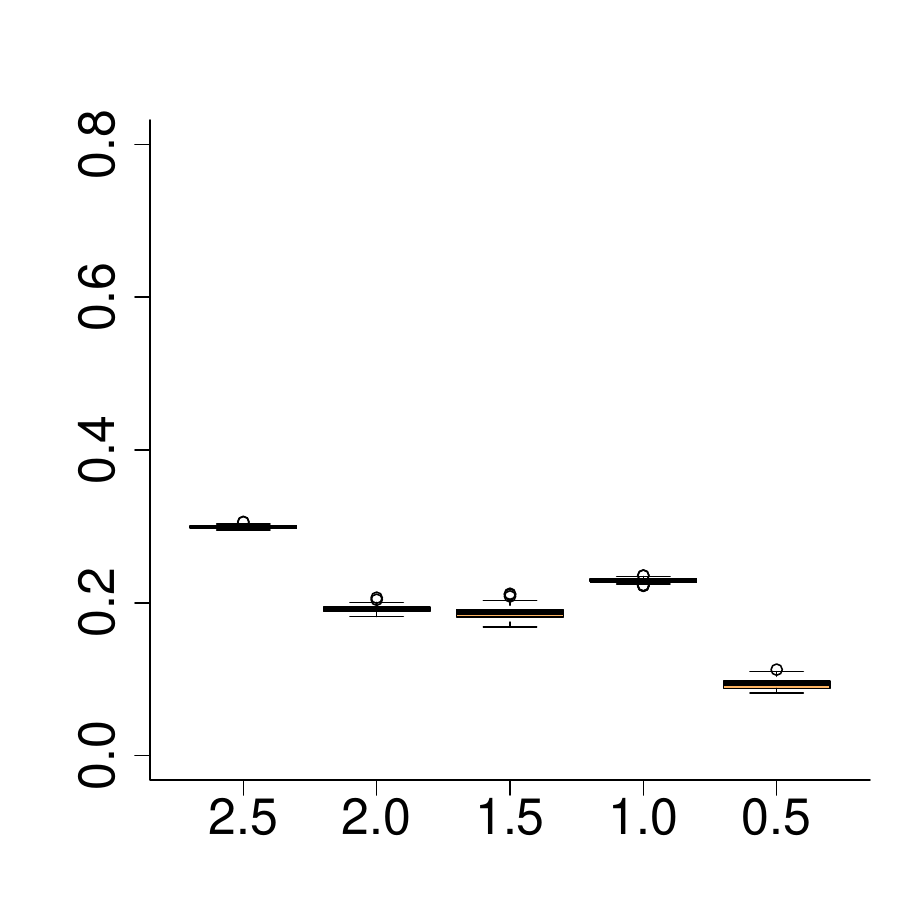}} &
        \subfloat{\includegraphics[width=0.19\textwidth,height=0.13\textheight,keepaspectratio]{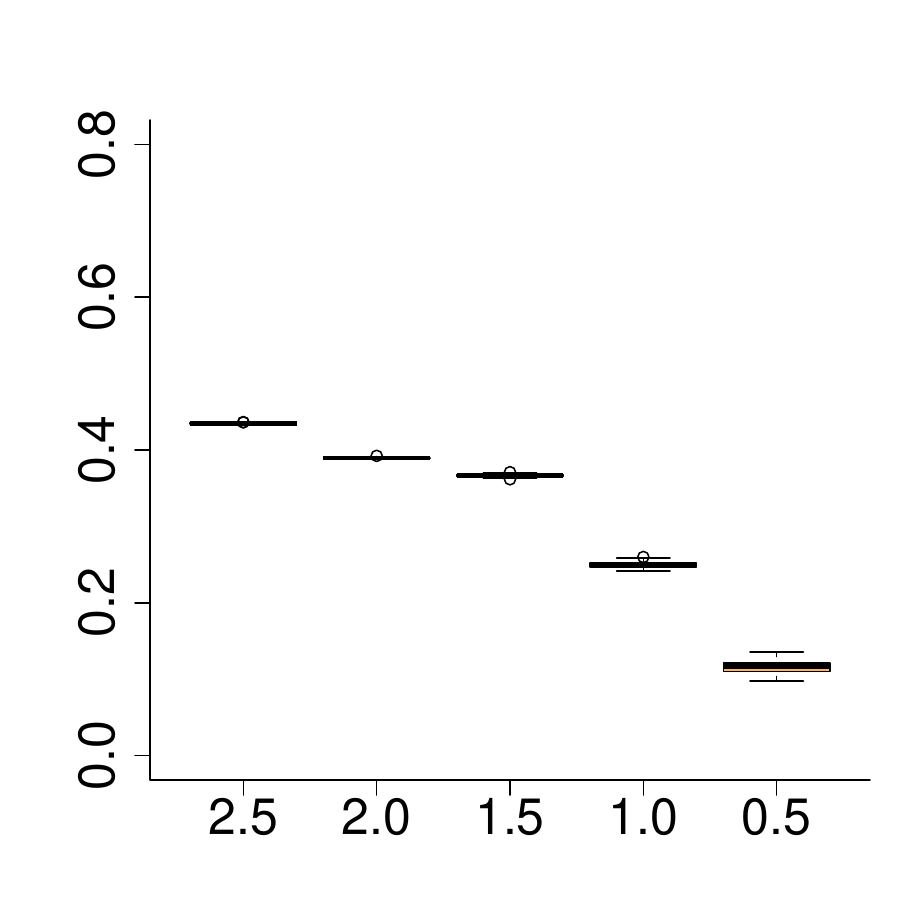}} \\[1ex]

        \raisebox{0.08\textwidth}{\rotatebox{90}{\small chi-square}} &
        \subfloat{\includegraphics[width=0.19\textwidth,height=0.13\textheight,keepaspectratio]{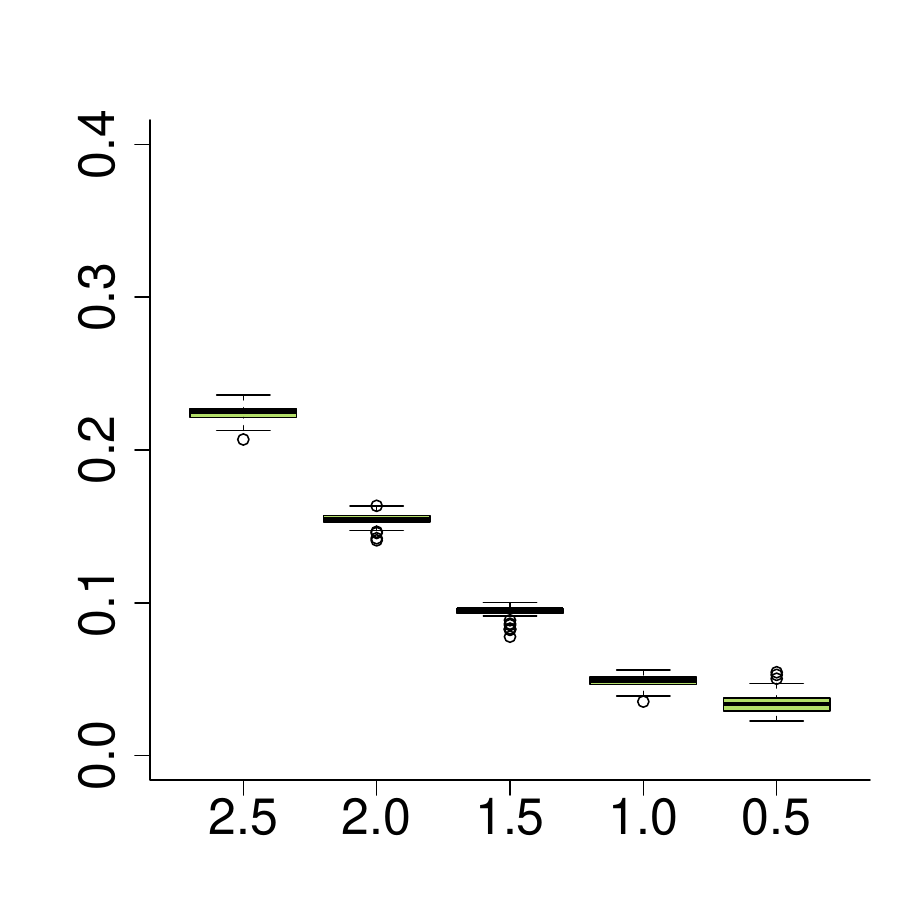}} &
        \subfloat{\includegraphics[width=0.19\textwidth,height=0.13\textheight,keepaspectratio]{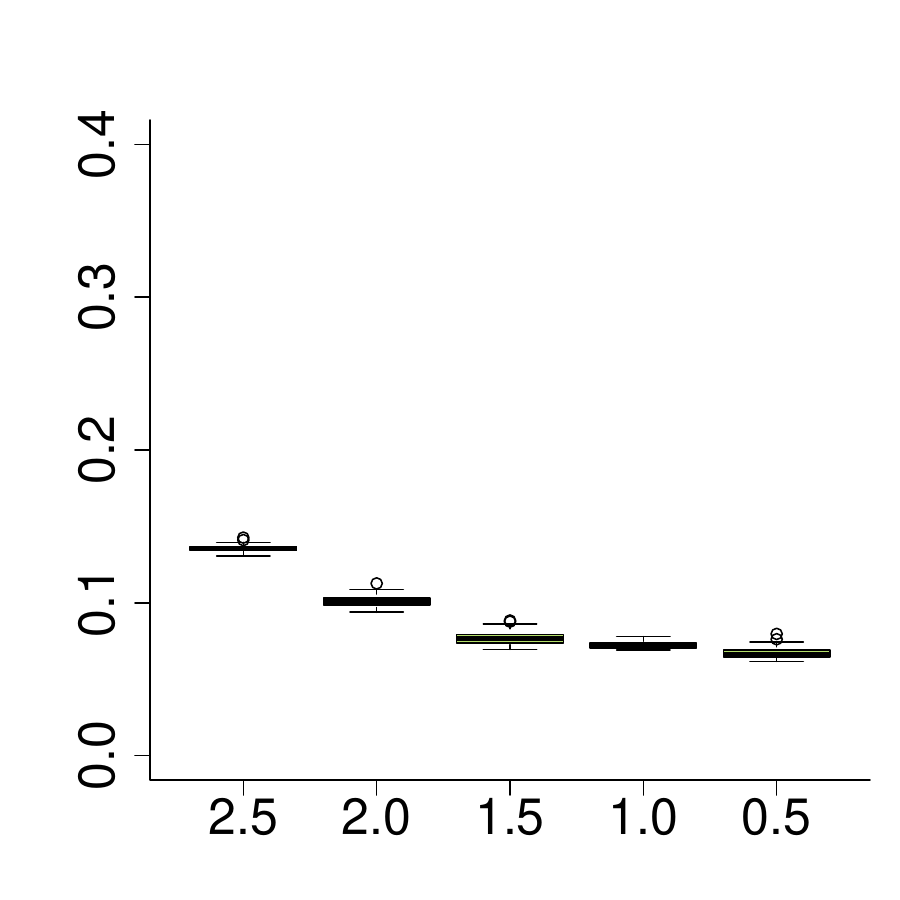}} &
        \subfloat{\includegraphics[width=0.19\textwidth,height=0.13\textheight,keepaspectratio]{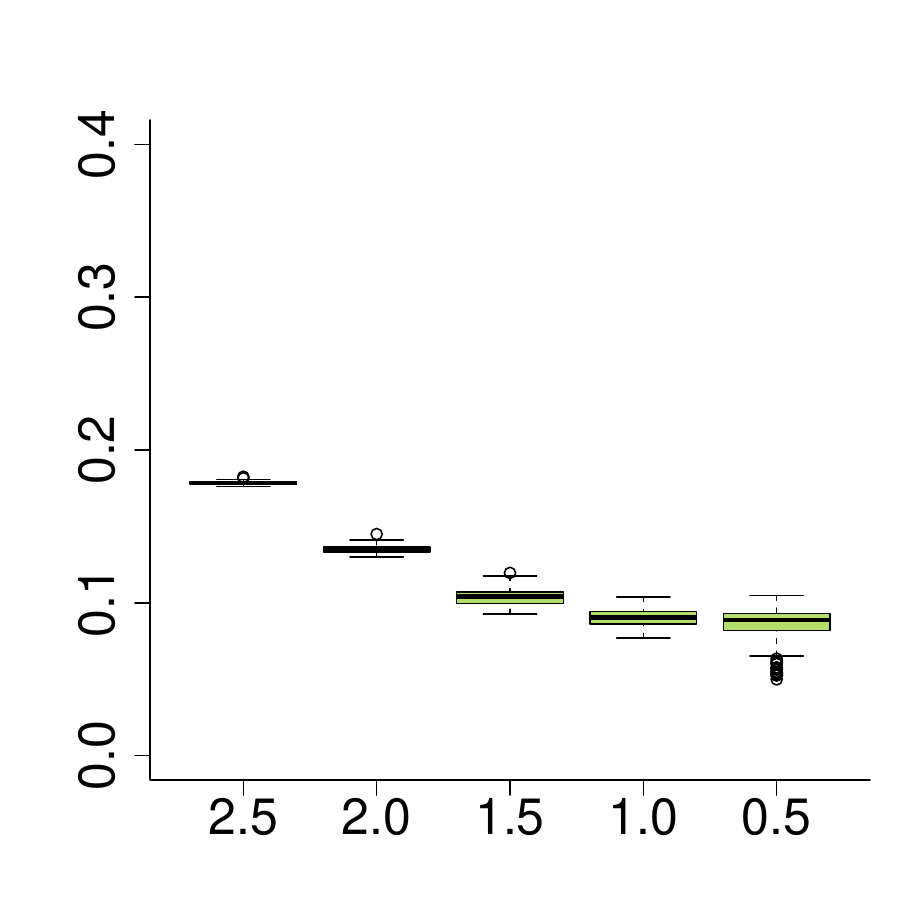}} &
        \subfloat{\includegraphics[width=0.19\textwidth,height=0.13\textheight,keepaspectratio]{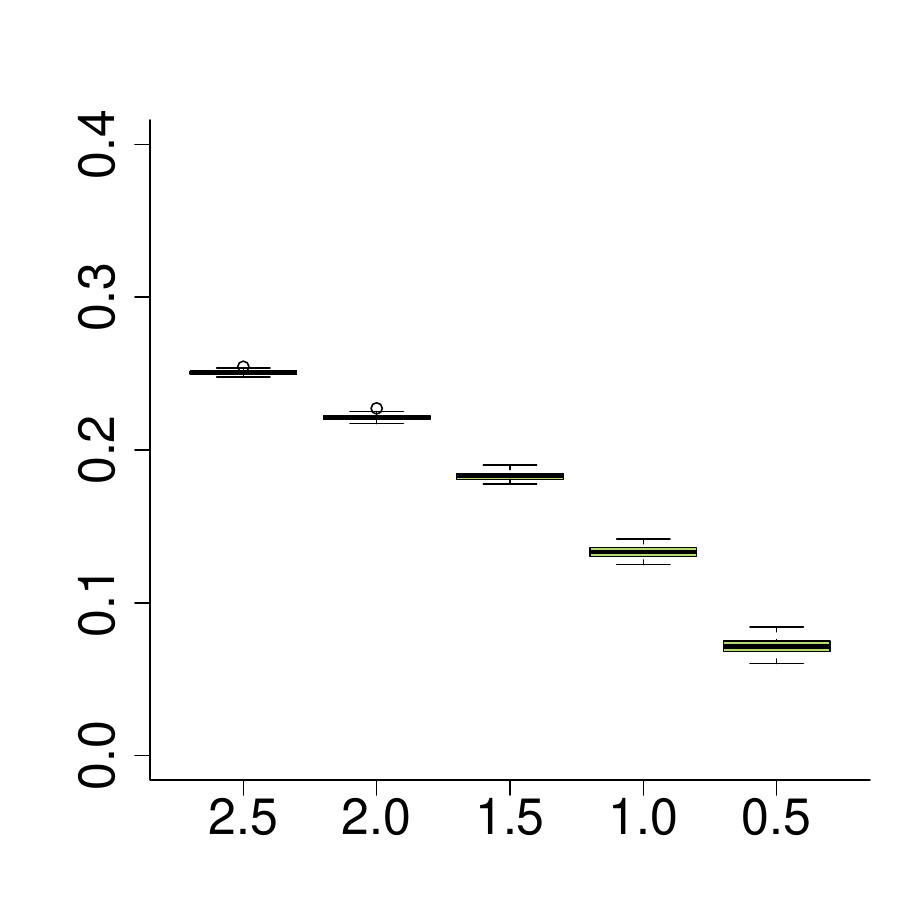}} \\[1ex]

        \raisebox{0.08\textwidth}{\rotatebox{90}{\small log-normal}} &
        \subfloat{\includegraphics[width=0.19\textwidth,height=0.13\textheight,keepaspectratio]{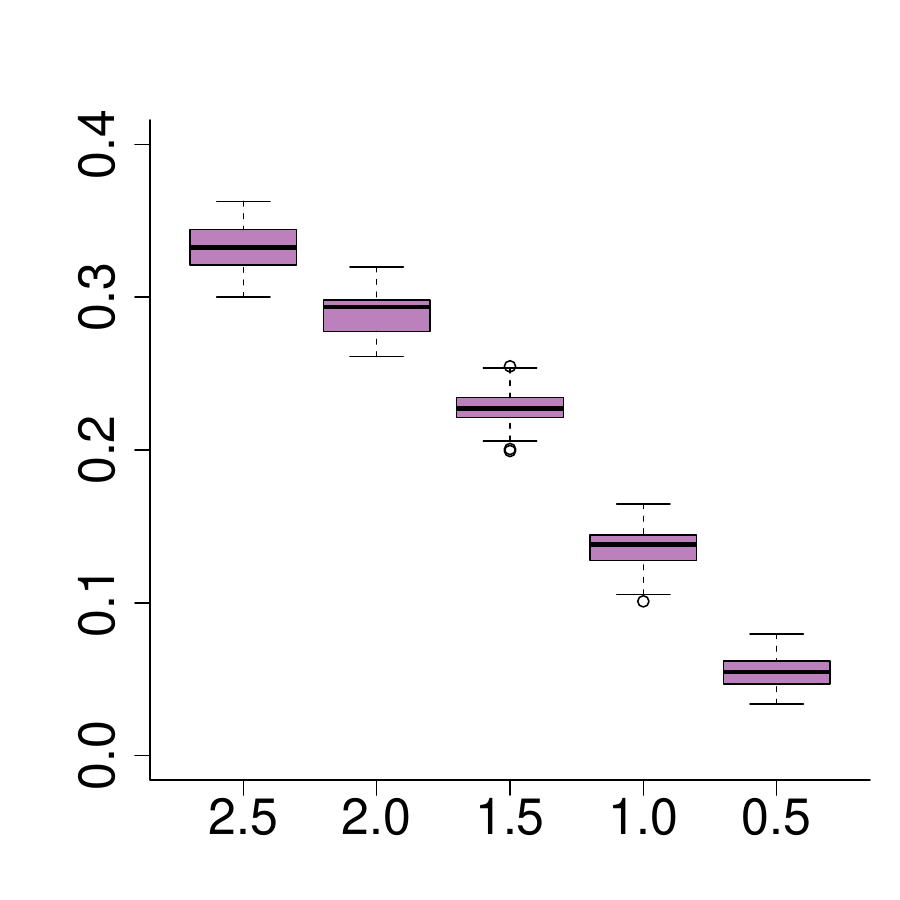}} &
        \subfloat{\includegraphics[width=0.19\textwidth,height=0.13\textheight,keepaspectratio]{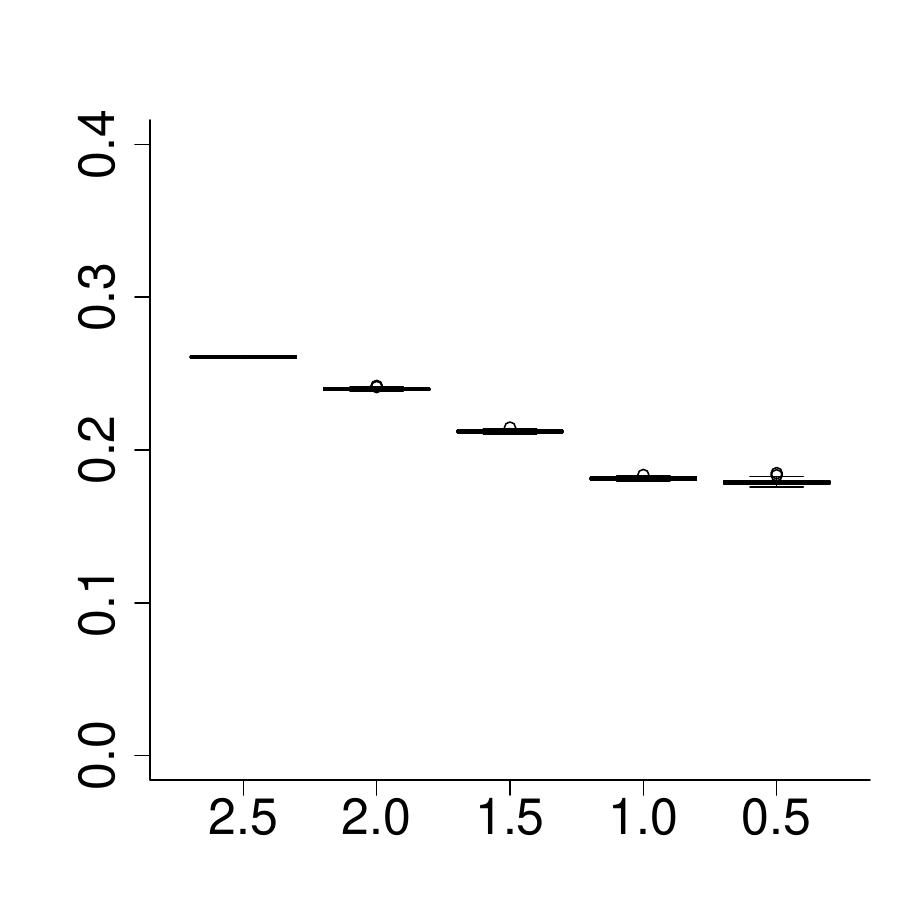}} &
        \subfloat{\includegraphics[width=0.19\textwidth,height=0.13\textheight,keepaspectratio]{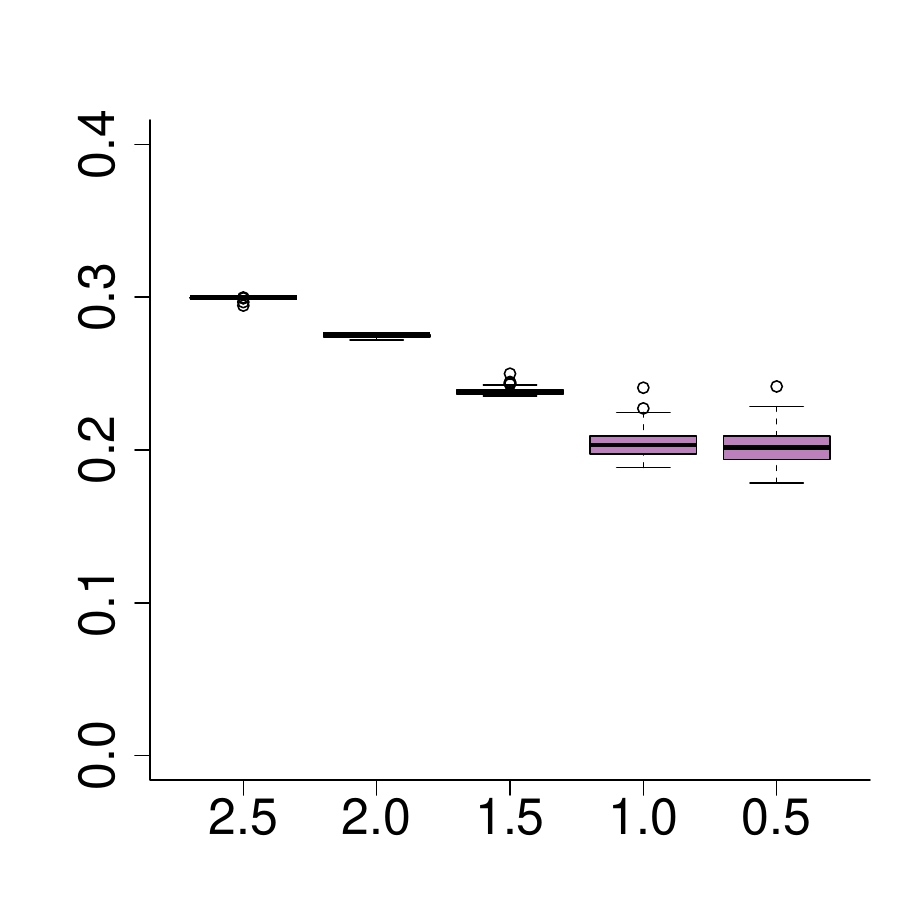}} &
        \subfloat{\includegraphics[width=0.19\textwidth,height=0.13\textheight,keepaspectratio]{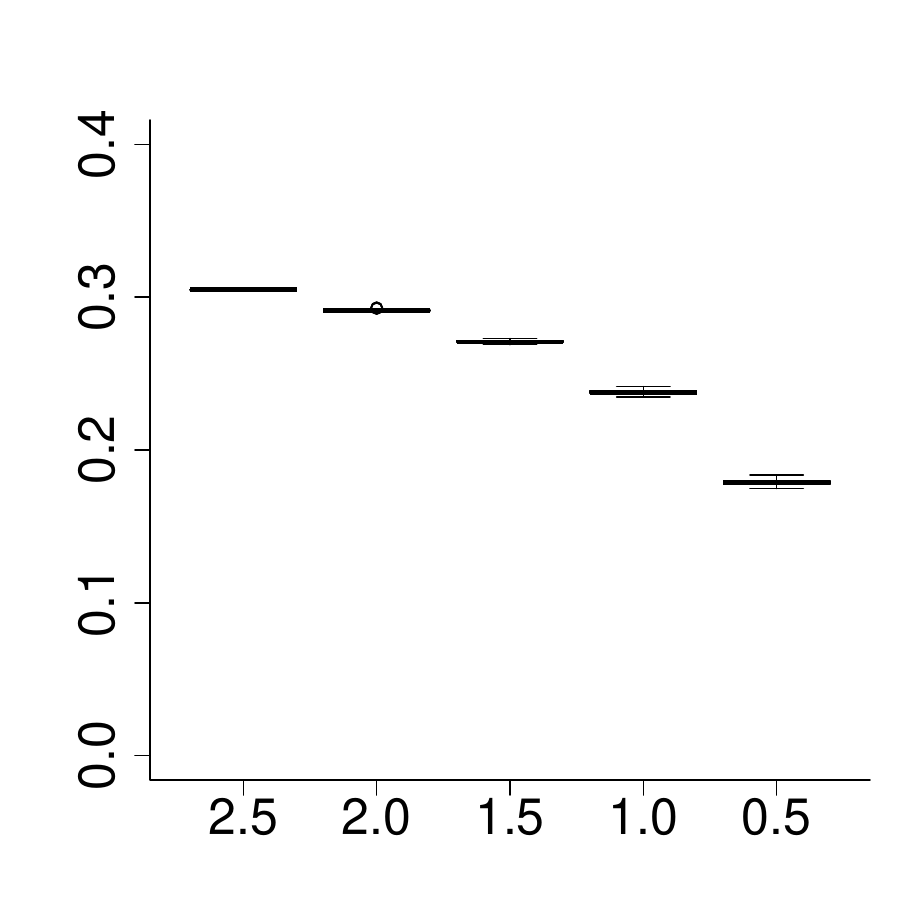}} \\[1ex]

        \raisebox{0.08\textwidth}{\rotatebox{90}{\small Weibull}} &
        \subfloat{\includegraphics[width=0.19\textwidth,height=0.13\textheight,keepaspectratio]{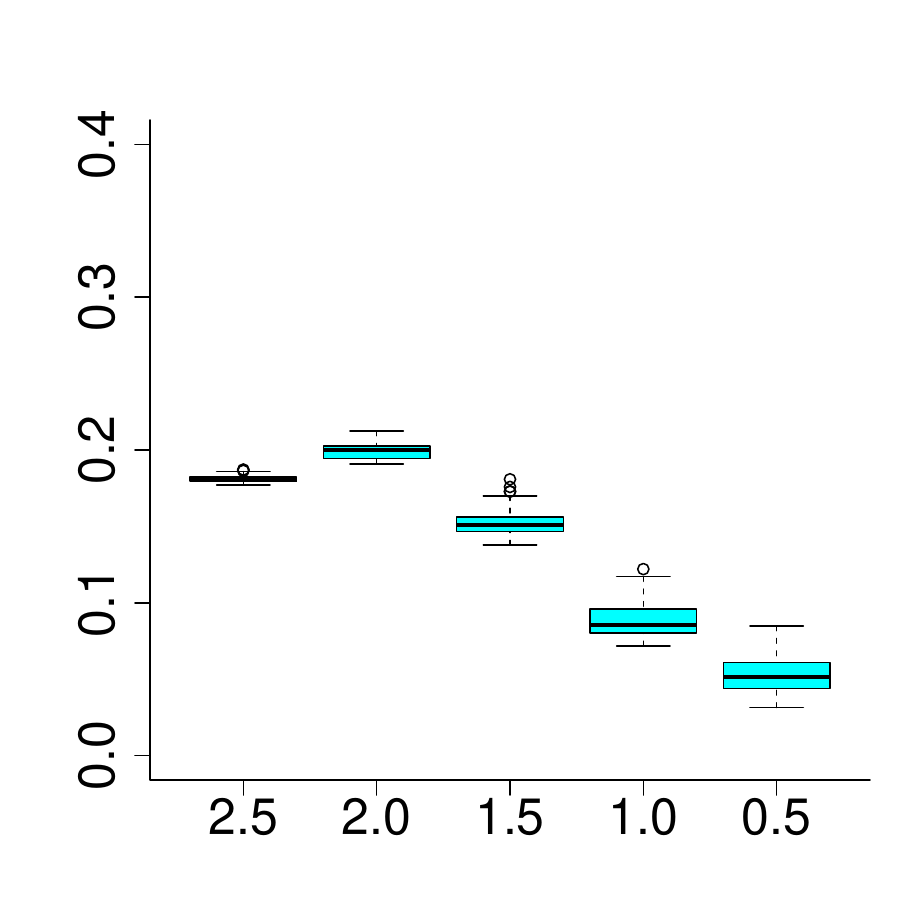}} &
        \subfloat{\includegraphics[width=0.19\textwidth,height=0.13\textheight,keepaspectratio]{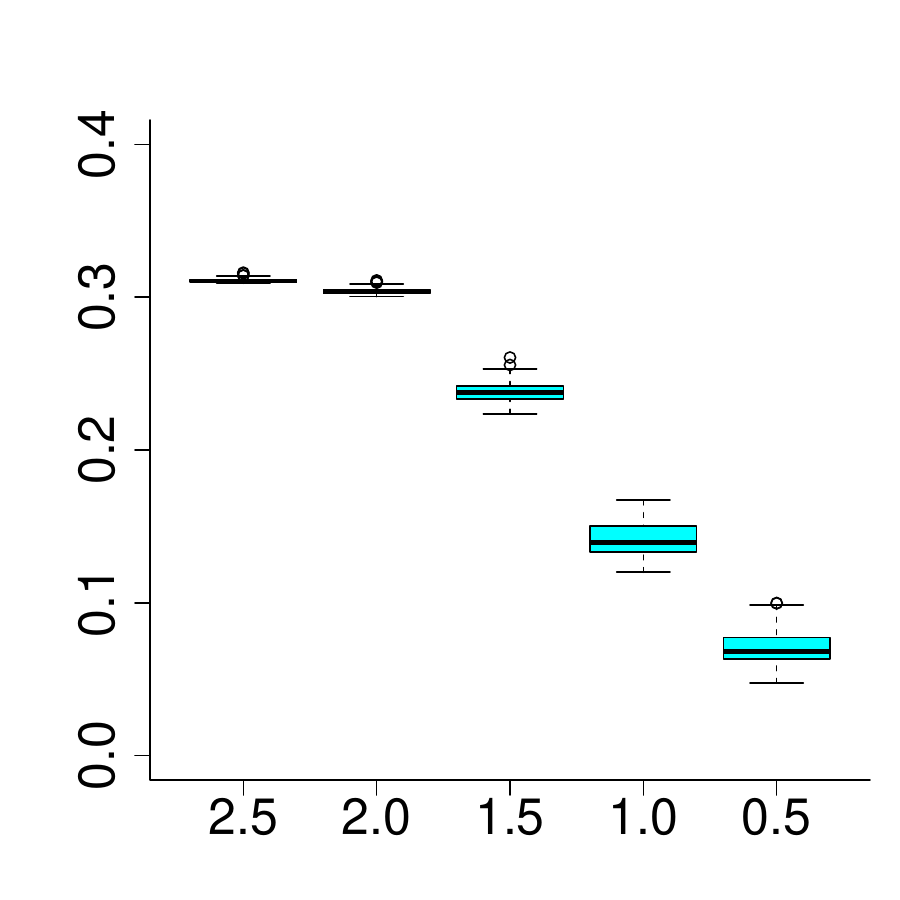}} &
        \subfloat{\includegraphics[width=0.19\textwidth,height=0.13\textheight,keepaspectratio]{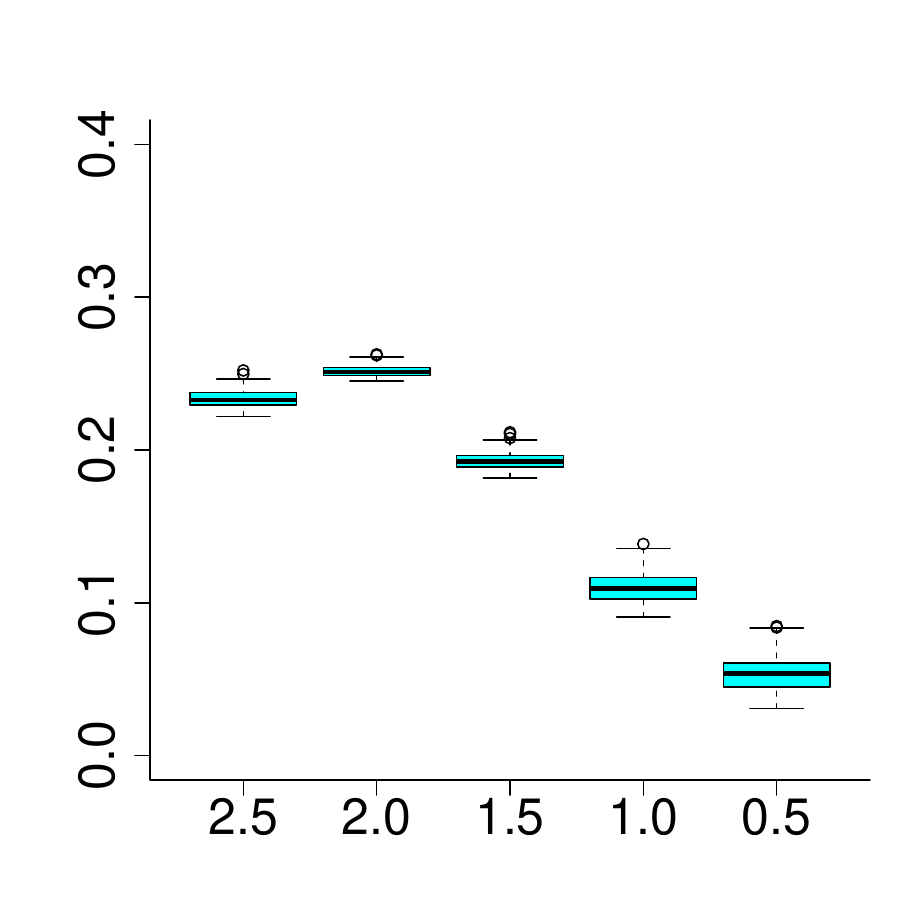}} &
        \subfloat{\includegraphics[width=0.19\textwidth,height=0.13\textheight,keepaspectratio]{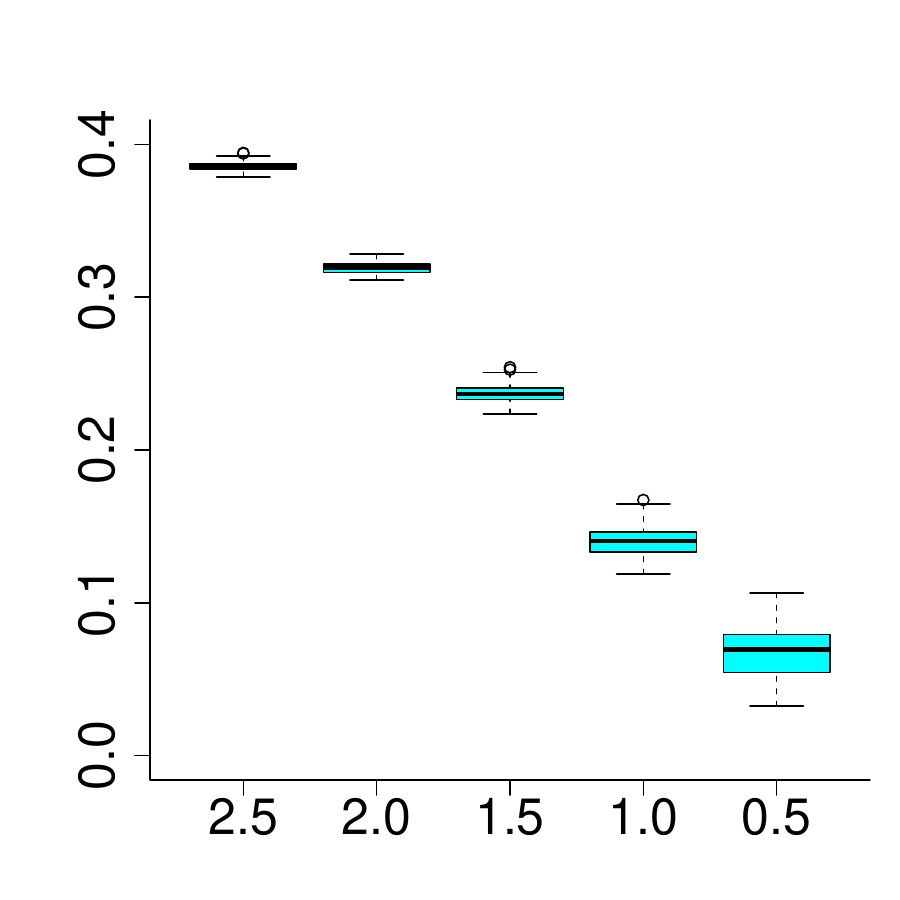}} \\[1ex]

        \raisebox{0.08\textwidth}{\rotatebox{90}{\small Pareto}} &
        \subfloat{\includegraphics[width=0.19\textwidth,height=0.13\textheight,keepaspectratio]{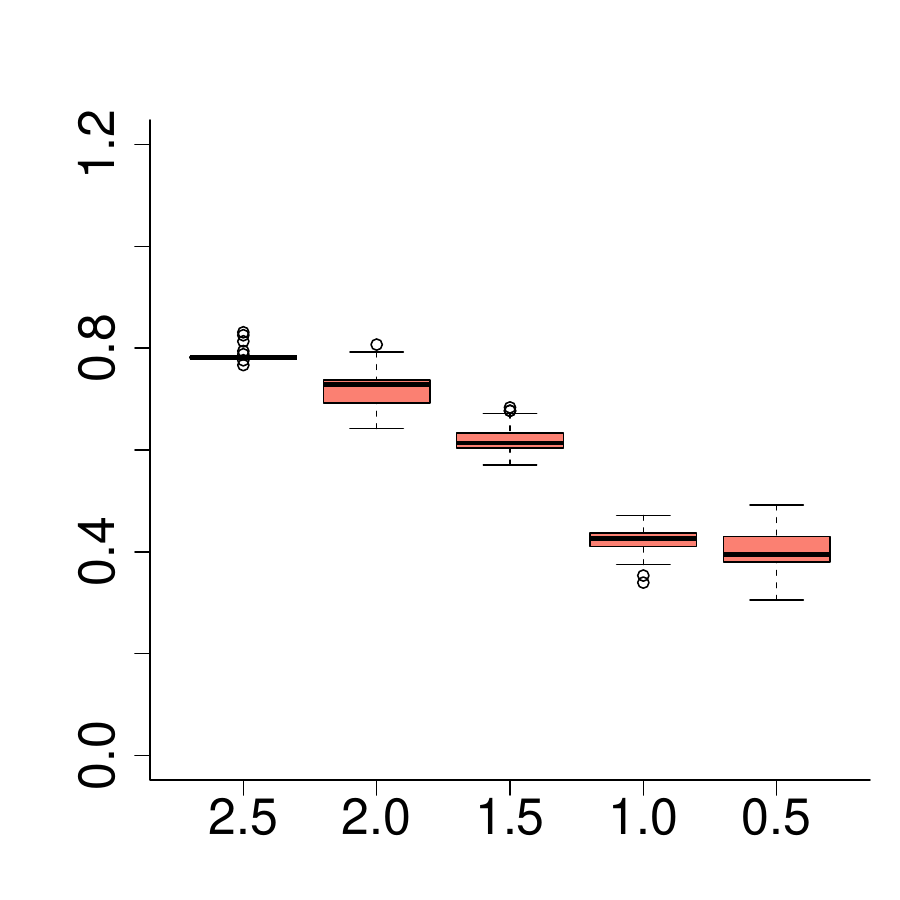}} &
        \subfloat{\includegraphics[width=0.19\textwidth,height=0.13\textheight,keepaspectratio]{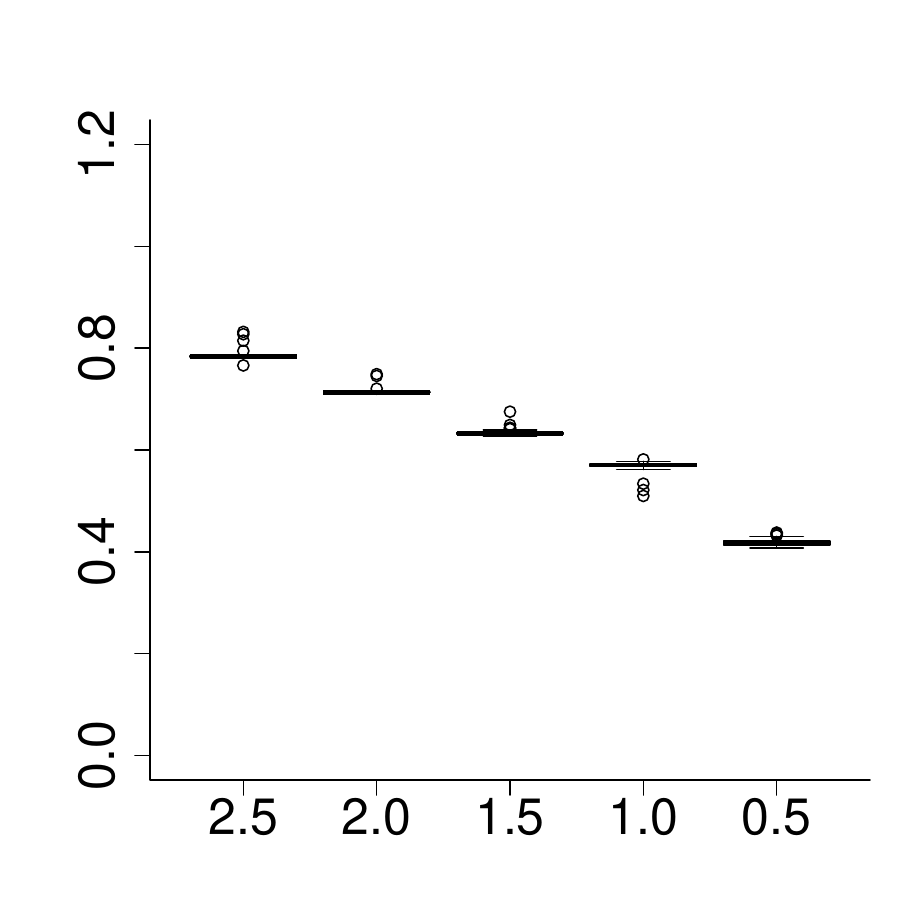}} &
        \subfloat{\includegraphics[width=0.19\textwidth,height=0.13\textheight,keepaspectratio]{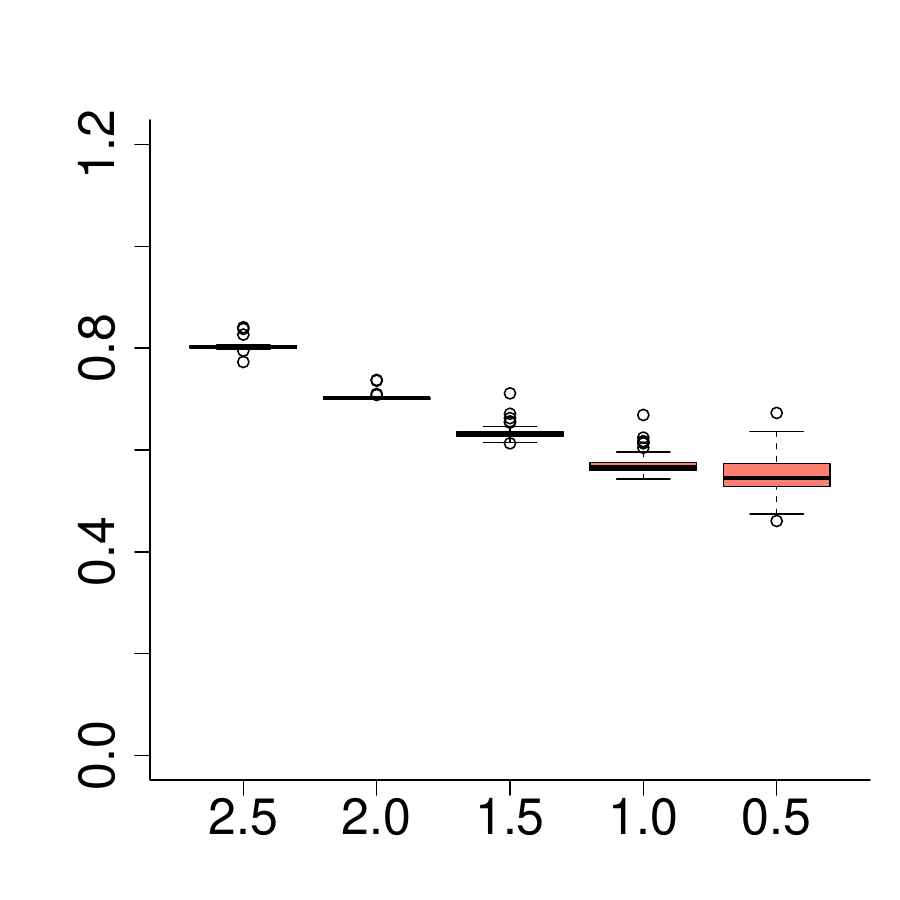}} &
        \subfloat{\includegraphics[width=0.19\textwidth,height=0.13\textheight,keepaspectratio]{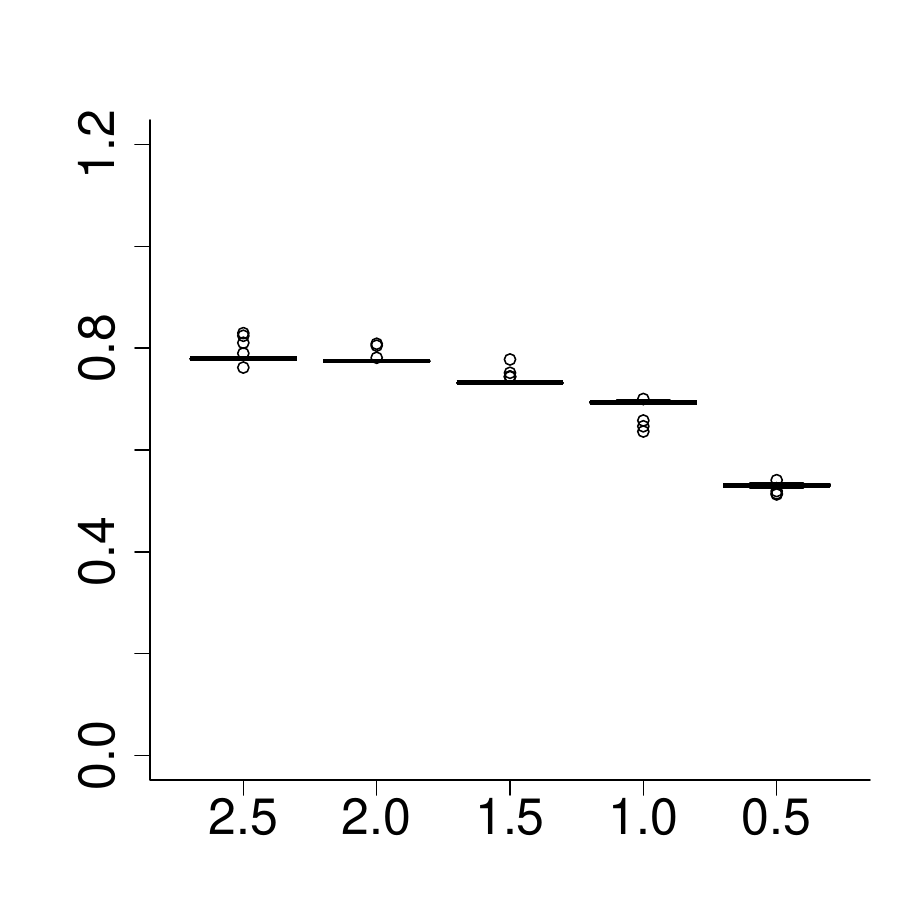}} \\
    \end{tabular}
    
    \caption{Simulation results under grid width variation ($\delta/\sigma \in \{0.5, 1.0, 1.5, 2.0, 2.5\}$) with a fixed sample size ($n = 10^3$). The plots display the $L_2$ norm between the true and estimated densities for the Laplace, chi-square, log-normal, Weibull, and Pareto distributions, using the proposed MALC and three kernel-based methods.}
    \label{fig:pdfgw2}
\end{figure}

Figures~\ref{fig:pdfgw1} and~\ref{fig:pdfgw2} report the performance of the proposed method and the kernel-based approaches as the (standardized) grid-width $\delta$ varies. Across the ten evaluated distributions, MALC is either the least sensitive method (i.e., its performance varies less across $\delta$ values than the other methods) or exhibits a similar level of sensitivity to the competing approaches in nine cases. Specifically, MALC shows the most stable performance for the log-concave beta and gamma distributions, comparable sensitivity for seven additional distributions, and greater sensitivity only for the non-log-concave log-normal distribution. These results suggest that MALC is a reliable and robust choice for density estimation across a wide range of grid-width settings.  

We end this section with two further observations. First, the MALC method allows for both the smoothed and non-smoothed density estimators. In our simulation results, we focus only on the non-smoothed version. When conducting simulations using the smoothed version, we observed a small improvement but also occasional numerical errors; therefore, to ensure stable, fully reproducible comparisons, we report only the non-smoothed version. Secondly, we note that MALC can achieve favourable performance even when the log-concavity assumption does not hold. This holds under small or medium sample sizes. Notably, similar behaviour has been observed previously; see, e.g., \citet{Cule08}.

\begin{remark}
It is natural to ask if it is possible to test if the underlying density is log-concave.  Due to the lack of identifiability, this is not easy.  However, one can use the test described in \citet{chenSam2013} to test if $p_0$ is log-concave or if $\widehat f_0$ is log-concave, giving some information in this direction.   
\end{remark}

\captionsetup[subfigure]{justification=centering}

\captionsetup[subfigure]{justification=centering}

\subsection{Data Analysis}
\label{subsec:dataanalysis}

Next, we apply our method to a survival data set obtained from the \cite{HMD2026}. 
Specifically, we analyzed age-specific mortality counts from 2010 to 2020 for six countries that span four continents: Canada, the United States, Japan, the Republic of Korea, Norway, and Australia. The data consist of grouped frequencies of deaths, rather than individual-level records, covering ages 0 through 110 in one-year intervals and stratified by sex. The Canadian dataset is presented as a demonstration of the structure in Table \ref{table3}.

\begin{table}[ht]
\centering
\caption{Human mortality data for Canada in $2010-2020$}
\begin{tabular}{l@{\hspace{3cm}}ccc}
\hline
\textbf{Age} & \textbf{Female} & \textbf{Male} & \textbf{Total} \\
\hline
$[0, 1)$ & 8065 & 9834 & 17899 \\
$[1, 2)$ & 519 & 601 & 1120 \\
$\vdots$ & $\vdots$ & $\vdots$ & $\vdots$ \\
$[109, 110)$ & 89 & 7 & 96 \\
\hline
\end{tabular}
\label{table3}
\end{table}

\begin{figure}[h!]
    \centering
    \newcommand{\figscale}{0.17}
    \newcommand{\colsep}{0.02\textwidth}

    \begin{minipage}{0.40\textwidth}
        \centering
        \subfloat[\footnotesize{Canada}\label{bs1:canada}]{\includegraphics[scale=\figscale]{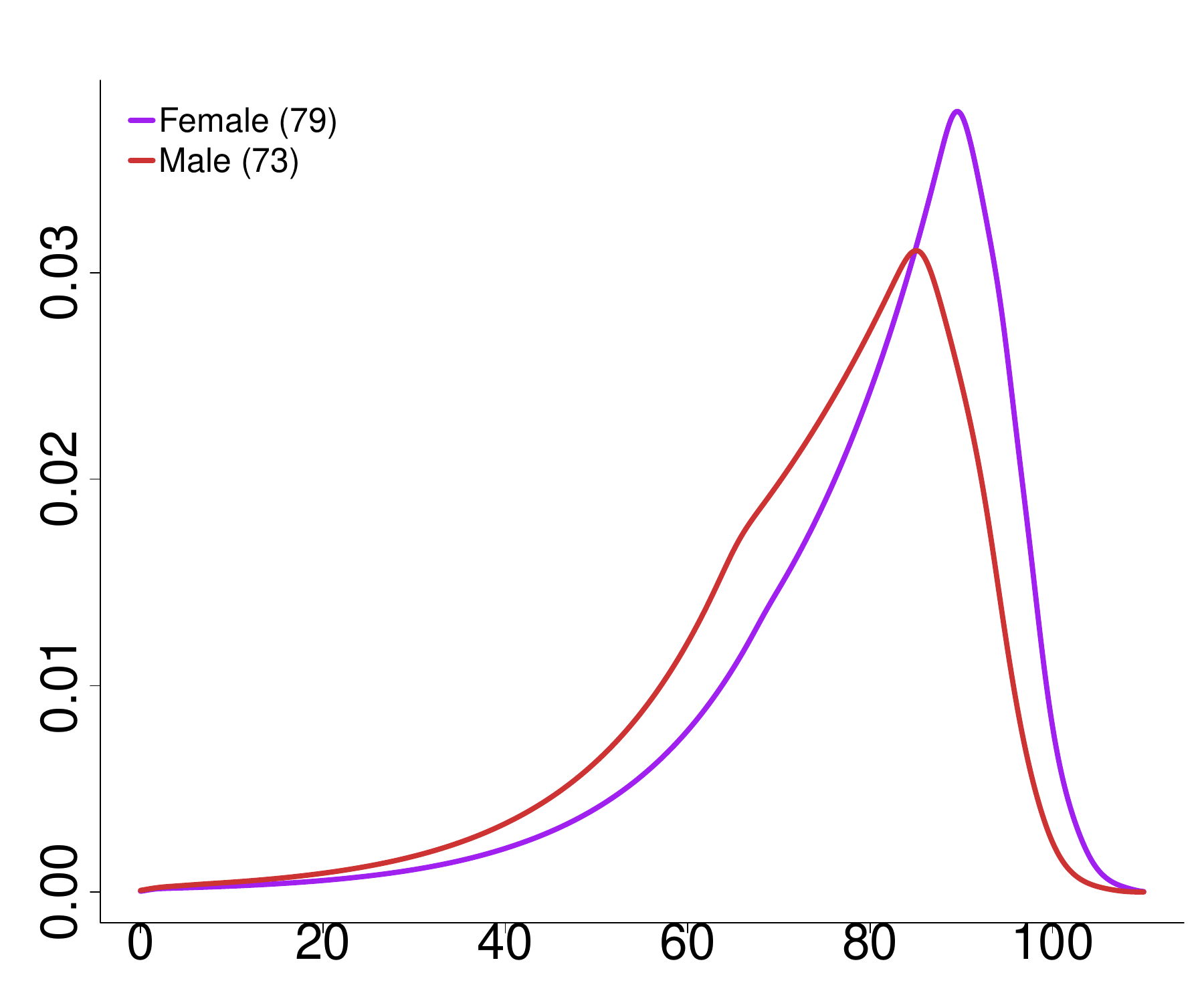}}
    \end{minipage}
    \hspace{\colsep}
    \begin{minipage}{0.40\textwidth}
        \centering
        \subfloat[\footnotesize{USA}\label{bs1:usa}]{\includegraphics[scale=\figscale]{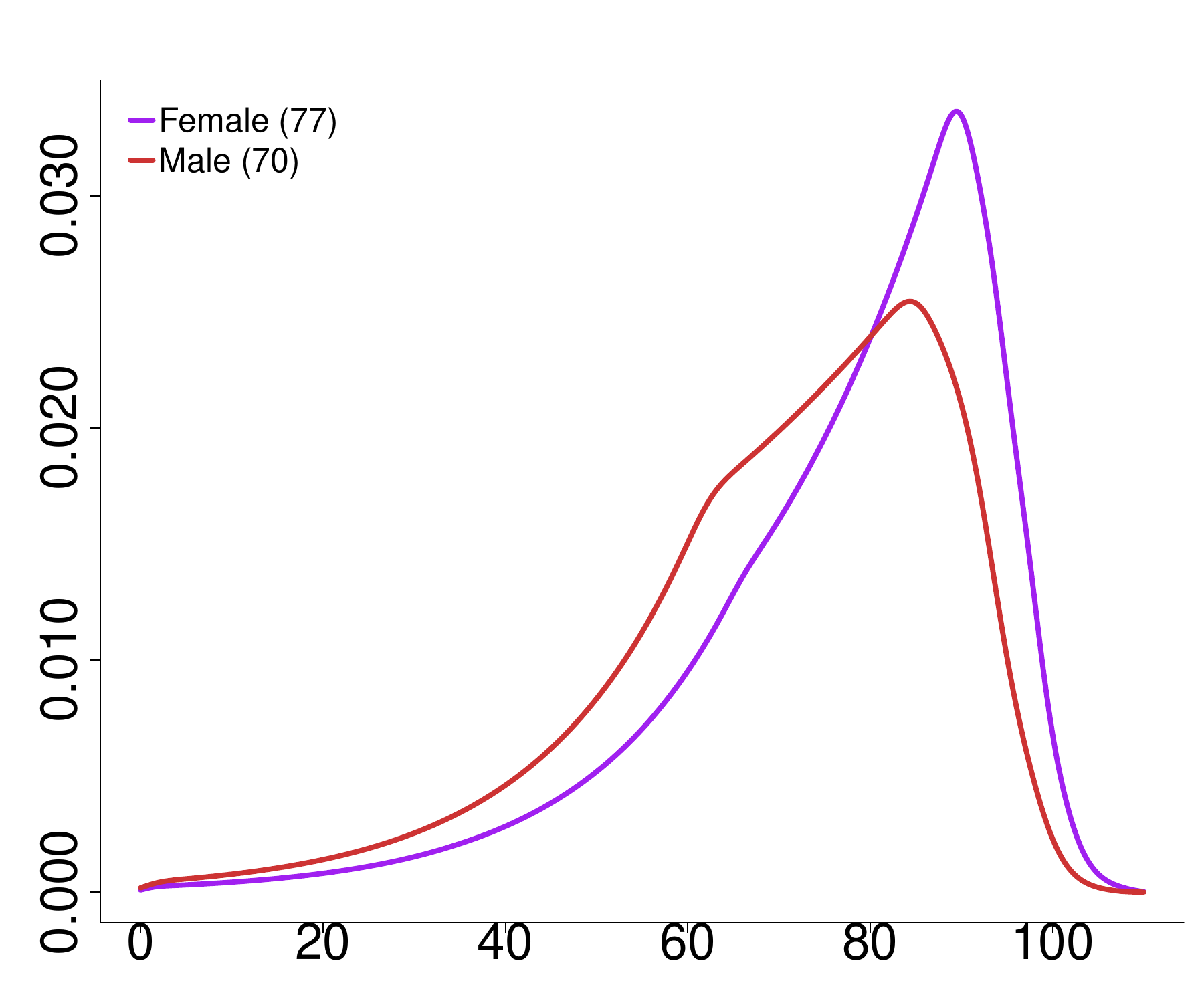}}
    \end{minipage}

    \vspace{0.4em}

    \begin{minipage}{0.40\textwidth}
        \centering
        \subfloat[\footnotesize{Japan}\label{bs1:japan}]{\includegraphics[scale=\figscale]{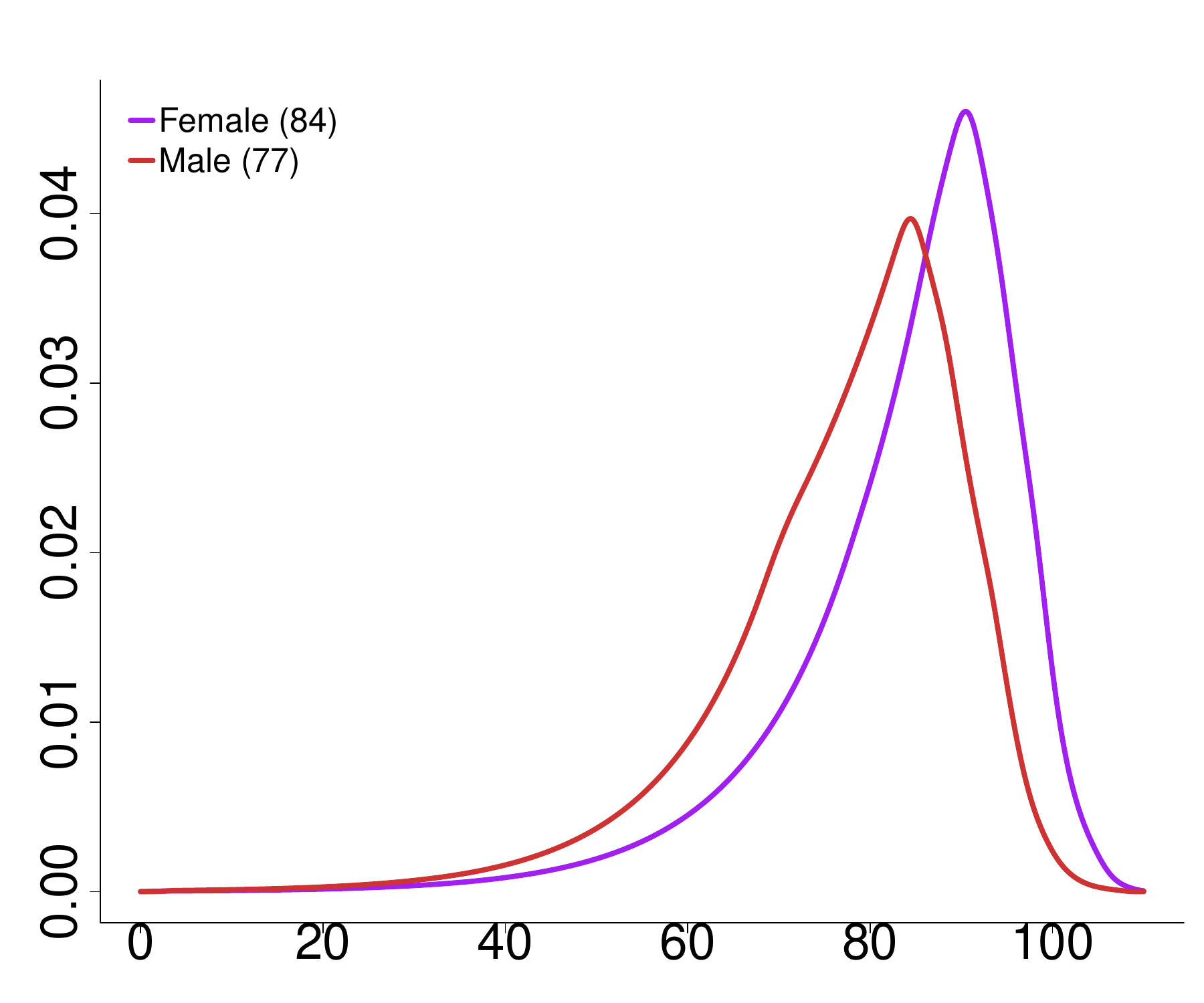}}
    \end{minipage}
    \hspace{\colsep}
    \begin{minipage}{0.40\textwidth}
        \centering
        \subfloat[\footnotesize{Republic of Korea}\label{bs1:korea}]{\includegraphics[scale=\figscale]{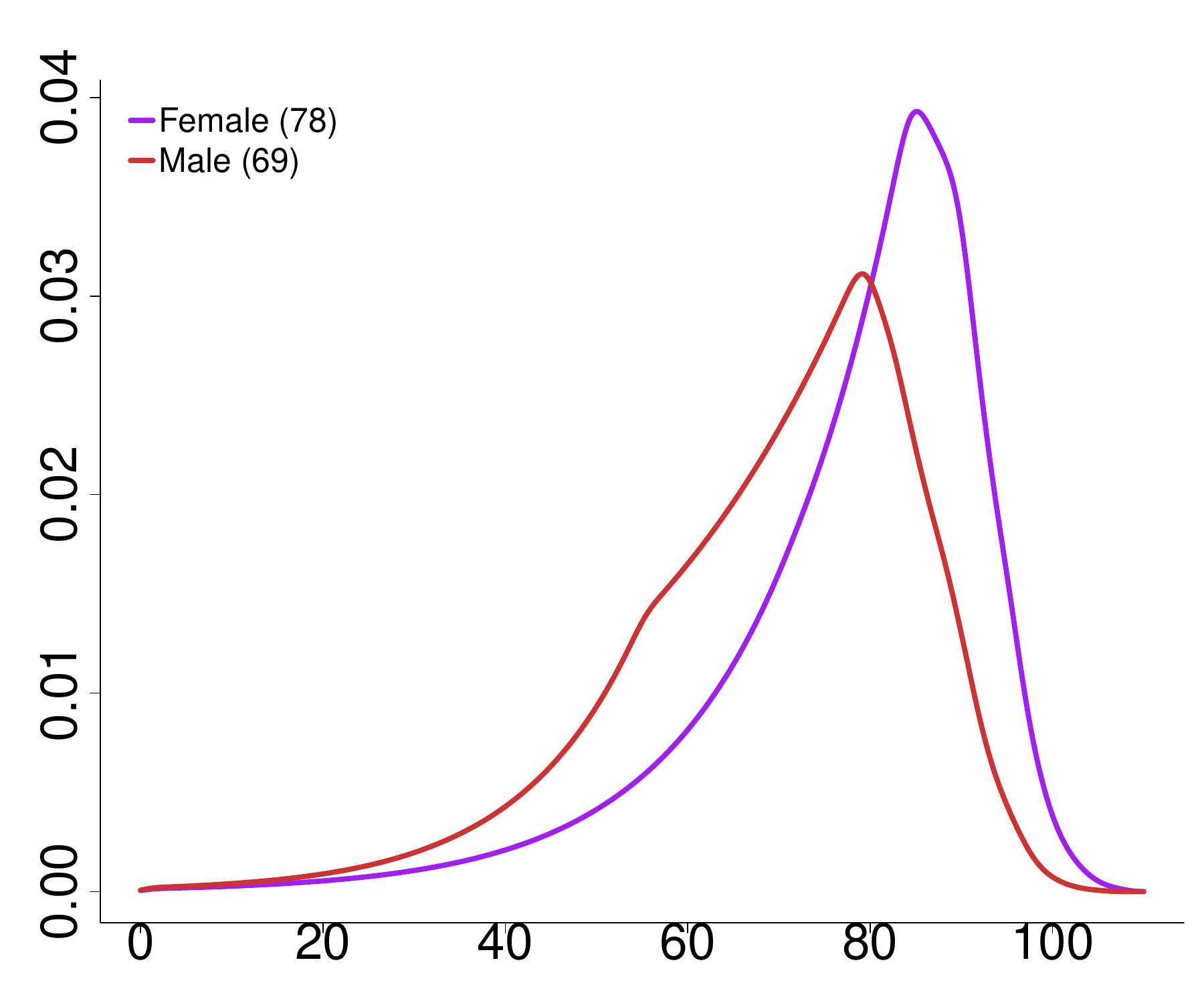}}
    \end{minipage}

    \vspace{0.4em}

    \begin{minipage}{0.40\textwidth}
        \centering
        \subfloat[\footnotesize{Norway}\label{bs1:norway}]{\includegraphics[scale=\figscale]{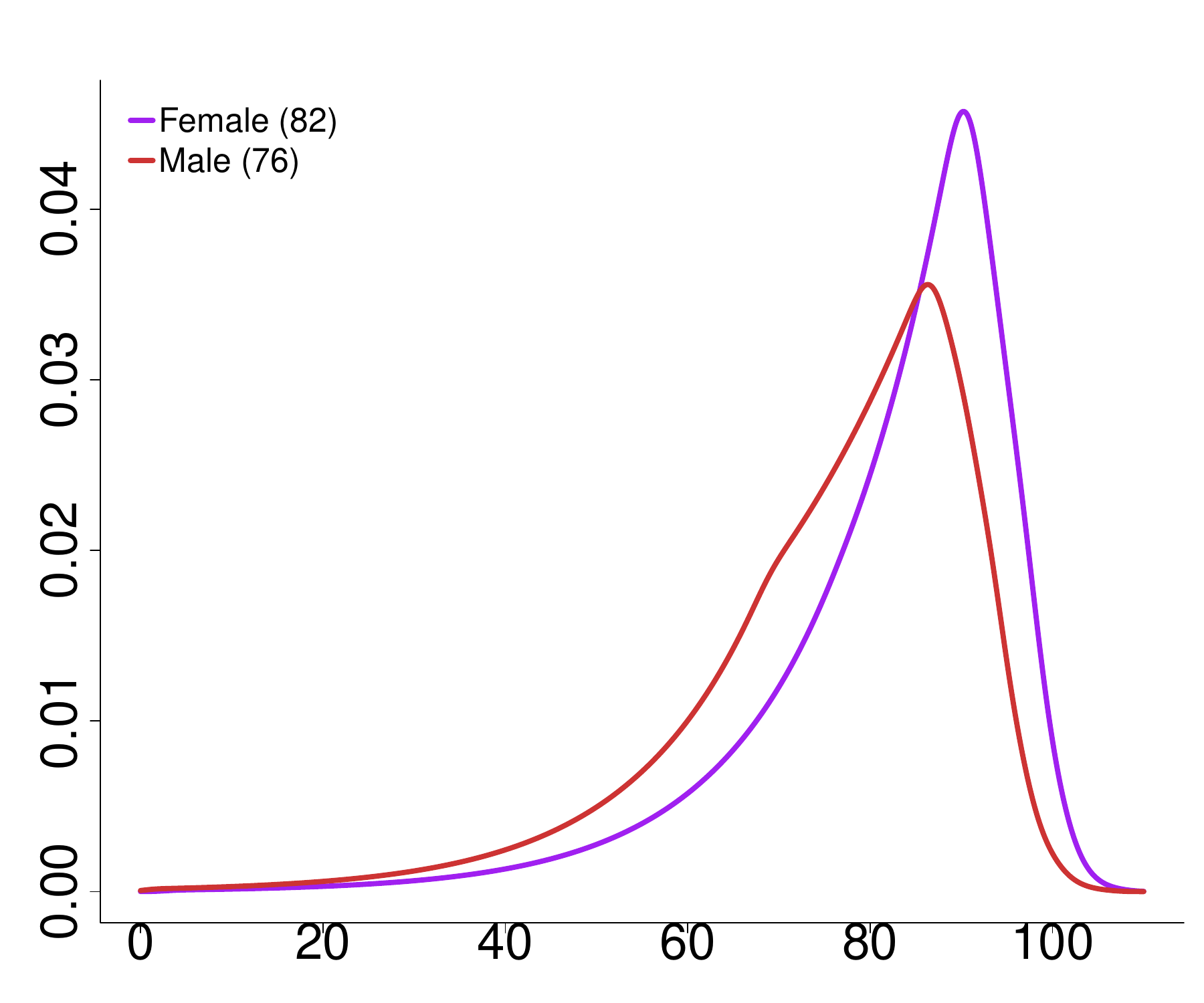}}
    \end{minipage}
    \hspace{\colsep}
    \begin{minipage}{0.40\textwidth}
        \centering
        \subfloat[\footnotesize{Australia}\label{bs1:australia}]{\includegraphics[scale=\figscale]{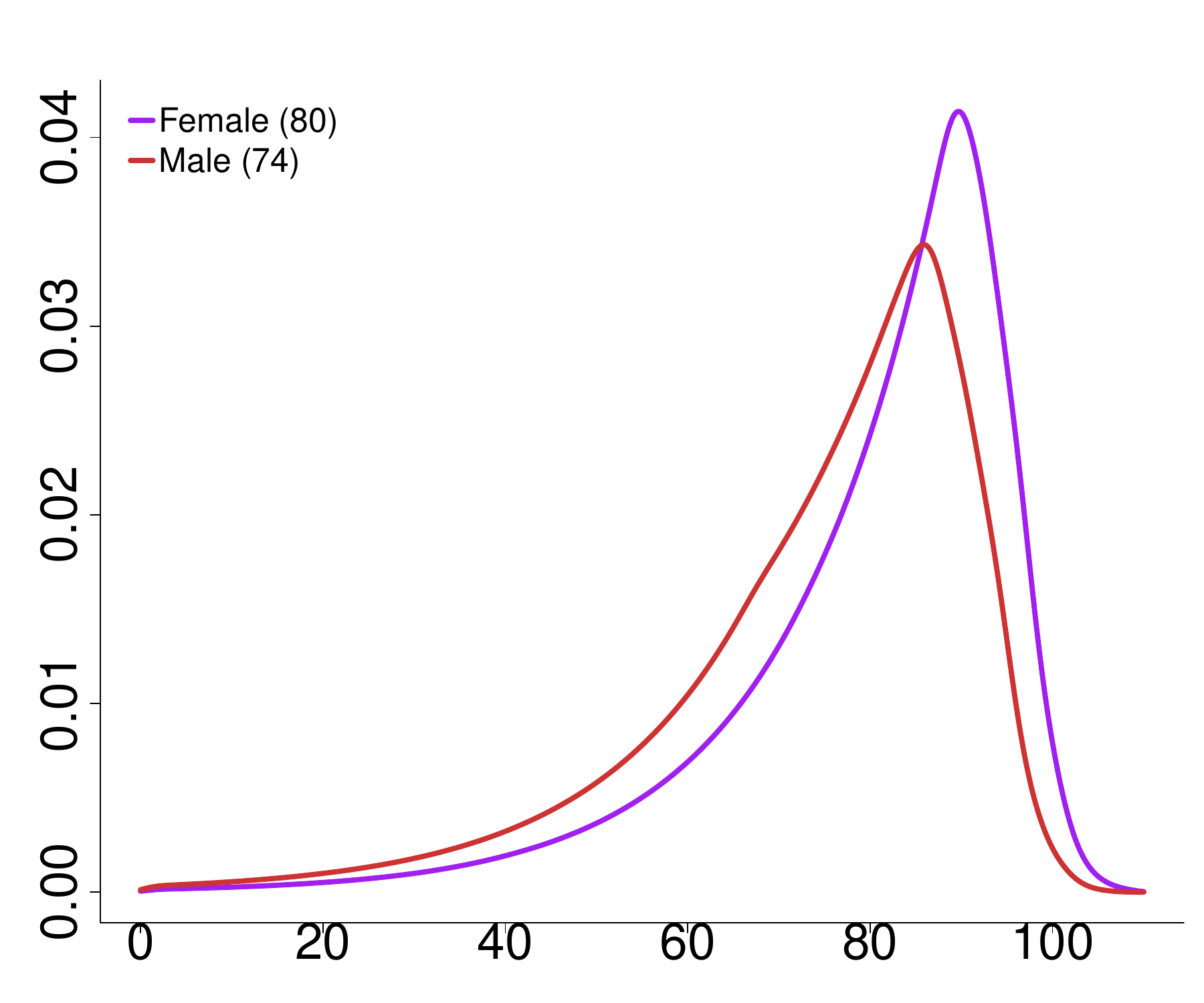}}
    \end{minipage}

    \caption{Estimated mortality densities for females and males in selected countries using HMD data from 2010 to 2020. The purple curve represents females and the red curve represents males.}
    \label{fig:malefemale}
\end{figure}

We first analyzed female and male mortality data separately to assess sex-based differences. Across all six countries, females exhibited higher mortality probabilities at older ages than males, reflecting their greater likelihood of survival to later ages. This is consistent with longer female life expectancy regardless of country characteristics. The corresponding estimates are reported in Figure \ref{fig:malefemale}.

\begin{figure}[!h]
    \centering
    \newcommand{\figscalehaz}{0.17}
    \newcommand{\colsephaz}{0.02\textwidth}

    \begin{minipage}{0.40\textwidth}
        \centering
        \subfloat[\footnotesize{Canada}\label{haz:canada}]{\includegraphics[scale=\figscalehaz]{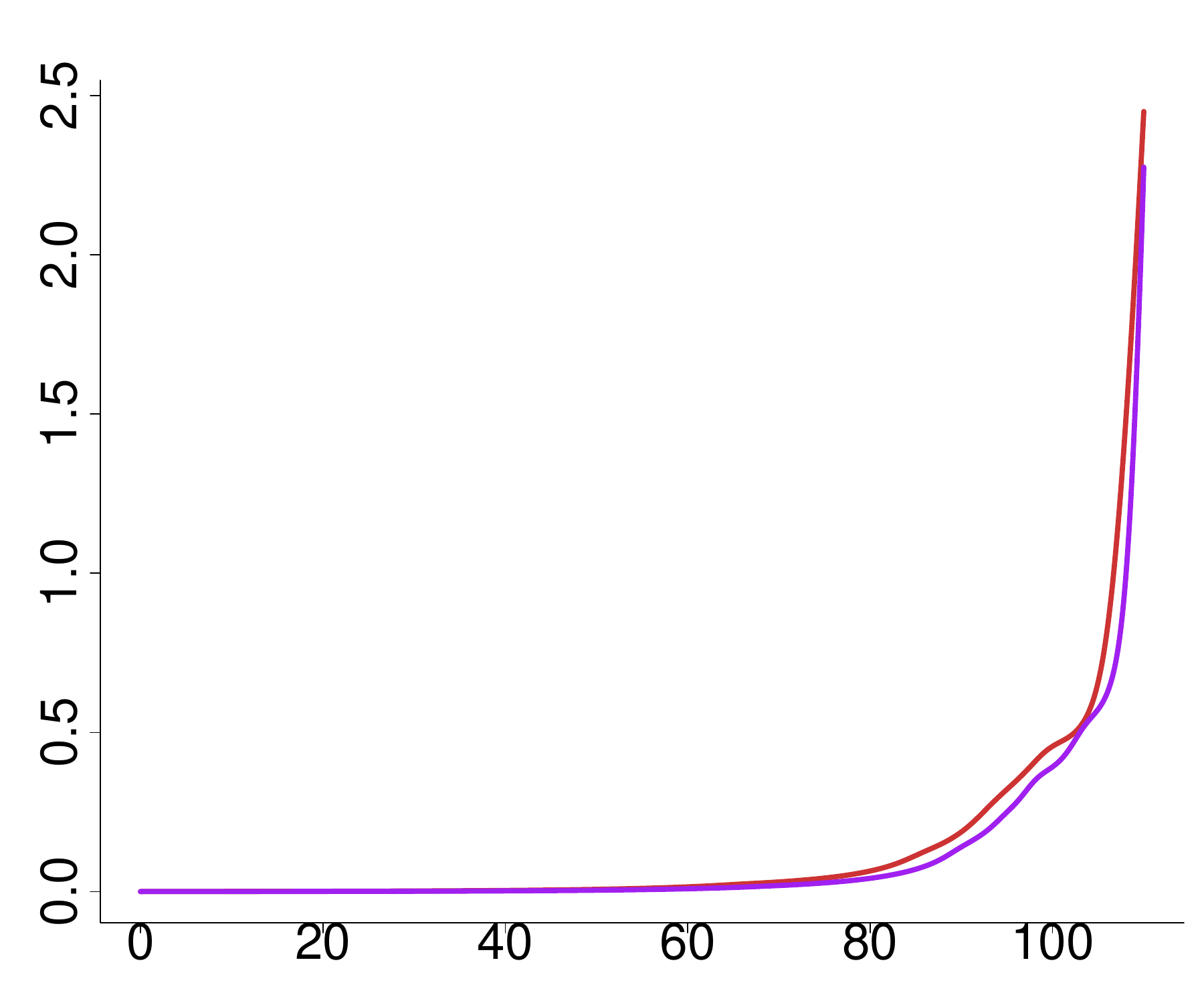}}
    \end{minipage}
    \hspace{\colsephaz}
    \begin{minipage}{0.40\textwidth}
        \centering
        \subfloat[\footnotesize{USA}\label{haz:usa}]{\includegraphics[scale=\figscalehaz]{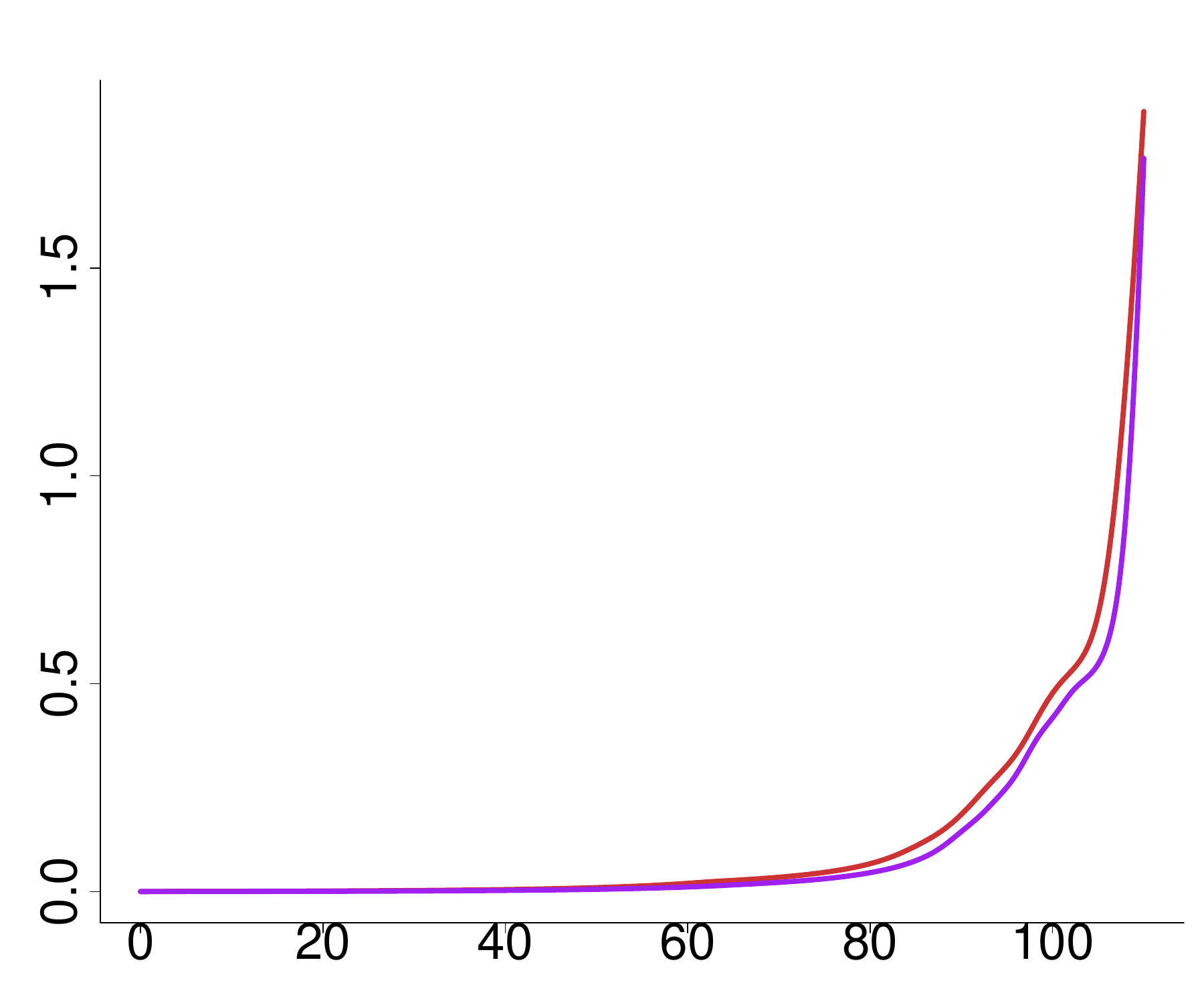}}
    \end{minipage}

    \vspace{0.4em}

    \begin{minipage}{0.40\textwidth}
        \centering
        \subfloat[\footnotesize{Japan}\label{haz:japan}]{\includegraphics[scale=\figscalehaz]{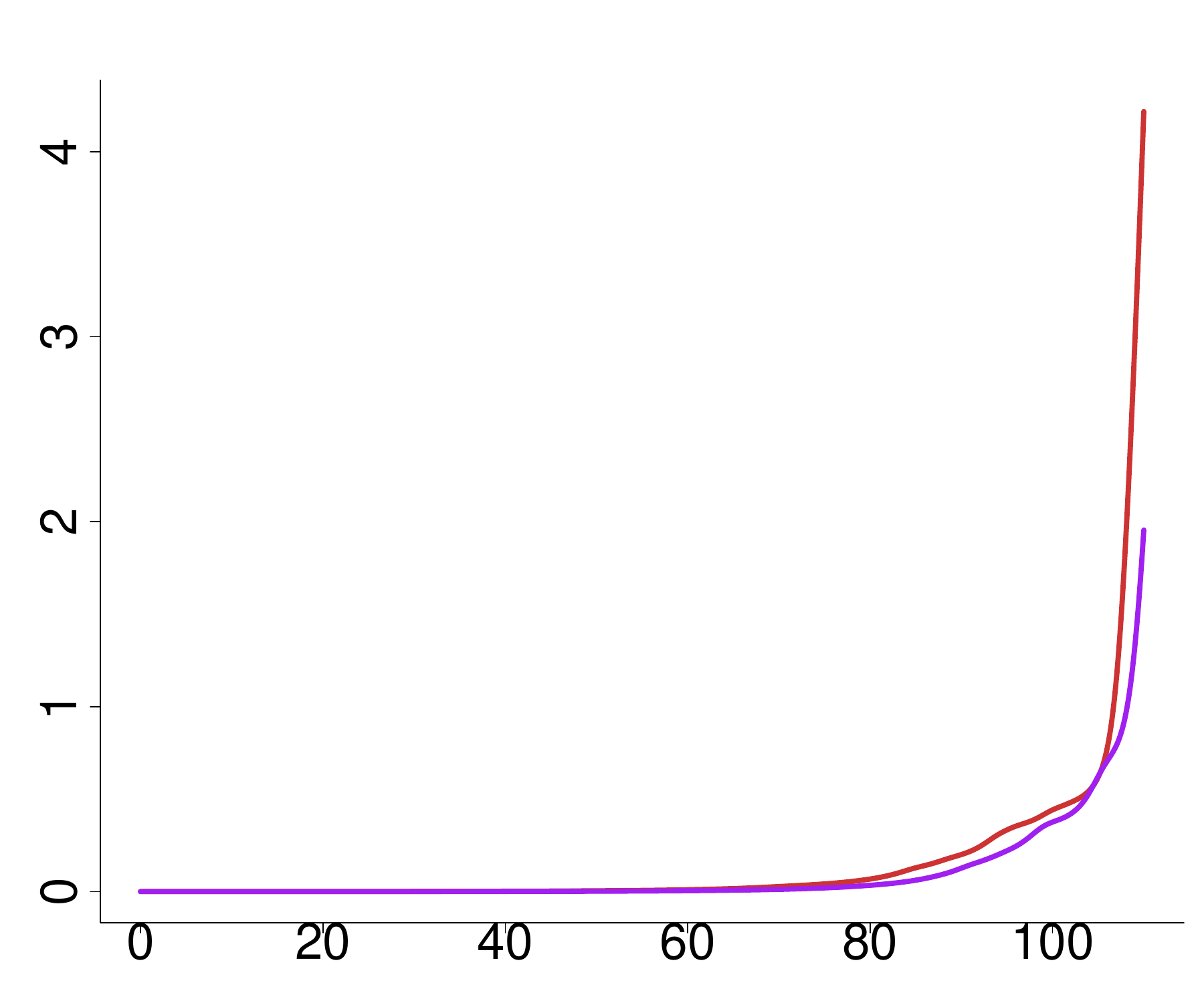}}
    \end{minipage}
    \hspace{\colsephaz}
    \begin{minipage}{0.40\textwidth}
        \centering
        \subfloat[\footnotesize{Republic of Korea}\label{haz:korea}]{\includegraphics[scale=\figscalehaz]{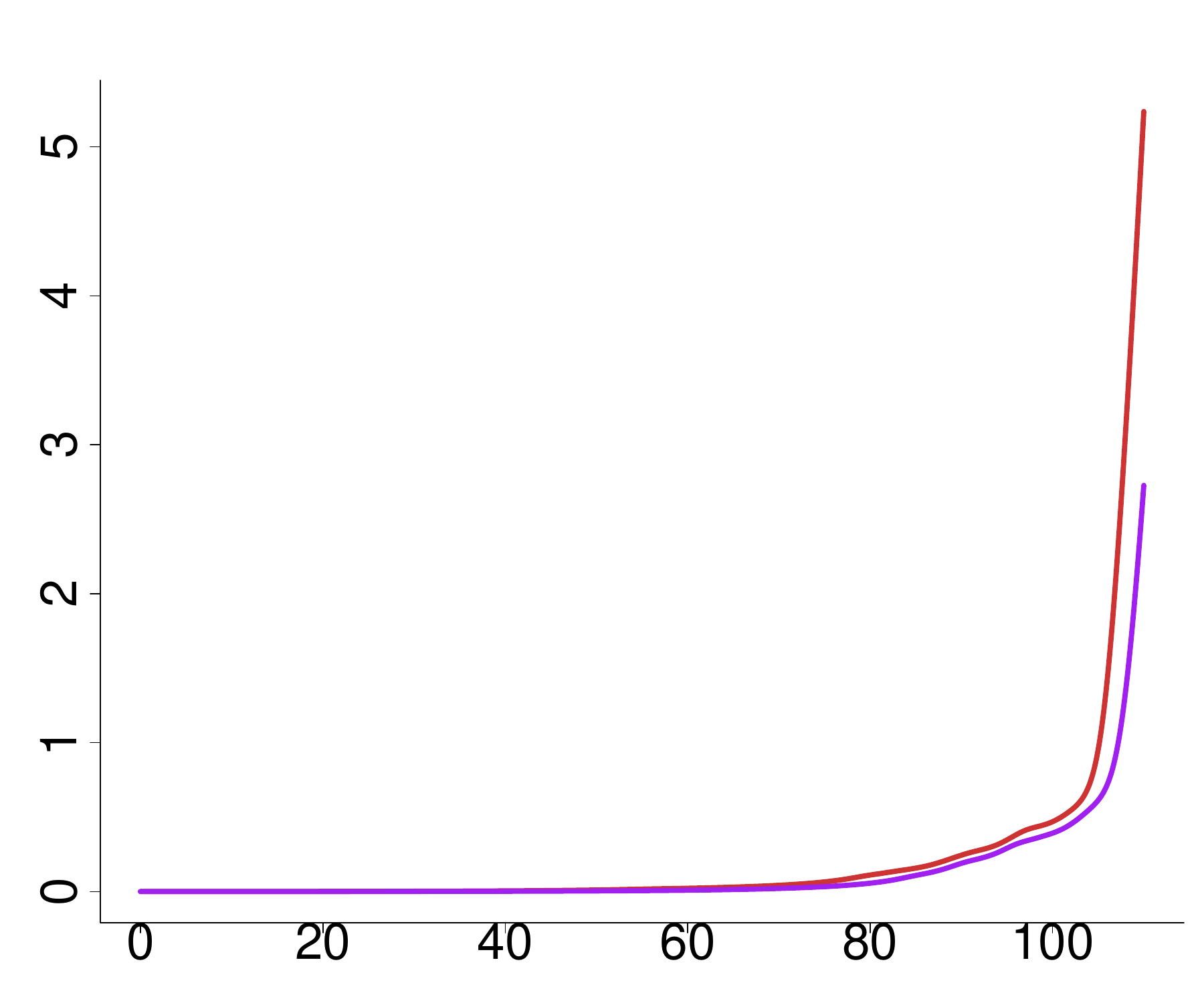}}
    \end{minipage}

    \vspace{0.4em}

    \begin{minipage}{0.40\textwidth}
        \centering
        \subfloat[\footnotesize{Norway}\label{haz:norway}]{\includegraphics[scale=\figscalehaz]{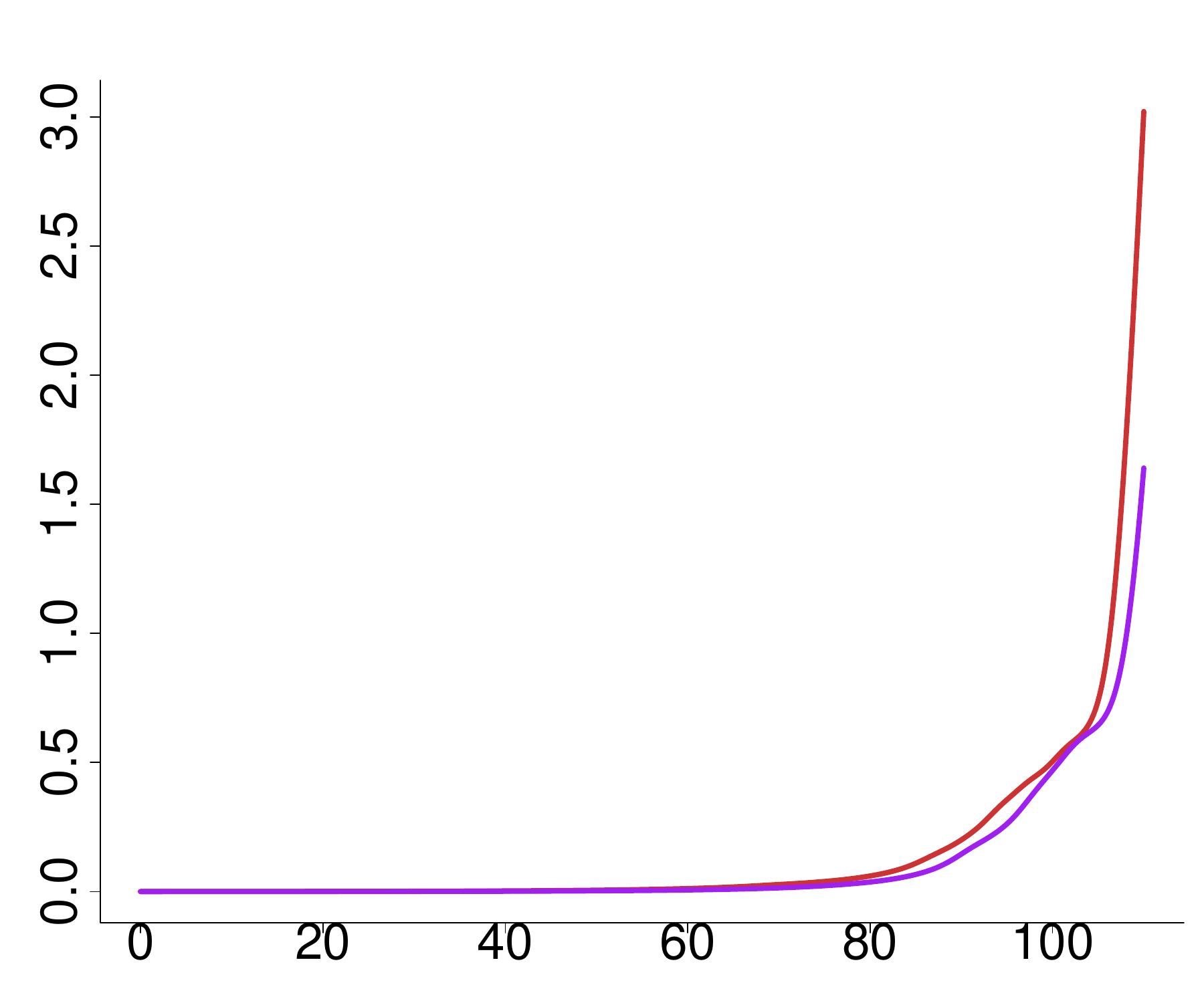}}
    \end{minipage}
    \hspace{\colsephaz}
    \begin{minipage}{0.40\textwidth}
        \centering
        \subfloat[\footnotesize{Australia}\label{haz:australia}]{\includegraphics[scale=\figscalehaz]{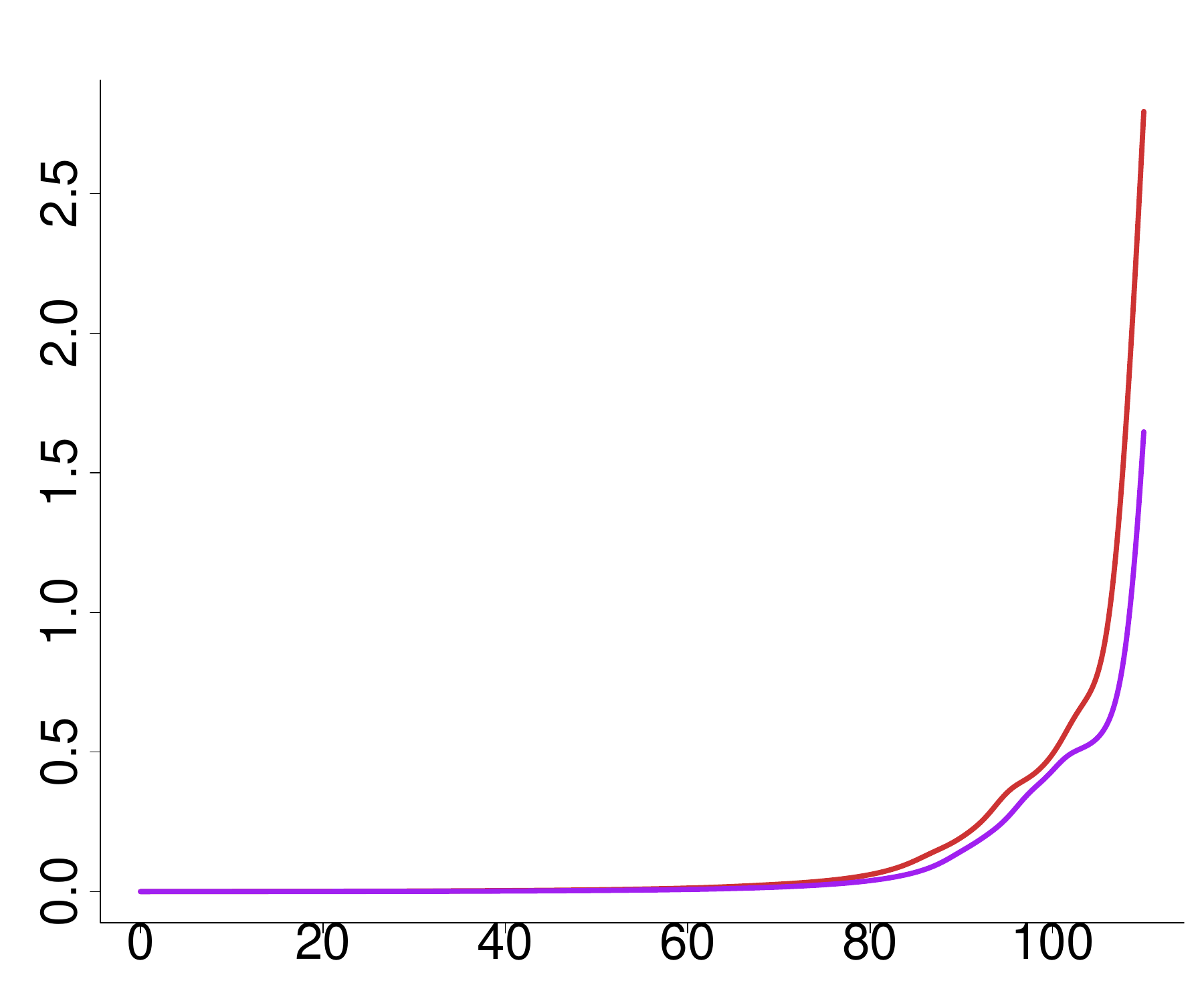}}
    \end{minipage}

    \caption{Estimated mortality hazard functions for females and males in selected countries using Human Mortality Data from 2010 to 2020. The purple curve represents females and the red curve represents males.}
    \label{fig:hazard_malefemale}
\end{figure}

{One important feature of the hazard function of a log-concave density is that it is necessarily non-decreasing.  As such, our method also provides a smooth estimate of a non-decreasing hazard.   The hazard functions corresponding to the densities in Figure~\ref{fig:malefemale} are shown in Figure~\ref{fig:hazard_malefemale}.  An increasing hazard function for lifetimes is common, and is known to hold for human lifetimes but only after some initial period.  As such, we do not promote this method if one is interested in early ages, and using the method condition on survival to -- say, age 5 -- would be warranted.}

\begin{figure}[h!]
    \centering
        \centering
        \includegraphics[scale = 0.2]{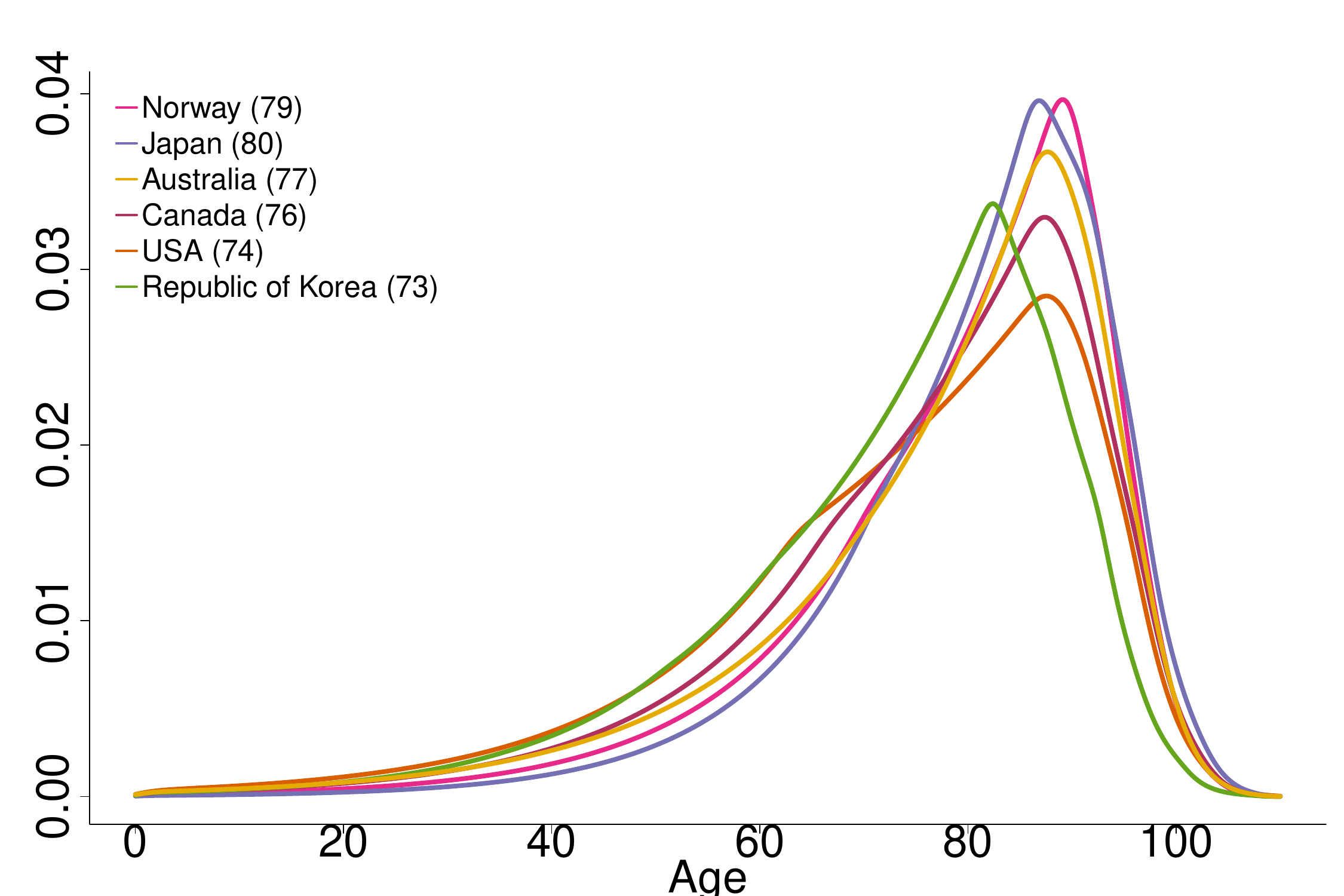}
    \caption{Estimated mortality densities for the total population in selected countries using Human Mortality Data from 2010 to 2020.}
    \label{fig:total}
\end{figure}

After examining sex-based differences, we compared the estimated densities across the countries, with the results shown in Figure \ref{fig:total}. At the population level, the distribution of deaths in Japan and Norway peaked at higher ages, with mean ages at death of approximately 80 and 79, respectively. These were followed by Australia, Canada, the Republic of Korea, and the United States, where the density of deaths concentrated at comparatively younger ages. This means, Japan and Norway has the highest life expectancy, whereas United States has the lowest among the compared countries. 

\section{Discussion and conclusion}
\label{sec:discussion}

We have proposed an log-concave density estimator for grouped data when the underlying distribution is assumed to be log-concave. Therefore, in scientific experiments, medical investigations, demographical or social studies, and survival analysis \cite{ferson2004summary, lindsey1998methods, osegueda2002non, zhang2010interval}, where the information about the distribution is limited or lacking completely, our method provides an efficient and robust way to handle grouped data with a broader range of applications.

The main advantage of the proposed estimator is that it is fully automatic, as no bandwidth selection is required.   Notably, performance of the MALC is favourable even when the true distribution is misspecified.

The log-concave assumption is a natural one.   Many common distribution families are log-concave (e.g.  Gaussian, gamma, Weibull).  The main restrictions of this family are two-fold:  first, the tails of the distribution are at most exponential, and therefore heavy-tailed distributions are excluded.   Secondly, all log-concave distributions are unimodal.   When the true density is multimodal and complete data is observed, one can use a simple mixture approach to obtain a multimodal estimator.  

Our simulations in Section~\ref{subsec:simresults} show that the MALC performs as well or outperforms the main non-parametric competitors.   All of the competitors are kernel-based and require bandwidth selection, whereas the MALC is fully automatic.   The current implementation of the MALC is limited to the univariate case with a uniform grid-width.   These extensions will be the subject of future work.  There is some hope that an appropriate implementation for a non-uniform grid-width can also be used to improved the performance of the MALC presented here when the grid with is large, but implementing an augmentation strategy.

\bibliographystyle{cas-model2-names}
\bibliography{references2}

\bigskip
\noindent\textbf{Supplementary Material.} The alternative methods, proofs of all theoretical results, and a review of competing nonparametric methods are provided in the Supplementary Material.

\end{document}